%% file: manuscrit.tex
\documentclass[10pt,a4paper]{article}

\usepackage{a4wide,multicol}
\usepackage{pdfpages}
\usepackage{import}
\usepackage{minitoc,url}
\usepackage{amsfonts,amsmath,amssymb,amsthm,enumerate,bbm,twoopt,tabularx,ifthen,color}
\usepackage[utf8]{inputenc}
\usepackage[T1]{fontenc}
\usepackage{lmodern}
\usepackage[french,english]{babel}
\usepackage[cyr]{aeguill}
\usepackage{xspace}
\usepackage[dvips]{epsfig}
\usepackage[%
backend=bibtex,%
bibstyle=authoryear,%
citestyle=authoryear-comp,%
natbib,%
language=auto,%
block=space,%
isbn=false,%
url=false,%
doi=false,%
eprint=false,%
firstinits=true,%
autopunct=true,%
mcite,%
sortcites,%
autopunct,%
maxcitenames=2,%
mincitenames=1,%
maxbibnames=25,%
]{biblatex}
\usepackage[autostyle,babel]{csquotes}
\usepackage{xpatch}
\usepackage{fancyhdr,framed}
\usepackage{pstricks,pst-fill,pst-grad}
\usepackage[draft]{fixme}
\usepackage{graphicx}
\usepackage{capt-of}
\usepackage{placeins}
\usepackage[french,onelanguage]{algorithm2e}
\usepackage{appendix}
\usepackage{chngcntr}
\usepackage{etoolbox}
\usepackage{lipsum}
\usepackage{mathdots}
\usepackage{MnSymbol}
\usepackage[nottoc]{tocbibind}

\graphicspath{{figures/}{figures_simuls/}}
\include{def_these}

\makeatletter
\def\blx@bblfile@bibtex{
  \blx@secinit
  \begingroup
  \blx@bblstart
\begingroup
\makeatletter
\@ifundefined{ver@biblatex.sty}
  {\@latex@error
     {Missing 'biblatex' package}
     {The bibliography requires the 'biblatex' package.}
      \aftergroup\endinput}
  {}
\endgroup

\entry{arcones1994limit}{article}{}
  \name{author}{1}{}{%
    {{}%
     {Arcones}{A.}%
     {Miguel~A.}{M.~A.}%
     {}{}%
     {}{}}%
  }
  \list{publisher}{1}{%
    {JSTOR}%
  }
  \strng{namehash}{AMA1}
  \strng{fullhash}{AMA1}
  \field{labelyear}{1994}
  \field{sortinit}{A}
  \field{pages}{2242\bibrangedash 2274}
  \field{title}{Limit theorems for nonlinear functionals of a stationary
  Gaussian sequence of vectors}
  \field{journaltitle}{The Annals of Probability}
  \field{year}{1994}
\endentry

\entry{AugerLawrence}{article}{}
  \name{author}{2}{}{%
    {{}%
     {Auger}{A.}%
     {Ivan~E.}{I.~E.}%
     {}{}%
     {}{}}%
    {{}%
     {Lawrence}{L.}%
     {Charles~E.}{C.~E.}%
     {}{}%
     {}{}}%
  }
  \list{publisher}{1}{%
    {Elsevier}%
  }
  \strng{namehash}{AIELCE1}
  \strng{fullhash}{AIELCE1}
  \field{labelyear}{1989}
  \field{sortinit}{A}
  \field{number}{1}
  \field{pages}{39\bibrangedash 54}
  \field{title}{Algorithms for the optimal identification of segment
  neighborhoods}
  \field{volume}{51}
  \field{journaltitle}{Bulletin of Mathematical Biology}
  \field{year}{1989}
\endentry

\entry{azencott}{book}{}
  \name{author}{2}{}{%
    {{}%
     {Azencott}{A.}%
     {Robert}{R.}%
     {}{}%
     {}{}}%
    {{}%
     {Dacunha-Castelle}{D.-C.}%
     {Didier}{D.}%
     {}{}%
     {}{}}%
  }
  \name{editorb}{2}{}{%
    {{}%
     {Gani}{G.}%
     {J.}{J.}%
     {}{}%
     {}{}}%
    {{}%
     {Heyde}{H.}%
     {C.~C.}{C.~C.}%
     {}{}%
     {}{}}%
  }
  \field{editorbtype}{redactor}
  \list{publisher}{1}{%
    {Springer New York}%
  }
  \strng{namehash}{ARDCD1}
  \strng{fullhash}{ARDCD1}
  \field{labelyear}{1986}
  \field{sortinit}{A}
  \field{isbn}{978-1-4612-9357-6, 978-1-4612-4912-2}
  \field{series}{Applied Probability}
  \field{title}{Series of Irregular Observations. Forecasting and model
  building.}
  \field{volume}{2}
  \list{location}{1}{%
    {New York}%
  }
  \field{year}{1986}
  \field{urlday}{03}
  \field{urlmonth}{06}
  \field{urlyear}{2015}
\endentry

\entry{BKW10}{article}{}
  \name{author}{3}{}{%
    {{}%
     {Bardet}{B.}%
     {Jean-Marc}{J.-M.}%
     {}{}%
     {}{}}%
    {{}%
     {Kengne}{K.}%
     {William~C.}{W.~C.}%
     {}{}%
     {}{}}%
    {{}%
     {Wintenberger}{W.}%
     {Olivier}{O.}%
     {}{}%
     {}{}}%
  }
  \strng{namehash}{BJM+1}
  \strng{fullhash}{BJMKWCWO1}
  \field{labelyear}{2012}
  \field{sortinit}{B}
  \field{pages}{1\bibrangedash 50}
  \field{title}{Detecting multiple change-points in general causal time series
  using penalized quasi-likelihood}
  \field{journaltitle}{Electronic Journal of Statistics}
  \field{year}{2012}
\endentry

\entry{billingsley1995probability}{book}{}
  \name{author}{1}{}{%
    {{}%
     {Billingsley}{B.}%
     {Patrick}{P.}%
     {}{}%
     {}{}}%
  }
  \list{publisher}{1}{%
    {Wiley}%
  }
  \strng{namehash}{BP1}
  \strng{fullhash}{BP1}
  \field{labelyear}{1995}
  \field{sortinit}{B}
  \field{isbn}{9780471007104}
  \field{series}{Wiley Series in Probability and Statistics}
  \field{title}{Probability and Measure}
  \field{year}{1995}
\endentry

\entry{braun1998statistical}{article}{}
  \name{author}{2}{}{%
    {{}%
     {Braun}{B.}%
     {Jerome~V.}{J.~V.}%
     {}{}%
     {}{}}%
    {{}%
     {Müller}{M.}%
     {Hans-Georg}{H.-G.}%
     {}{}%
     {}{}}%
  }
  \list{publisher}{1}{%
    {JSTOR}%
  }
  \strng{namehash}{BJVMHG1}
  \strng{fullhash}{BJVMHG1}
  \field{labelyear}{1998}
  \field{sortinit}{B}
  \field{pages}{142\bibrangedash 162}
  \field{title}{Statistical methods for DNA sequence segmentation}
  \field{journaltitle}{Statistical Science}
  \field{year}{1998}
\endentry

\entry{braun2000multiple}{article}{}
  \name{author}{3}{}{%
    {{}%
     {Braun}{B.}%
     {Jerome~V.}{J.~V.}%
     {}{}%
     {}{}}%
    {{}%
     {Braun}{B.}%
     {Rudolf~K.}{R.~K.}%
     {}{}%
     {}{}}%
    {{}%
     {Müller}{M.}%
     {Hans-Georg}{H.-G.}%
     {}{}%
     {}{}}%
  }
  \list{publisher}{1}{%
    {Biometrika Trust}%
  }
  \strng{namehash}{BJV+1}
  \strng{fullhash}{BJVBRKMHG1}
  \field{labelyear}{2000}
  \field{sortinit}{B}
  \field{number}{2}
  \field{pages}{301\bibrangedash 314}
  \field{title}{Multiple changepoint fitting via quasilikelihood, with
  application to DNA sequence segmentation}
  \field{volume}{87}
  \field{journaltitle}{Biometrika}
  \field{year}{2000}
\endentry

\entry{brockwell}{book}{}
  \name{author}{2}{}{%
    {{}%
     {Brockwell}{B.}%
     {Peter~J.}{P.~J.}%
     {}{}%
     {}{}}%
    {{}%
     {Davis}{D.}%
     {Richard~A.}{R.~A.}%
     {}{}%
     {}{}}%
  }
  \list{publisher}{1}{%
    {Springer-Verlag}%
  }
  \strng{namehash}{BPJDRA1}
  \strng{fullhash}{BPJDRA1}
  \field{labelyear}{1991}
  \field{sortinit}{B}
  \field{isbn}{9780387974293}
  \field{series}{Springer series in statistics}
  \field{title}{Time Series: Theory and Methods}
  \field{year}{1991}
\endentry

\entry{ChakarAR1}{article}{}
  \name{author}{4}{}{%
    {{}%
     {Chakar}{C.}%
     {Souhil}{S.}%
     {}{}%
     {}{}}%
    {{}%
     {Lebarbier}{L.}%
     {\'Emilie}{E.}%
     {}{}%
     {}{}}%
    {{}%
     {Lévy-Leduc}{L.-L.}%
     {Céline}{C.}%
     {}{}%
     {}{}}%
    {{}%
     {Robin}{R.}%
     {Stéphane}{S.}%
     {}{}%
     {}{}}%
  }
  \keyw{Mathematics - Statistics Theory, Statistics - Methodology, 62M10
  (Primary), 62F12, 62F35 (Secondary)}
  \strng{namehash}{CS+1}
  \strng{fullhash}{CSLELLCRS1}
  \field{labelyear}{2014}
  \field{sortinit}{C}
  \field{title}{A robust approach for estimating change-points in the mean of
  an AR(1) process}
  \field{journaltitle}{arXiv eprint 1403.1958}
  \field{eprinttype}{arXiv}
  \field{eprintclass}{math.ST}
  \field{year}{2014}
\endentry

\entry{frick2014multiscale}{article}{}
  \name{author}{3}{}{%
    {{}%
     {Frick}{F.}%
     {Klaus}{K.}%
     {}{}%
     {}{}}%
    {{}%
     {Munk}{M.}%
     {Axel}{A.}%
     {}{}%
     {}{}}%
    {{}%
     {Sieling}{S.}%
     {Hannes}{H.}%
     {}{}%
     {}{}}%
  }
  \list{publisher}{1}{%
    {Wiley Online Library}%
  }
  \strng{namehash}{FK+1}
  \strng{fullhash}{FKMASH1}
  \field{labelyear}{2014}
  \field{sortinit}{F}
  \field{number}{3}
  \field{pages}{495\bibrangedash 580}
  \field{title}{Multiscale change point inference}
  \field{volume}{76}
  \field{journaltitle}{Journal of the Royal Statistical Society: Series B
  (Statistical Methodology)}
  \field{year}{2014}
\endentry

\entry{Gazeaux2013}{article}{}
  \name{author}{13}{}{%
    {{}%
     {Gazeaux}{G.}%
     {Julien}{J.}%
     {}{}%
     {}{}}%
    {{}%
     {Williams}{W.}%
     {Simon}{S.}%
     {}{}%
     {}{}}%
    {{}%
     {King}{K.}%
     {Matt}{M.}%
     {}{}%
     {}{}}%
    {{}%
     {Bos}{B.}%
     {Machiel}{M.}%
     {}{}%
     {}{}}%
    {{}%
     {Dach}{D.}%
     {Rolf}{R.}%
     {}{}%
     {}{}}%
    {{}%
     {Deo}{D.}%
     {Manoj}{M.}%
     {}{}%
     {}{}}%
    {{}%
     {Moore}{M.}%
     {Angelyn~W.}{A.~W.}%
     {}{}%
     {}{}}%
    {{}%
     {Ostini}{O.}%
     {Luca}{L.}%
     {}{}%
     {}{}}%
    {{}%
     {Petrie}{P.}%
     {Elizabeth}{E.}%
     {}{}%
     {}{}}%
    {{}%
     {Roggero}{R.}%
     {Marco}{M.}%
     {}{}%
     {}{}}%
    {{}%
     {Teferle}{T.}%
     {Felix~N.}{F.~N.}%
     {}{}%
     {}{}}%
    {{}%
     {Olivares}{O.}%
     {German}{G.}%
     {}{}%
     {}{}}%
    {{}%
     {Webb}{W.}%
     {Frank~H.}{F.~H.}%
     {}{}%
     {}{}}%
  }
  \strng{namehash}{GJ+1}
  \strng{fullhash}{GJWSKMBMDRDMMAWOLPERMTFNOGWFH1}
  \field{labelyear}{2013}
  \field{sortinit}{G}
  \field{number}{5}
  \field{title}{Detecting offsets in GPS time series: First results from the
  detection of offsets in {GPS} experiment}
  \field{volume}{118}
  \field{journaltitle}{Journal of Geophysical Research (Solid Earth)}
  \field{year}{2013}
\endentry

\entry{killick2012optimal}{article}{}
  \name{author}{3}{}{%
    {{}%
     {Killick}{K.}%
     {Rebecca}{R.}%
     {}{}%
     {}{}}%
    {{}%
     {Fearnhead}{F.}%
     {Paul}{P.}%
     {}{}%
     {}{}}%
    {{}%
     {Eckley}{E.}%
     {Idris~A.}{I.~A.}%
     {}{}%
     {}{}}%
  }
  \list{publisher}{1}{%
    {Taylor \& Francis}%
  }
  \strng{namehash}{KR+1}
  \strng{fullhash}{KRFPEIA1}
  \field{labelyear}{2012}
  \field{sortinit}{K}
  \field{number}{500}
  \field{pages}{1590\bibrangedash 1598}
  \field{title}{Optimal detection of changepoints with a linear computational
  cost}
  \field{volume}{107}
  \field{journaltitle}{Journal of The American Statistical Association}
  \field{year}{2012}
\endentry

\entry{lavielle1999}{article}{}
  \name{author}{1}{}{%
    {{}%
     {Lavielle}{L.}%
     {Marc}{M.}%
     {}{}%
     {}{}}%
  }
  \strng{namehash}{LM1}
  \strng{fullhash}{LM1}
  \field{labelyear}{1999}
  \field{sortinit}{L}
  \field{number}{1}
  \field{pages}{79\bibrangedash 102}
  \field{title}{Detection of multiple changes in a sequence of dependent
  variables}
  \field{volume}{83}
  \field{journaltitle}{Stochastic Processes and their Applications}
  \field{year}{1999}
\endentry

\entry{lavielle2005using}{article}{}
  \name{author}{1}{}{%
    {{}%
     {Lavielle}{L.}%
     {Marc}{M.}%
     {}{}%
     {}{}}%
  }
  \list{publisher}{1}{%
    {Elsevier}%
  }
  \strng{namehash}{LM1}
  \strng{fullhash}{LM1}
  \field{labelyear}{2005}
  \field{sortinit}{L}
  \field{number}{8}
  \field{pages}{1501\bibrangedash 1510}
  \field{title}{Using penalized contrasts for the change-point problem}
  \field{volume}{85}
  \field{journaltitle}{Signal Processing}
  \field{year}{2005}
\endentry

\entry{LM}{article}{}
  \name{author}{2}{}{%
    {{}%
     {Lavielle}{L.}%
     {Marc}{M.}%
     {}{}%
     {}{}}%
    {{}%
     {Moulines}{M.}%
     {\'Eric}{E.}%
     {}{}%
     {}{}}%
  }
  \strng{namehash}{LMME1}
  \strng{fullhash}{LMME1}
  \field{labelyear}{2000}
  \field{sortinit}{L}
  \field{number}{1}
  \field{pages}{33\bibrangedash 59}
  \field{title}{Least-squares Estimation of an Unknown Number of Shifts in a
  Time Series}
  \field{volume}{21}
  \field{journaltitle}{Journal of Time Series Analysis}
  \field{year}{2000}
\endentry

\entry{climat}{article}{}
  \name{author}{3}{}{%
    {{}%
     {Lu}{L.}%
     {QiQi}{Q.}%
     {}{}%
     {}{}}%
    {{}%
     {Lund}{L.}%
     {Robert}{R.}%
     {}{}%
     {}{}}%
    {{}%
     {Lee}{L.}%
     {Thomas C.~M.}{T.~C.~M.}%
     {}{}%
     {}{}}%
  }
  \list{publisher}{1}{%
    {Institute of Mathematical Statistics}%
  }
  \strng{namehash}{LQ+1}
  \strng{fullhash}{LQLRLTCM1}
  \field{labelyear}{2010}
  \field{sortinit}{L}
  \field{number}{1}
  \field{pages}{299\bibrangedash 319}
  \field{title}{An MDL approach to the climate segmentation problem}
  \field{volume}{4}
  \field{journaltitle}{The Annals of Applied Statistics}
  \field{year}{2010}
\endentry

\entry{levy2011robust}{article}{}
  \name{author}{5}{}{%
    {{}%
     {Lévy-Leduc}{L.-L.}%
     {Céline}{C.}%
     {}{}%
     {}{}}%
    {{}%
     {Boistard}{B.}%
     {Hélène}{H.}%
     {}{}%
     {}{}}%
    {{}%
     {Moulines}{M.}%
     {\'Eric}{E.}%
     {}{}%
     {}{}}%
    {{}%
     {Taqqu}{T.}%
     {Murad~S.}{M.~S.}%
     {}{}%
     {}{}}%
    {{}%
     {Reisen}{R.}%
     {Valderio~A.}{V.~A.}%
     {}{}%
     {}{}}%
  }
  \list{publisher}{1}{%
    {Wiley Online Library}%
  }
  \strng{namehash}{LLC+1}
  \strng{fullhash}{LLCBHMETMSRVA1}
  \field{labelyear}{2011}
  \field{sortinit}{L}
  \field{number}{2}
  \field{pages}{135\bibrangedash 156}
  \field{title}{Robust estimation of the scale and of the autocovariance
  function of Gaussian short-and long-range dependent processes}
  \field{volume}{32}
  \field{journaltitle}{Journal of Time Series Analysis}
  \field{year}{2011}
\endentry

\entry{MG}{article}{}
  \name{author}{2}{}{%
    {{}%
     {Ma}{M.}%
     {Yanyuan}{Y.}%
     {}{}%
     {}{}}%
    {{}%
     {Genton}{G.}%
     {Marc~G.}{M.~G.}%
     {}{}%
     {}{}}%
  }
  \list{publisher}{1}{%
    {Blackwell Publishers Ltd}%
  }
  \keyw{autocovariance, breakdown point, influence function, robustness, scale
  estimation}
  \strng{namehash}{MYGMG1}
  \strng{fullhash}{MYGMG1}
  \field{labelyear}{2000}
  \field{sortinit}{M}
  \field{issn}{1467-9892}
  \field{number}{6}
  \field{pages}{663\bibrangedash 684}
  \field{title}{Highly Robust Estimation of the Autocovariance Function}
  \field{volume}{21}
  \field{journaltitle}{Journal of Time Series Analysis}
  \field{year}{2000}
\endentry

\entry{mestre2000methodes}{thesis}{}
  \name{author}{1}{}{%
    {{}%
     {Mestre}{M.}%
     {Olivier}{O.}%
     {}{}%
     {}{}}%
  }
  \strng{namehash}{MO1}
  \strng{fullhash}{MO1}
  \field{labelyear}{2000}
  \field{sortinit}{M}
  \field{title}{Méthodes statistiques pour l'homogénéisation de longues
  séries climatiques}
  \list{institution}{1}{%
    {Université Toulouse 3}%
  }
  \field{type}{phdthesis}
  \field{year}{2000}
\endentry

\entry{picard2005statistical}{article}{}
  \name{author}{5}{}{%
    {{}%
     {Picard}{P.}%
     {Franck}{F.}%
     {}{}%
     {}{}}%
    {{}%
     {Robin}{R.}%
     {Stéphane}{S.}%
     {}{}%
     {}{}}%
    {{}%
     {Lavielle}{L.}%
     {Marc}{M.}%
     {}{}%
     {}{}}%
    {{}%
     {Vaisse}{V.}%
     {Christian}{C.}%
     {}{}%
     {}{}}%
    {{}%
     {Daudin}{D.}%
     {Jean-Jacques}{J.-J.}%
     {}{}%
     {}{}}%
  }
  \list{publisher}{1}{%
    {BioMed Central Ltd}%
  }
  \strng{namehash}{PF+1}
  \strng{fullhash}{PFRSLMVCDJJ1}
  \field{labelyear}{2005}
  \field{sortinit}{P}
  \field{number}{1}
  \field{pages}{27}
  \field{title}{A statistical approach for array CGH data analysis}
  \field{volume}{6}
  \field{journaltitle}{BMC bioinformatics}
  \field{year}{2005}
\endentry

\entry{pruned}{article}{}
  \name{author}{1}{}{%
    {{}%
     {Rigaill}{R.}%
     {Guillem}{G.}%
     {}{}%
     {}{}}%
  }
  \keyw{Statistics - Computation}
  \strng{namehash}{RG1}
  \strng{fullhash}{RG1}
  \field{labelyear}{2010}
  \field{sortinit}{R}
  \field{title}{Pruned dynamic programming for optimal multiple change-point
  detection}
  \field{journaltitle}{arXiv eprint 1004.0887}
  \field{eprinttype}{arXiv}
  \field{eprintclass}{stat.CO}
  \field{year}{2010}
\endentry

\entry{CR}{article}{}
  \name{author}{2}{}{%
    {{}%
     {Rousseeuw}{R.}%
     {Peter~J.}{P.~J.}%
     {}{}%
     {}{}}%
    {{}%
     {Croux}{C.}%
     {Christophe}{C.}%
     {}{}%
     {}{}}%
  }
  \strng{namehash}{RPJCC1}
  \strng{fullhash}{RPJCC1}
  \field{labelyear}{1993}
  \field{sortinit}{R}
  \field{number}{424}
  \field{pages}{1273\bibrangedash 1283}
  \field{title}{Alternatives to the Median Absolute Deviation}
  \field{volume}{88}
  \field{journaltitle}{Journal of The American Statistical Association}
  \field{year}{1993}
\endentry

\entry{schwarz}{article}{}
  \name{author}{1}{}{%
    {{}%
     {Schwarz}{S.}%
     {Gideon}{G.}%
     {}{}%
     {}{}}%
  }
  \strng{namehash}{SG1}
  \strng{fullhash}{SG1}
  \field{labelyear}{1978}
  \field{sortinit}{S}
  \field{pages}{461\bibrangedash 464}
  \field{title}{Estimating the dimension of a model}
  \field{journaltitle}{The Annals of Statistics}
  \field{year}{1978}
\endentry

\entry{sheather1991}{article}{}
  \name{author}{2}{}{%
    {{}%
     {Sheather}{S.}%
     {Simon~J.}{S.~J.}%
     {}{}%
     {}{}}%
    {{}%
     {Jones}{J.}%
     {M.~C.}{M.~C.}%
     {}{}%
     {}{}}%
  }
  \list{publisher}{1}{%
    {JSTOR}%
  }
  \strng{namehash}{SSJJMC1}
  \strng{fullhash}{SSJJMC1}
  \field{labelyear}{1991}
  \field{sortinit}{S}
  \field{pages}{683\bibrangedash 690}
  \field{title}{A reliable data-based bandwidth selection method for kernel
  density estimation}
  \field{journaltitle}{Journal of the Royal Statistical Society: Series B
  (Statistical Methodology)}
  \field{year}{1991}
\endentry

\entry{Williams2003}{article}{}
  \name{author}{1}{}{%
    {{}%
     {Williams}{W.}%
     {Simon}{S.}%
     {}{}%
     {}{}}%
  }
  \strng{namehash}{WS1}
  \strng{fullhash}{WS1}
  \field{labelyear}{2003}
  \field{sortinit}{W}
  \field{title}{Offsets in Global Positioning System time series}
  \field{volume}{108}
  \field{journaltitle}{Journal of Geophysical Research (Solid Earth)}
  \field{year}{2003}
\endentry

\entry{yao}{article}{}
  \name{author}{1}{}{%
    {{}%
     {Yao}{Y.}%
     {Yi-Ching}{Y.-C.}%
     {}{}%
     {}{}}%
  }
  \strng{namehash}{YYC1}
  \strng{fullhash}{YYC1}
  \field{labelyear}{1988}
  \field{sortinit}{Y}
  \field{number}{3}
  \field{pages}{181\bibrangedash 189}
  \field{title}{Estimating the number of change-points via Schwarz' criterion}
  \field{volume}{6}
  \field{journaltitle}{Statistics \& Probability Letters}
  \field{year}{1988}
\endentry

\entry{Zha05}{thesis}{}
  \name{author}{1}{}{%
    {{}%
     {Zhang}{Z.}%
     {Nancy~R.}{N.~R.}%
     {}{}%
     {}{}}%
  }
  \strng{namehash}{ZNR1}
  \strng{fullhash}{ZNR1}
  \field{labelyear}{2005}
  \field{sortinit}{Z}
  \field{title}{Change-point detection and sequence alignment: statistical
  problems of genomics}
  \list{institution}{1}{%
    {Stanford University}%
  }
  \field{type}{phdthesis}
  \field{year}{2005}
\endentry

\entry{ZS}{article}{}
  \name{author}{2}{}{%
    {{}%
     {Zhang}{Z.}%
     {Nancy~R.}{N.~R.}%
     {}{}%
     {}{}}%
    {{}%
     {Siegmund}{S.}%
     {David~O.}{D.~O.}%
     {}{}%
     {}{}}%
  }
  \strng{namehash}{ZNRSDO1}
  \strng{fullhash}{ZNRSDO1}
  \field{labelyear}{2007}
  \field{sortinit}{Z}
  \field{number}{1}
  \field{pages}{22\bibrangedash 32}
  \field{title}{A modified Bayes Information Criterion with applications to the
  analysis of Comparative Genomic Hybridization data}
  \field{volume}{63}
  \field{journaltitle}{Biometrics}
  \field{year}{2007}
\endentry

\lossort
\endlossort

  \blx@bblend
  \endgroup
  \csnumgdef{blx@labelnumber@\the\c@refsection}{0}%
  \iftoggle{blx@reencode}{\blx@reencode}{}}
  \makeatother

\begin{document}

\title{A robust approach for estimating change-points in the mean of an $AR(p)$ process}
\author{Souhil Chakar\footnote{INP Grenoble, Laboratoire Jean Kuntzmann,
Grenoble cedex 09, France.}  \footnote{INRA-AgroParisTech, UMR 518 MIA-Paris,
Paris cedex 05, France.}\\
\tt{souhil.chakar@imag.fr}}
\maketitle
\input{./chapitres/arp.tex}

%
%
%
%
%
%

\printbibliography[heading=bibintoc]
%
%
%

\end{document}

%% file: def_these.tex
\definecolor{gris}{gray}{0.8}

\newcounter{hypA}

\def\rmd{\mathrm{d}}

\def\PE{\mathbb{E}}

\DeclareMathOperator{\argmin}{arg\,min}
\DeclareMathOperator{\argmax}{arg\,max}
\newcommandtwoopt{\QRC}[4][][]
{\ifthenelse{\equal{#1}{}}
{\ifthenelse{\equal{#2}{}}{\mathrm{Q}_{#3}\left( #4\right)}{\mathrm{Q}_{#3}\left(#4,#2\right)}}
{\ifthenelse{\equal{#2}{}}{\mathrm{Q}^{#1}_{#3}\left(#4\right)}{\mathrm{Q}^{#1}_{#3}\left(#4,#2\right)}
}
}
\def\IF{\mathrm{IF}}

\newtheorem{prop}{Proposition}[section]
\newtheorem{lemma}{Lemma}[section]

\theoremstyle{remark}
\newtheorem{remark}{Remark}[section]

\def \1{\mathbbm{1}}

  {\end{enumerate}}

  {\end{enumerate}}

\def\PE{\mathbb{E}} 

\newcommand{\AVvar}[3][]
{
\ifthenelse{\equal{#1}{}}{\mathbf{V}_{#3}(#2)}{\mathbf{V}_{#3}(#2,#1)}}
\newcommand{\AVvarJoint}[3][]
{
\ifthenelse{\equal{#1}{}}{\mathbf{W}_{#3}(#2)}{\mathbf{W}_{#3}(#2,#1)}}

\newcommand{\AVvarInv}[3][]
{\ifthenelse{\equal{#1}{}}{\mathbf{V}^{-1}_{#3}(#2)}{\mathbf{V}^{-1}_{#3}(#2,#1)}}
\newcommand{\sigmaasymp}[2][]
{
\ifthenelse{\equal{#1}{}}{\sigma(#2)}{\sigma(#2,#1)}}

\def\vjsymb{\sigma}
\newcommand{\vj}[4][]{%
\ifthenelse{\equal{#1}{}}{\vjsymb^{#4}_{#2}(#3)}{\vjsymb^{#4}_{#2}(#3,#1)}}
\newcommand{\stdj}[3][]{%
\ifthenelse{\equal{#1}{}}{\vjsymb_{#2}(#3)}{\vjsymb_{#2}(#3,#1)}}

\newcommand{\hvj}[3][]{%
\ifthenelse{\equal{#1}{}}{\hat{\vjsymb}^2_{#2}}{\hat{\vjsymb}^2_{#2}(#1)}}

\def\l{\mathbf{\ell}}

\renewcommand{\hat}{\widehat}
\renewcommand{\tilde}{\widetilde}

\def \2{2^{J_2-j}}
\def \2p{2^{J_2-j'}}

\renewcommand{\hat}{\widehat}

\def\IF{\mathrm{IF}}

\AtBeginEnvironment{subappendices}{%
\section*{Appendix}
\addcontentsline{toc}{section}{Appendices}{}{}
\addcontentsline{minitoc}{section}{Appendices}{}{}
\counterwithin{figure}{section}
\counterwithin{table}{section}
}

\def\rset{\mathbb{R}}

\def\zset{\mathbb{Z}}

\definecolor{darkgreen}{rgb}{0,0.55,0}

\makeatletter
\if@titlepage
  \renewenvironment{abstract}{%
      \null\vfil
      \@beginparpenalty\@lowpenalty
      \begin{center}%
        \bfseries \abstractname
        \@endparpenalty\@M
      \end{center}}%
     {\par\vfil\null}
\else
  \renewenvironment{abstract}{%
      \if@twocolumn
        \section*{\abstractname}%
      \else
        \small
        \begin{center}%
          {\bfseries \abstractname\vspace{-.5em}\vspace{\z@}}%
        \end{center}%
        \quotation
      \fi}
      {\if@twocolumn\else\endquotation\fi}
\fi
\makeatother

%% file: chapitres/arp.tex
\input{./chapitres/arp/arp.tex}

%% file: chapitres/arp/arp.tex
\selectlanguage{english}
\begin{abstract}
\input{./chapitres/arp/Abstract.tex}
\end{abstract}


\input{./chapitres/arp/Intro.tex}

\input{./chapitres/arp/Correlation.tex}

\input{./chapitres/arp/Segment.tex}

\input{./chapitres/arp/Selection.tex}

\input{./chapitres/arp/Simul.tex}

\input{./chapitres/arp/Appendix.tex}



%% file: chapitres/arp/Abstract.tex
\vspace{-0.2cm}
We consider the problem of multiple change-point estimation in the mean of an AR(p) process. Taking into account the dependence structure does not allow us to use the inference approach of the independent case. Especially, the dynamic programming algorithm giving the optimal solution in the independent case cannot be used anymore. We propose a two-step method, based on the preliminary estimation of the autoregression parameters. It is based on robust statistics techniques, since our estimator has to be robust to the change-points if we do not want to estimate them before. Then, we propose to follow the classical inference approach, by plugging this estimator in the criterion used for change-point estimation, which is equivalent to decorrelate the series using the estimated autoregression parameters. We show that the asymptotic properties of these change-point location and mean estimators are the same as those of the classical estimators in the independent framework. The same plug-in approach is then used to approximate 
the modified BIC and choose the number of segments, and to derive a heuristic BIC criterion to select both the number of changes and the order of the autoregression. Finally, we show, in the simulation section, that for finite sample size taking into account the dependence structure improves the statistical
performance of the change-point estimators and of the selection criterion.

%% file: chapitres/arp/Intro.tex
\section{Introduction}\label{arp:sec:intro}

Change-point detection problems arise in many fields, such as genomics~\parencites{braun1998statistical}{braun2000multiple}{picard2005statistical}, medical imaging~\citep{lavielle2005using}, earth sciences~\parencites{Williams2003}{Gazeaux2013} or climate~\parencites{mestre2000methodes}{climat}. In many of these problems, the observations cannot be assumed to be independent.

In the literature, in the frequentist framework, there is two ways to deal with the dependence structure of series affected by multiple change-points:
\begin{itemize}
\item Apply the methodology of the independent case, and prove that asymptotic results are not affected by dependence under some conditions~\parencites{LM}{lavielle1999}.
\item Consider that all the parameters of the model can change at each change-point \citep{BKW10}. In fact, inference is performed by considering also possible changes in the spectrum of the process.
\end{itemize}
The parameters of the dependence structure are here assumed to be global (i.e. not depending on the segments) nuisance parameters that have to be estimated to perform the change-points inference.

In this paper, we consider the segmentation of an AR($p$) process with homogeneous autoregression coefficients:
\begin{equation}\label{eq:modele_new}
y_i=\mu_k^\star+\eta_i\;,\;
t_{n,k}^\star+1\leq i\leq t_{n,k+1}^\star\;,\; 0\leq k\leq m^\star\;,\; 1\leq
i\leq n\;,
\end{equation}
where $(\eta_i)_{i\in \mathbb{Z}}$ is a zero-mean stationary AR($p$) process. That is, it is a stationary solution of
\begin{equation}\label{eq:arp}
\forall i \in \mathbb{Z}, \eta_i - \sum_{r=1}^p \phi_r^\star \eta_{i-r} = \epsilon_i\;,
\end{equation}
where the $\epsilon_i$'s are uncorrelated zero-mean rv's with variance $\sigma^2$ and $\Phi^\star = \left(\phi_r^\star\right)_{1\leq r\leq p}\in \mathbb{R}^p$ is such that a stationary solution to \eqref{eq:arp} exists. In \eqref{eq:modele_new}, we use the following conventions: $t_{n,0}^\star=0,t_{n,m^\star+1}^\star=n$.

Our aim is to propose a methodology for estimating both the change-point locations $\boldsymbol{t_n^\star} = (t_{n,k}^\star)_{1\leq k\leq m^\star}$ and the means $\boldsymbol{\mu}^\star=(\mu_k^\star)_{0\leq k\leq m^\star}$, accounting for the existence of $\Phi^\star$. Moreover, we propose a criterion to choose the number of change-points $m^\star$.

In the sequel, we shall assume that there exists $\boldsymbol{\tau^\star}=(\tau_k^\star)_{0\leq k\leq m+1}$ such that, for $0\leq k\leq m+1$ $t_{n,k}^\star=\lfloor n\tau_k^\star\rfloor$, $\lfloor x\rfloor$ denoting the integer part of $x$. Consequently, $\tau_0^\star=0$ and $\tau_{m^\star+1}^\star=1$.

A first idea is to use the maximum-likelihood approach to estimate the parameters of the model. However, maximizing the likelihood function, especially in the change-point location parameters $\boldsymbol{\tau}$, leads to a complex discrete optimization problem in an algorithmic point of view.

When the observations are independent, the optimal segmentation (e.g. in the maximum-likelihood sense) can be recovered via the dynamic programming (DP) algorithm introduced by \textcite{AugerLawrence}. The computational complexity of this algorithm is quadratic relatively to the length of the series. This algorithm and some of its improvements \parencites(such as these proposed by)(){pruned}[or][]{killick2012optimal} are the only one that provide exactly the optimal change-point location estimators. However, the DP algorithm only applies when ($i$) the loss function (e.g. the negative log-likelihood) is additive with respect to the segments and when ($ii$) no parameter to be estimated is common to several segments.

In the autoregressive case, the likelihood function violates both requirements~($i$) and~($ii$). Indeed the log-likelihood is not additive with respect to the segments because of the dependence that exists between data from neighbor segments and the unknown coefficients $\Phi^\star$ needs to be estimated jointly over all segments. Even if $\Phi^\star$ was known, the DP principle would not apply to the log-likelihood of Model~\eqref{eq:modele_new} as it will still not be additive. We introduce an alternative criterion, based on the quasi-likelihood described by \textcite{BKW10}.
This criterion is equivalent to the classic least-squares applied to a decorrelated version of the series, computed with an estimated $\widehat{\Phi}$. To achieve this decorrelation, we shall provide an estimator of $\Phi^\star$.

We shall prove that, under mild asymptotic assumptions on the estimator of $\Phi^\star$, the resulting change-point estimators satisfy the same rate of convergence as those proposed by \textcites{LM}{BKW10}.

We show that such an estimator of $\Phi^\star$ exists and can be computed before segmenting the series. In order to estimate $\Phi^\star$, we first differentiate the series of observations $(y_i)$ and work on
$x_i=y_i-y_{i-1}$ which satisfies
\begin{equation}\label{eq:def_x}
x_i=\nu_i \textrm{ except for } i=t_{n,k}^\star +1, \textrm{ where } 
x_{t_{n,k}^\star +1}=(\mu_k^\star-\mu_{k-1}^\star)+\nu_{t_{n,k}^\star +1},\; 0\leq k\leq m^\star\;,
\end{equation}
where $\nu_i=\eta_i-\eta_{i-1}$ is an ARMA($p$,1) defined from $(\eta_i)$ by
\begin{equation}\label{eq:def_nu}
\nu_i- \sum_{r=1}^p \phi_r^\star \nu_{i-r} = \epsilon_i-\epsilon_{i-1}\;.
\end{equation}
To this aim, we borrow techniques from robust estimation \citep{MG}. Briefly speaking, we consider the data observed at the change-point locations as outliers and propose an estimator of $\Phi^\star$ that is robust to the presence of such outliers. We shall prove that the estimator that we propose is consistent and asymptotically Gaussian. Moreover, we propose a model selection criterion on the number of changes, the order of the autoregression being fixed, inspired by the one proposed by \textcite{ZS} and prove some asymptotic properties of this criterion. Finally, we discuss the problem of selecting jointly the number of changes and the order of the autoregression and propose a practical solution to this problem based on a Bayesian heuristic.


%% file: chapitres/arp/Correlation.tex
\section{Robust estimation of the autoregression coefficients}\label{sec:correlation}

The aim of this section is to provide an estimator of $\Phi^\star$ which can deal with the
presence of change-points in the data. In the absence of change-points~($m^\star=0$ in \eqref{eq:modele_new}), the estimation of $\Phi^\star$ is a well-known issue \parencite[see][for a comprehensive introduction]{brockwell} and a consistent, asymptotically Gaussian estimator is given by the Yule-Walker method. We aim to adapt this method to our framework.
 
Since change-points can be seen as outliers in the AR($p$) process, we shall propose a robust approach
for estimating $\Phi^\star$. Our approach is based on the estimator of the autocorrelation function of a stationary time series proposed by \textcite{MG} which is based on the robust scale estimator of \textcite{CR}. More precisely, let us define $\widetilde{\Phi}_n^{(p)}$ by
\begin{equation}\label{eq:Phihat}
\widetilde{\Phi}_n^{(p)} = \widetilde{R}_{n,p}^{-1} \boldsymbol{\widetilde{\rho}}_{n,2:(p+1)}^T \; ,
\end{equation}
where for $i<j$ integers, 
\begin{eqnarray}
\boldsymbol{\widetilde{\rho}}_{n,i:j} & = & \left(\widetilde{\rho}_n(h)\right)_{i\leq h\leq j}\;, \label{eq:rhohat}
\end{eqnarray}
is an estimator of $\boldsymbol{\rho}_{i:j}$ defined by
\begin{eqnarray}
\boldsymbol{\rho}_{i:j} & = & \left(\rho(h)\right)_{i\leq h\leq j} \; , \label{eq:rho}
\end{eqnarray}
where $\rho(h)$ denotes the autocorrelation of the process $(x_i)$ defined in (\ref{eq:def_x}) at lag $h$, and $\cdot^T$ state for the transpose.
In \eqref{eq:Phihat}, $\widetilde{R}_{n,p}$ denotes the following matrix
\begin{eqnarray}
\widetilde{R}_{n,p} & = & \left(\widetilde{\rho}_n\left(j-i-1\right)\right)_{1\leq i, j\leq p} \; , \label{eq:matrixRhat}
\end{eqnarray}
which is an estimator of 
\begin{eqnarray}
R_p & = & \left(\rho\left(j-i-1\right)\right)_{1\leq i, j\leq p}\;. \label{eq:matrixR}
\end{eqnarray}
Moreover, for all $h$ in $\zset$,
\begin{equation}\label{eq:def:ma_genton}
\widetilde{\rho}_n(h)=\dfrac{Q_{n}^2 \left(x^+_h\right) - Q_{n}^2 \left(x^-_h\right)}{Q_{n}^2 \left(x^+_h\right) + Q_{n}^2 \left(x^-_h\right)}\;,
\end{equation}
where $x^+_h = (x_{i+h}+x_i)_{0\leq i\leq n-h}$, $x^-_h = (x_{i+h}-x_i)_{0\leq i\leq n-h}$ 
and $Q_n$ is the scale estimator of \textcite{CR} which is such that $Q_n \left(z\right)$
is proportional to the first quartile of $\left\lbrace \vert z_i -
z_j \vert ; 0\leq i<j\leq n \right\rbrace$.

\begin{prop}\label{prop:corr_tcl}
Let $\boldsymbol{\widetilde{\rho}}_{n,i:j}  =  \left(\widetilde{\rho}_n (h)\right)_{i\leq h\leq j}$ and
$\boldsymbol{\rho}_{i:j} = \left(\rho(h)\right)_{i\leq h\leq j}$ be defined in \eqref{eq:rhohat}, \eqref{eq:rho} and \eqref{eq:def:ma_genton}. Under the assumption that $\nu_i$ in \eqref{eq:def_x} is Gaussian, we have
\begin{equation}\label{eq:ANrho}
n^{1/2}\left(\boldsymbol{\widetilde{\rho}}_{n,1:(p+1)} - \boldsymbol{\rho}_{1:(p+1)}\right) \underset{n\rightarrow\infty}{\longrightarrow} \mathcal{N}\left(0, V\right)
\end{equation}
in distribution, where $\mathcal{N}\left(0, V\right)$ is the $(p+1)-$ dimensional centered Gaussian distribution with covariance matrix $V$ and
 \begin{equation}\label{eq:ANphi}
 n^{1/2}\left(\widetilde{\Phi}_{n}^{(p)} - \Phi^\star\right) \underset{n\rightarrow\infty}{\longrightarrow} \mathcal{N}\left(0, M^T \left(R_p^{-1}\right)^T V R_p^{-1} M\right) \; ,
 \end{equation}
 in distribution, where
 \begin{equation}\label{eq:Jacobian}
 M = \left(\phi_{i-j+1}^\star\mathbf{1}_{i\geq j} + \phi_{i+j+1}^\star\mathbf{1}_{i+j\leq p-1} - \mathbf{1}_{j=i+1}\right)_{1\leq i\leq p, 1\leq j\leq p+1} \; .
 \end{equation}
\end{prop}
The proof of Proposition \ref{prop:corr_tcl} is given in Section \ref{App:Proof}.

\begin{remark}\label{arp:remark:regularization}
Proposition \ref{prop:corr_tcl} argues for the possibility to estimate robustly $\Phi^\star$. Equations~\eqref{eq:ANphi} and~\eqref{eq:Jacobian} show that the variance of $\widehat{\Phi}$ depends on $\Phi^\star$. This variance may be large for some $\Phi^\star$. One can deal with this problem by preferring a regularized version of this estimator, that is
\begin{equation}\label{eq:regularisation}
\left( \widetilde{R}_{n,p}^{T} \widetilde{R}_{n,p} + S_{n,p} \right)^{-1} \widetilde{R}_{n,p}^{T} \boldsymbol{\widetilde{\rho}}_{n,2:(p+1)}^T \; ,
\end{equation}
where $S_{n,p}$ is a $p\times p$ positive semi-definite symmetric real matrix. If $S_{n,p} = o_P \left( n^{-1/2} \right)$, then Proposition~\ref{prop:corr_tcl} remains true with $\widetilde{\Phi}_n^{(p)}$ being replaced by the expression of Equation~\eqref{eq:regularisation}.
\end{remark}


%% file: chapitres/arp/Segment.tex
\section{Change-points and expectations estimation}\label{sec:change-points}

In this section, the number of change-points $m^\star$ is assumed to be known. In the sequel, for notational simplicity, $m^\star$ will be denoted by $m$.
Our goal is to estimate both the change-points and the means in Model~\eqref{eq:modele_new}. A first idea consists in maximizing the Gaussian quasi-likelihood conditioned on $y_0,\dots , y_{1-p}$ and to minimize it with respect to $\Phi$. Due to a quadratic term that involves both $\delta_{k-1}$ and $\delta_k$, this criterion cannot be efficiently minimized. Therefore, we propose to use an alternative criterion defined as follows:
\begin{equation*}
SS_m\left(y,\Phi, \boldsymbol{\delta},\boldsymbol{t}\right) = \sum_{k=0}^m \sum_{i=t_k +1}^{t_{k+1}}
\left(y_i-\sum_{r=1}^p \phi_r y_{i-r}- \delta_k\right)^2\;.
\end{equation*}
Note that minimizing $SS_m\left(z,\Phi, \left(1-\sum_{r=1}^p\phi_r\right) \boldsymbol{\mu},\boldsymbol{t}\right)$ is equivalent to maximizing the Gaussian log-likelihood, conditioned on $z_0,\dots ,z_{1-p}$, of the following model maximized with respect to $\sigma$:
\begin{eqnarray}\label{eq:bkw}
z_i - \mu_k^\star & = & \sum_{r=1}^p \phi_r \left(z_{i-r} - \mu_k^\star\right) +\epsilon_i\;,\; t_{n,k}^\star +1\leq i\leq t_{n,k+1}^\star \;,\; 0\leq k\leq m\;,\; 1\leq i\leq n\;,
\end{eqnarray}
where the $\epsilon_i$'s are defined as in Model~\eqref{eq:modele_new}. In this model, as in the models considered in \textcite{BKW10}, the expectation changes are not abrupt anymore as in Model~\eqref{eq:modele_new}.

\begin{prop}\label{Prop:Segment}
Let $\left(\overline{\Phi}_n\right)$ be a sequence of rv's on $\mathbb{R}^p$ and $z = \left(z_{1-p} ,\dots ,z_n\right)$ a finite sequence of real-valued rv's satisfying~\eqref{eq:bkw}. Let $\boldsymbol{\widehat{\delta}}_n(z, \overline{\Phi}_n)$ and $\boldsymbol{\widehat{t}}_n(z, \overline{\Phi}_n)$ be defined by 
\begin{eqnarray}
\left(\boldsymbol{\widehat{\delta}}_n(z, \overline{\Phi}_n), \boldsymbol{\widehat{t}}_n(z, \overline{\Phi}_n)\right) & = & \underset{\left(\boldsymbol{\delta},\boldsymbol{t}\right)\in \mathbb{R}^{m+1}\times \mathcal{A}_{n,m}}{\argmin} SS_m\left(z,\overline{\Phi}_n , \boldsymbol{\delta},\boldsymbol{t}\right) \ \textit{,}\label{eq:delta_n,t_n}\\
\boldsymbol{\widehat{\tau}}_n(z, \overline{\Phi}_n) & = & \frac{1}{n} \boldsymbol{\widehat{t}}_n(z, \overline{\Phi}_n),\label{eq:tau_n}
\end{eqnarray}
\noindent where
\begin{equation}\label{eq:Anm}
\mathcal{A}_{n,m} = \left\lbrace \left(t_0,\dots ,t_{m+1}\right);t_0=0<t_1<\dots <t_m<t_{m+1}=n, \forall k=1,\dots ,m+1, t_k-t_{k-1}\geq\Delta_n \right\rbrace
\end{equation}
and where $\left(\Delta_n\right)$ is a real sequence such that
$n^{-\alpha}\Delta_n \longrightarrow \infty$, as $n\to\infty$ and $\alpha>0$.
Assume that
\begin{equation}\label{eq:hypRhoRate}
\left(\overline{\Phi}_n-\Phi^\star\right) = O_P\left(n^{-1/2}\right),
\end{equation}
as $n$ tends to infinity. 
Then, 
\begin{equation*}
\| \boldsymbol{\widehat{\tau}}_n(z, \overline{\Phi}_n) - \boldsymbol{\tau^\star} \| = O_P\left(n^{-1}\right), \qquad \| \boldsymbol{\widehat{\delta}}_n(z, \overline{\Phi}_n) - \boldsymbol{\delta^\star} \|= O_P\left(n^{-1/2}\right),
\end{equation*}
where $\|\cdot\|$ is the Euclidian norm.
\end{prop}

\begin{prop}\label{Prop:Segment2}
The result of Proposition \ref{Prop:Segment} still holds under the same assumptions 
when $z$ is replaced with $y$
satisfying \eqref{eq:modele_new}.
\end{prop}

The proofs of Propositions \ref{Prop:Segment} and \ref{Prop:Segment2} are given in Section \ref{subsec:prop:Segment} and \ref{subsec:prop:Segment2}, respectively.
Note that the estimators defined in these propositions have the same asymptotic properties 
as those of the estimators proposed by \textcite{LM}. Also, the estimator $\widetilde{\Phi}_n^{(p)}$ defined in Section \ref{sec:correlation} satisfies the same properties as $\overline{\Phi}_n$ and can thus be used in the criterion $SS_m$ for providing consistent estimators of the change-points and of the means.


%% file: chapitres/arp/Selection.tex
\section{Selecting the number of change-points} \label{sec:selection}

\subsection{Criterion to select the number of change-points, the order of the autoregression being fixed}
In this section, we propose to adapt the modified Bayesian information criterion \parencite[mBIC,][]{ZS} to our autoregressive noise framework. mBIC was proposed to select the number $m$ of change-points in the mean in the particular case of segmentation of an independent Gaussian process $x$. This criterion is derived from an $O_P(1)$ approximation of the Bayes factor between models with $m$ and $0$ change-points, respectively. Its performance for model selection is assessed by simulation studies \parencites{Zha05}{frick2014multiscale}.

The mBIC selection procedure consists in choosing the number of change-points as:
\begin{equation}\label{eq:Criterion_formula}
\widehat{m} = \underset{m}{\argmax} C_m(x, 0) \; ,
\end{equation}
where the criterion $C_m(y, \Phi)$ is defined for a process $y$ as
\begin{multline}\label{eq:Cm}
  C_m(y, \Phi) = \\
- \left(\frac{n-m+1}{2}\right) \log SS_m(y, \Phi) + \log
\Gamma\left(\frac{n-m+1}{2}\right) -\frac{1}{2} \sum_{k=0}^m \log n_k(\boldsymbol{\widehat{t}}(y, \Phi)) - m \log n .
\end{multline}
In the latter equation
\begin{equation} \label{Eq:SSm}
SS_m(y, \Phi) = \min_{\boldsymbol{\delta}, \boldsymbol{t}} SS_m(y, \Phi, \boldsymbol{\delta}, \boldsymbol{t}),
\end{equation}
where the minimization with respect to $\boldsymbol{t}$ is performed in $\mathcal{A}_{n,m}$ defined in~\eqref{eq:Anm}, and 
\begin{equation*}
n_k(\boldsymbol{\widehat{t}}(y, \Phi))=\widehat{t}_{k+1}(y, \Phi)-\widehat{t}_{k}(y, \Phi),
\end{equation*}
where $\boldsymbol{\widehat{t}}(y, \Phi)=(\widehat{t}_1(y, \Phi), \dots, \widehat{t}_m(y, \Phi))$ is defined as $\boldsymbol{\widehat{t}}(y, \Phi)= \underset{ \boldsymbol{t}\in\mathcal{A}_{n,m}}{\argmin}\; \underset{\delta}{\min}\; SS_m(y, \Phi, \delta, \boldsymbol{t})$.

Note that, in Model \eqref{eq:bkw}, the criterion could be directly applied to the decorrelated series $v^\star = \left(v^\star_i\right)_{1\leq i\leq n} = \left(y_i - \sum_{r=1}^p\phi_r^\star y_{i-r}\right)_{1\leq i\leq n}$ since
$$
C_m(y, \Phi^\star) = C_m(v^\star, 0).
$$ 
We propose to use the same selection criterion, replacing $\Phi^\star$ by some relevant estimator $\overline{\Phi}_n$. 
The following two propositions show that this plug-in approach results in the same asymptotic properties under both Models~\eqref{eq:bkw} and~\eqref{eq:modele_new}. 

\begin{prop} \label{Prop:mBICBardet}
For any positive $m$, for a process $z$ satisfying \eqref{eq:bkw} and under the assumptions of Proposition \ref{Prop:Segment}, we have
\begin{eqnarray*}
C_m(z, \overline{\Phi}_n) & = & C_m(z, \Phi^\star) + O_P(1),\textrm{ as } n\to\infty\;.
\end{eqnarray*}
\end{prop}

\begin{prop} \label{Prop:mBIC}
For any positive $m$, for a process $y$ satisfying \eqref{eq:modele_new} and under the assumptions of Proposition \ref{Prop:Segment2} , we have
\begin{eqnarray*}
C_m(y, \overline{\Phi}_n) & = & C_m(y, \Phi^\star) + O_P(1),\textrm{ as } n\to\infty\;.
\end{eqnarray*}
\end{prop}

The proofs of Propositions \ref{Prop:mBICBardet} and \ref{Prop:mBIC} are similar to \textcite[Propositions 6 and 7]{ChakarAR1}.

In practice, we propose to take $\overline{\Phi}_n = \widetilde{\Phi}_n^{(p)}$ as defined in Section \ref{sec:correlation}, which satisfies the conditions of Proposition \ref{Prop:mBIC} to estimate the number of segments by
\begin{eqnarray} \label{Eq:BIC}
\widehat{m} & = & \underset{0\leq m \leq m_{\max} }{\argmax} C_m\left(y,\widetilde{\Phi}_n^{(p)}\right) \; ,
\end{eqnarray}
where $C_m$ is defined in \eqref{eq:Cm}, and for a given maximum number of changes $m_{\max}$.

\subsection{Heuristic to select both the number of changes and the autoregression order}\label{subs:heuristic}

Applying \textcite[Theorem 2.2]{Zha05} to the series $v^\star = \left(v^\star_i\right)_{1\leq i\leq n} = \left(z_i - \sum_{r=1}^p\phi_r^\star z_{i-r}\right)_{1\leq i\leq n}$, for a process $z$ satisfying \eqref{eq:bkw} and with the corresponding priors, gives:

\begin{equation*}
\log P(m | v^\star ) = C_m (v^\star,0) - C_0 (v^\star,0) + \log P(m=0 | v^\star ) + O_P (1) \; ,
\end{equation*}
that is
\begin{equation*}
 P(m | z,p,\Phi )  =  C_m (z,\Phi) - C_0 (v^\star,0) + \log P(m=0 | v^\star ) + O_P (1) \; ,
\end{equation*}

From \textcite{schwarz}, we approximate then $\log P(m,p | z )$, up to a constant, by 
$$\log P\left(m \left| z,p,\widehat{\Phi}_{ML}^{(p)} \right. \right) - \frac{p}{2}\log n \; , $$
and then, up to a constant, by 
$$C_m \left(z,\widehat{\Phi}_{ML}^{(p)}\right) - \frac{p}{2}\log n\; , $$
 where $\widehat{\Phi}_{ML}^{(p)}$ is the maximum likelihood estimator of $\Phi^\star$ in the model with $m$ changes and autoregression order $p$. Replacing $\widehat{\Phi}_{ML}^{(p)}$ by $\widetilde{\Phi}_n^{(p)}$, and $z$ by $y$ satisfying \eqref{eq:modele_new}, we propose to select $m$ and $p$ by
 \begin{equation}\label{eq:Sel_mp}
\left(\widehat{m}^\prime , \widehat{p}^\prime \right) = \underset{0\leq m \leq m_{\max}, 0\leq p \leq p_{\max}}{\argmax} \left\lbrace C_m\left(y,\widetilde{\Phi}_n^{(p)}\right) - \frac{p}{2} \log n \right\rbrace \; ,
 \end{equation}

with given $m_{\max}$ and $p_{\max}$. Even if we do not aim to estimate the order $p$ of the autoregression, this criterion is interesting by being more flexible than \eqref{Eq:BIC}. Indeed, if at the true order $p$, $\widetilde{\Phi}_n^{(p)}$ provides a poor estimate of $\Phi^\star$, a different order (e.g. $p+1$) can lead to a better fitting of the model and then a better estimate of the number of changes.


%% file: chapitres/arp/Simul.tex
\section{Numerical experiments}\label{sec:simul}

\subsection{Practical implementation}\label{sec:post-proc}

Our decorrelation procedure introduces spurious change-points in the series, at distance $\leq p$ of the true change-points. When the length of the series tends to infinity, the effect of these artefacts on estimates vanishes, but with a finite length series our procedure may be affected. To fix this point, we propose a post-processing to the estimated change-points $\boldsymbol{\widehat{t}_n}$, which consists in removing segments of length smaller than $p$:
\begin{multline*}
PP\left( \boldsymbol{\widehat{t}}_n ,p \right) = \\
\left\lbrace \widehat{t}_{n,k}\in \boldsymbol{\widehat{t}}_n \right\rbrace \setminus \left\lbrace \widehat{t}_{n,i} \textit{ such that } \exists j , \widehat{t}_{n,i} - p \leq \widehat{t}_{n,j} < \widehat{t}_{n,i} \textit{ and } (j=1 \textit{ or } \widehat{t}_{n,j-1} < \widehat{t}_{n,j} - p ) \right\rbrace \; .
\end{multline*}

\subsection{Simulation design}\label{subs:simul_design}

We simulated Gaussian $AR(p)$ series $y_1,\dots ,y_n$ of length  $n=7200$ and $n=14400$, with $p=2$ or $p=5$. Additionally, the observations $y_{-19},\dots ,y_0,$ are simulated and used for conditioning the quasi-likelihood. All series were affected by $m^\star = 6$ change-points located at fractions $1/6 \pm 1/36, 3/6 \pm 2/36, 5/6 \pm 3/36$ of their length. The mean within each segment alternates between 0 and 1, starting with $\mu_1 = 0$. We considered autoregression parameters that verify the assumptions to get a stationary causal process, see \textcite[~Theorem 3.1.1]{brockwell}. We focused our attention on some parameters for which the computed estimators have a typical behavior and are interesting to illustrate our method.

\begin{description}
\item[For $p=2$] the parameters are the following
 $$\mkern-18mu \left(\phi_1^\star,\phi_2^\star,\sigma^\star\right) \in \left\lbrace (-1.2,-0.4,0.4),(1.6,-0.8,0.4),(0.2,0.2,0.4), (0.2,0.6,0.4), (0.4,0.2,0.2) \right\rbrace \; . $$
\item[For $p=5$] the parameters are the following
 $$ \left(\phi_1^\star,\phi_2^\star,\phi_3^\star,\phi_4^\star,\phi_5^\star,\sigma^\star\right) \in \left\lbrace (0.5,0,0,0.5,-0.5,0.4), (0.5,0,0,0,-0.5,0.4)\right\rbrace \; . $$
\end{description}
Each combination was replicated $S=100$ times.

\subsection{Quality criteria}
 To assess the quality of the estimation of the autoregression parameters, we used the Root-Mean-Square Errors (RMSE) of  $\widehat{\Phi}_n^{(p)}$ defined in \eqref{eq:Phihat}.
 
 To study the performances of our proposed model selection criteria, we computed and compared:
\begin{itemize}
\item $\widehat{m}^0 = \underset{m}{\argmax} C_m (y, 0)$, the criterion $C_m$ without any decorrelation procedure,
\item $\widehat{m}^0_{Y}$, the estimated number of changes derived from the BIC-type penalized criterion defined by \textcite{yao},
\item $\widehat{m}^\star = \underset{m}{\argmax} C_m (y, \Phi^\star)$, the criterion $C_m$ where the series is exactly decorrelated,
\item $\widehat{m}^\star_{PP}$, the post-processed number of changes, that is the number of changes of $ PP(\boldsymbol{\widehat{t}}_n , p)$ if $\boldsymbol{\widehat{t}}_n$ is the vector of the estimated change-point locations obtained by the minimization~\eqref{Eq:SSm} with $\Phi = \Phi^\star$,
\item $\widehat{m} = \underset{m}{\argmax} C_m (y, \widehat{\Phi}_n^{(p)})$, and the result of post-processing $\widehat{m}_{PP}$,
\item $\left(\widehat{m}^\prime , \widehat{p}^\prime \right)$ defined in \eqref{eq:Sel_mp}. The post-processing, giving $\widehat{m}^\prime_{PP}$, is performed with $PP\left(\cdot ,\widehat{p}^\prime \right)$,
\end{itemize}
where $C_m$ is defined in \eqref{eq:Cm}. We were particularly interested in the comparison of:
\begin{itemize}
\item $\widehat{m}^0_{Y}$ and $\widehat{m}^0$ to the other estimates to illustrate how much our method improves the estimation of the number of changes,
\item $\widehat{m}^\star$ and $\widehat{m}^\star_{PP}$ are compared to $\widehat{m}$ and $\widehat{m}_{PP}$ to identify the errors coming from the estimation of $\Phi^\star$ by $\widetilde{\Phi}_n$,
\item the post-processed estimates are compared to the non post-processed estimates to assess the usefulness of this finite-sample correction of our method.
\end{itemize}

In order to measure the performance of change-point location estimators, we plotted their frequencies. In particular we are interested in the following change-point estimates:
\begin{itemize}
\item $\widehat{t}_n^0$, the minimizer on $\mathcal{A}_{n,\widehat{m}^0}$ of $\underset{\boldsymbol{\delta}}{\min} SS_{\widehat{m}^0}\left(\boldsymbol{y},0,\boldsymbol{\delta},\cdot \right)$.
\item $\widehat{t}_{n, PP} = PP\left(\widehat{t}_{n},p\right)$, where $\widehat{t}_{n}$ minimizes $\underset{\boldsymbol{\delta}}{\min} SS_{\widehat{m}}\left(\boldsymbol{y},\widetilde{\Phi}_n^{(p)},\boldsymbol{\delta},\cdot \right)$ on  $\mathcal{A}_{n,\widehat{m}}$.
\item $\widehat{t}_{n, PP}^\prime = PP\left(\widehat{t}_{n}^\prime,p^\prime\right)$, where $\widehat{t}_{n}^\prime$ minimizes $\underset{\boldsymbol{\delta}}{\min} SS_{\widehat{m}^\prime}\left(\boldsymbol{y},\widetilde{\Phi}_n^{\left(p^\prime\right)},\boldsymbol{\delta},\cdot \right)$ on  $\mathcal{A}_{n,\widehat{m}^\prime}$.
\end{itemize}
To highlight the peaks, we plotted also the Gaussian kernel density estimate. The bandwith is selected following the method of~\textcite{sheather1991}.

\subsection{Results}
\begin{description}
\item[For $p=2$] the simulation results suggest that the decorrelation procedure is not necessary in all cases. If $\phi_1^\star$ and $\phi_2^\star$ are negative, $\widehat{m}^0 = 6$ almost all the times~(see Table~\ref{table:phi1_-12_phi2_-04_sigma04} and Figures~\ref{fig:box:phi1_-12_phi2_-04_sigma04} and~\ref{fig:dens:phi1_-12_phi2_-04_sigma04}). If only one of the parameters is negative, problems can arise without decorrelation if the other parameter is positively large~(see Table~\ref{table:phi1_16_phi2_-08_sigma04} and Figures~\ref{fig:box:phi1_16_phi2_-08_sigma04} and~\ref{fig:dens:phi1_16_phi2_-08_sigma04}). The core of the problems with $\widehat{m}^0$ is located at $\phi_1^\star>0, \phi_2^\star>0$. In these cases, our decorrelation procedure is required~(see Tables~\ref{table:phi1_02_phi2_02_sigma04},~\ref{table:phi1_02_phi2_06_sigma04}, and~\ref{table:phi1_04_phi2_02_sigma02}, and Figures~\ref{fig:box:phi1_02_phi2_02_sigma04},~\ref{fig:dens:phi1_02_phi2_02_sigma04},~\ref{fig:box:phi1_02_phi2_06_sigma04},~\ref{fig:dens:phi1_02_phi2_06_sigma04},~\ref{fig:box:phi1_04_phi2_02_sigma02}, and~\ref{fig:dens:phi1_04_phi2_02_sigma02}).
$\widehat{m}$ can provide poor estimates of $m$ because of a poor preliminary estimate of $\Phi^\star$. However, in this case, $\widehat{m}^\prime$ can provide better estimates, thanks to an overestimation of $p$ by $\widehat{p}^\prime$, as noticed in Section~\ref{subs:heuristic}~(see Tables~\ref{table:phi1_02_phi2_06_sigma04},~\ref{table:phi1_04_phi2_02_sigma02}, and Figures~\ref{fig:dens:phi1_02_phi2_02_sigma04},~\ref{fig:box:phi1_02_phi2_06_sigma04},~\ref{fig:dens:phi1_02_phi2_06_sigma04},~\ref{fig:box:phi1_04_phi2_02_sigma02},~\ref{fig:dens:phi1_04_phi2_02_sigma02}).

Tables \ref{table:phi1_02_phi2_06_sigma04} and \ref{table:phi1_04_phi2_02_sigma02} illustrate the usefulness of post-processing.
\item[For $p=5$] we can see from Tables \ref{table:phi1_05_phi2_0_phi3_0_phi4_05_phi5_-05_sigma04} and \ref{table:phi1_05_phi2_0_phi3_0_phi4_0_phi5_-05_sigma04} that our method, when $p$ is known, has the same performance as the methodology which would have access to the autoregression parameters $\Phi^\star$, see the lines $\widehat{m}$ and $\widehat{m}^\star$. These performances are not altered by the choice of $p$, even if it is overestimated by our model selection approach, see the lines $\widehat{m}^\prime$ and $\widehat{p}^\prime$.  Moreover, our method outperforms a methodology which would ignore the dependence structure of the process, see the line $\widehat{m}^0$ of these tables. In addition, our method is not only able to select the true number of change-points, whatever $p$, but also the true change-point positions, as displayed in Figures \ref{fig:dens:phi1_05_phi2_0_phi3_0_phi4_05_phi5_-05_sigma04} and \ref{fig:dens:phi1_05_phi2_0_phi3_0_phi4_0_phi5_-05_sigma04}.
\end{description}

%

%% file: chapitres/arp/Appendix.tex
\section{Proofs} \label{App:Proof}

\input{./chapitres/arp/AppSec2.tex}

\input{./chapitres/arp/AppSec3.tex}
\newpage

\section{Tables and figures}
\noindent\begin{minipage}{\linewidth}
\centering
\begin{tabular}{|c||c|c|c|c|c||c|c|c|c|c|}
\hline
$n$ & \multicolumn{5}{c||}{7200} & \multicolumn{5}{c|}{14400} \\
\hline
\hline
estimate \textbackslash number of changes  & $<5$ & $5$ & $\mathbf{6} $ & $7$ & $>7$ & $<5$ & $5$ & $\mathbf{6} $ & $7$ & $>7$ \\
 \hline
 $\widehat{m}^0_Y$ & $0$ & $0$ & $\mathbf{100}$ & $0$ & $0$ & $0$ & $0$ & $\mathbf{100}$ & $0$ & $0$ \\
 \hline
$\widehat{m}^0$ & $0$ & $0$ & $\mathbf{100}$ & $0$ & $0$ & $0$ & $0$ & $\mathbf{100}$ & $0$ & $0$ \\
 \hline
$\widehat{m}$ & $0$ & $0$ & $\mathbf{97}$ & $3$ & $0$ & $0$ & $0$ & $\mathbf{100}$ & $0$ & $0$ \\
 \hline
$\widehat{m}_{PP}$ & $0$ & $0$ & $\mathbf{99}$ & $1$ & $0$ & $0$ & $0$ & $\mathbf{100}$ & $0$ & $0$ \\
 \hline
$\widehat{m}^\star$ & $0$ & $0$ & $\mathbf{97}$ & $3$ & $0$ & $0$ & $0$ & $\mathbf{100}$ & $0$ & $0$ \\
 \hline
$\widehat{m}_{PP}^\star$ & $0$ & $0$ & $\mathbf{99}$ & $1$ & $0$& $0$ & $0$ & $\mathbf{100}$ & $0$ & $0$ \\
 \hline
$\widehat{m}^\prime$ & $0$ & $0$ & $\mathbf{97}$ & $3$ & $0$ & $0$ & $0$ & $\mathbf{100}$ & $0$ & $0$ \\
 \hline
$\widehat{m}_{PP}^\prime$ & $0$ & $0$ & $\mathbf{99}$ & $1$ & $0$ & $0$ & $0$ & $\mathbf{100}$ & $0$ & $0$ \\
 \hline
 \hline
 estimate \textbackslash order of the autoregression & $0$ & $1$ & $\mathbf{2}$ & $3$ & $>3$ & $0$ & $1$ & $\mathbf{2}$ & $3$ & $>3$ \\
 \hline
 $\widehat{p}^\prime$  & $0$ & $0$ & $\mathbf{96}$ & $4$ & $0$ & $0$ & $0$ & $\mathbf{97}$ & $1$ & $2$ \\
 \hline
 \hline
 $\widetilde{\phi}_{n,1}^{(2)}$ RMSE & \multicolumn{5}{c||}{$1.99 \cdot 10^{-2}$} & \multicolumn{5}{c|}{$1.64 \cdot 10^{-2}$} \\
 \hline
 $\widetilde{\phi}_{n,2}^{(2)}$ RMSE & \multicolumn{5}{c||}{$1.80 \cdot 10^{-2}$} & \multicolumn{5}{c|}{$1.54 \cdot 10^{-2}$} \\
 \hline
\end{tabular}
\captionof{table}[Résultats dans le cas AR(2), $(\phi_1^\star, \phi_2^\star, \sigma^\star) = (-1.2,-0.4,0.4)$.]{Estimates of the number of changes, of the order of the autoregression, and RMSEs of the estimates of the autoregression parameters, for 100 AR(2) series with the parameters $(\phi_1^\star, \phi_2^\star, \sigma^\star) = (-1.2,-0.4,0.4)$.}\label{table:phi1_-12_phi2_-04_sigma04}
\vspace{2cm}
\includegraphics[width=.55\textwidth]{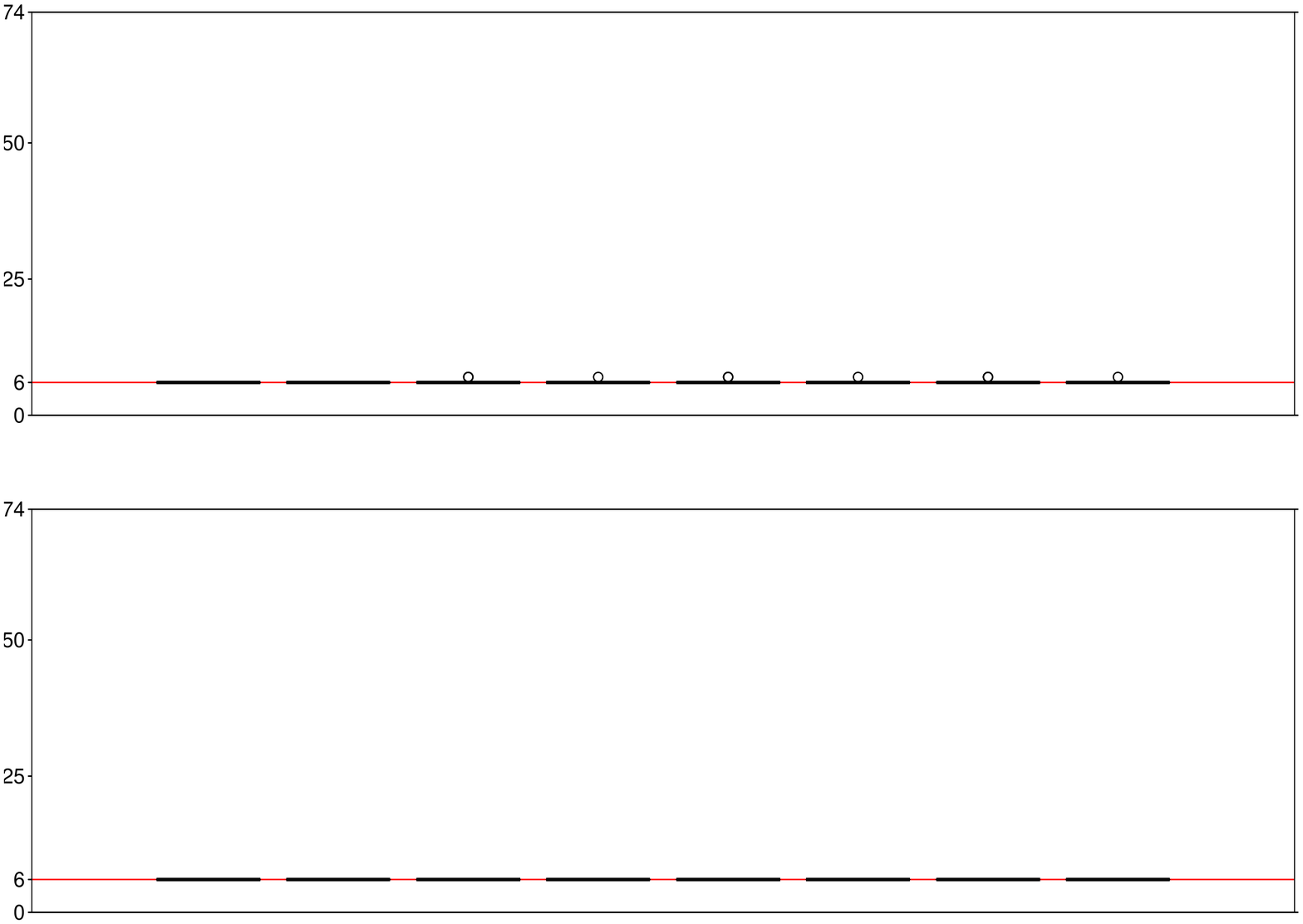}
\captionof{figure}[Boîtes à moustaches du nombre de ruptures estimé, cas AR(2), $(\phi_1^\star, \phi_2^\star, \sigma^\star) = (-1.2,-0.4,0.4)$.]{Boxplots of the estimates of the number of changes for 100 AR(2) series with the parameters $(\phi_1^\star, \phi_2^\star, \sigma^\star) = (-1.2,-0.4,0.4)$. $n=7200$ (top) or $14400$ (bottom). In each plot, the estimates boxplots are in the following order (from left to right): $\widehat{m}^0_Y $, $\widehat{m}^0 $, $\widehat{m} $, $\widehat{m}_{PP} $, $\widehat{m}^\star $, $\widehat{m}_{PP}^\star $, $\widehat{m}^\prime $, $\widehat{m}_{PP}^\prime $. The true number of changes is equal to 6 (red horizontal line).}\label{fig:box:phi1_-12_phi2_-04_sigma04}
\end{minipage} 
\begin{figure}[h]
\centering
\includegraphics[width=\textwidth]{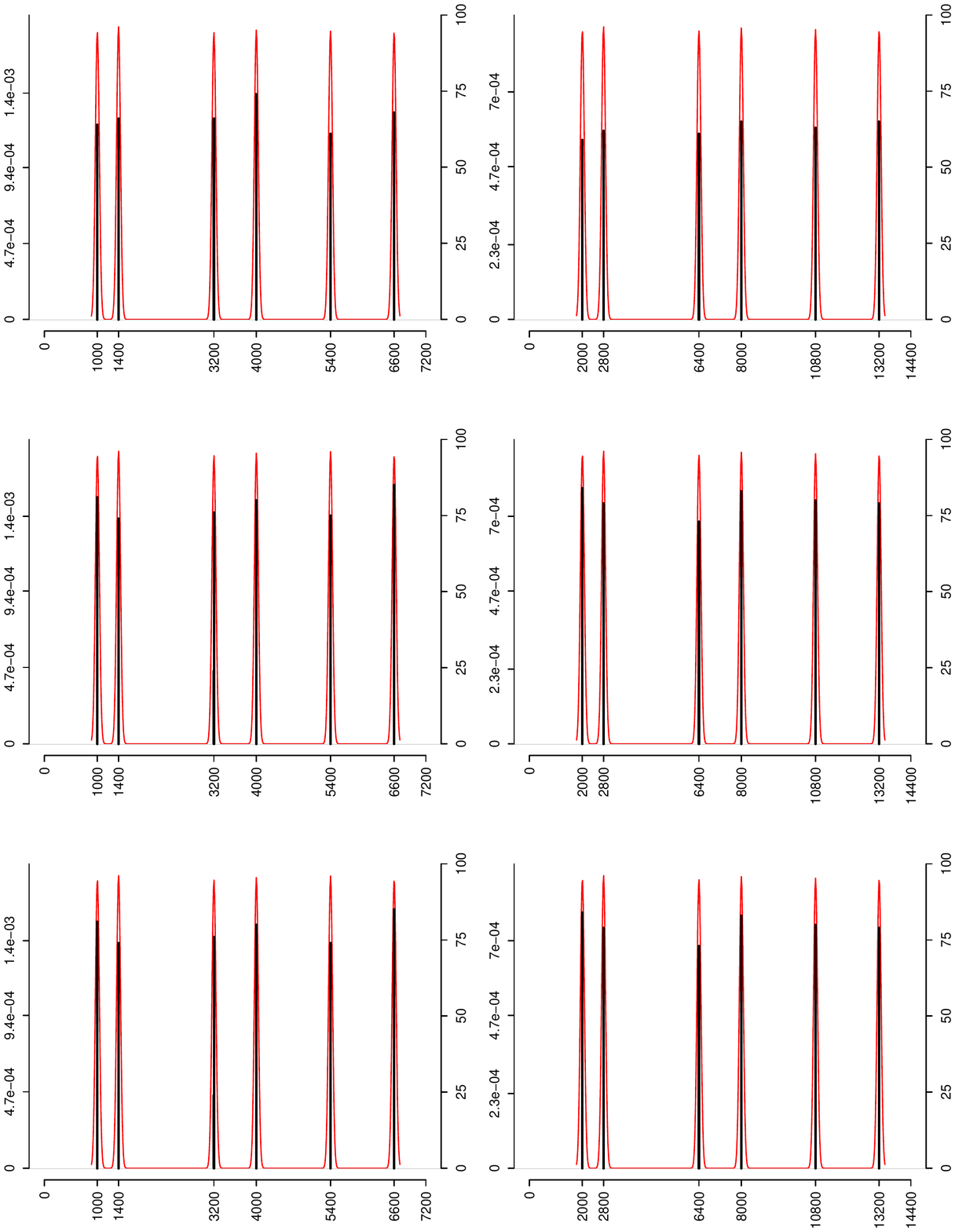}
\caption[Fréquence des instants de rupture estimés, cas AR(2), $(\phi_1^\star, \phi_2^\star, \sigma^\star) = (-1.2,-0.4,0.4)$.]{Frequency plots of the change-point location estimates for 100 AR(2) series with the parameters $(\phi_1^\star, \phi_2^\star, \sigma^\star) = (-1.2,-0.4,0.4)$. $n=7200$ (left) or $14400$ (right). Estimates: $\widehat{t}_n^0$ (top), $\widehat{t}_{n, PP}$ (middle), $\widehat{t}_{n, PP}^\prime$ (bottom). The black line represents the absolute frequency of each location between $1$ and $n$ in estimates (scale on right axis). The red line represents the Gaussian kernel density estimate of this dataset (scale on left axis).}\label{fig:dens:phi1_-12_phi2_-04_sigma04}
\end{figure}

\FloatBarrier
\begin{table}
\centering
\begin{tabular}{|c||c|c|c|c|c||c|c|c|c|c|}
\hline
$n$ & \multicolumn{5}{c||}{7200} & \multicolumn{5}{c|}{14400} \\
\hline
\hline
estimate \textbackslash number of changes  & $<5$ & $5$ & $\mathbf{6} $ & $7$ & $>7$ & $<5$ & $5$ & $\mathbf{6} $ & $7$ & $>7$ \\
 \hline
 $\widehat{m}^0_Y$ & $0$ & $0$ & $\mathbf{0}$ & $0$ & $100$ & $0$ & $0$ & $\mathbf{0}$ & $0$ & $100$ \\
 \hline
$\widehat{m}^0$ & $0$ & $0$ & $\mathbf{0}$ & $0$ & $100$ & $0$ & $0$ & $\mathbf{0}$ & $1$ & $99$ \\
 \hline
$\widehat{m}$ & $3$ & $0$ & $\mathbf{91}$ & $6$ & $0$ & $0$ & $0$ & $\mathbf{99}$ & $1$ & $0$ \\
 \hline
$\widehat{m}_{PP}$ & $3$ & $0$ & $\mathbf{97}$ & $0$ & $0$ & $0$ & $0$ & $\mathbf{100}$ & $0$ & $0$ \\
 \hline
$\widehat{m}^\star$ & $0$ & $0$ & $\mathbf{94}$ & $6$ & $0$ & $0$ & $0$ & $\mathbf{99}$ & $1$ & $0$ \\
 \hline
$\widehat{m}_{PP}^\star$ & $0$ & $0$ & $\mathbf{100}$ & $0$ & $0$ & $0$ & $0$ & $\mathbf{100}$ & $0$ & $0$ \\
 \hline
$\widehat{m}^\prime$ & $4$ & $0$ & $\mathbf{90}$ & $6$ & $0$ & $0$ & $0$ & $\mathbf{98}$ & $2$ & $0$ \\
 \hline
$\widehat{m}_{PP}^\prime$ & $4$ & $0$ & $\mathbf{96}$ & $0$ & $0$ & $0$ & $0$ & $\mathbf{99}$ & $1$ & $0$ \\
 \hline
 \hline
 estimate \textbackslash order of the autoregression & 0 & 1 & $\mathbf{2}$ & 3 & $>3$ & 0 & 1 & $\mathbf{2}$ & 3 & $>3$ \\
 \hline
 $\widehat{p}^\prime$  & $0$ & $0$ & $\mathbf{46}$ & $24$ & $30$ & $0$ & $0$ & $\mathbf{55}$ & $15$ & $30$ \\
 \hline
 \hline
 $\widetilde{\phi}_{n,1}^{(2)}$ RMSE & \multicolumn{5}{c||}{$4.93 \cdot 10^{-2}$} & \multicolumn{5}{c|}{$3.46 \cdot 10^{-2}$} \\
 \hline
 $\widetilde{\phi}_{n,2}^{(2)}$ RMSE & \multicolumn{5}{c||}{$3.13 \cdot 10^{-2}$} & \multicolumn{5}{c|}{$2.16 \cdot 10^{-2}$} \\
 \hline
\end{tabular}
\caption[Résultats dans le cas AR(2), $(\phi_1^\star, \phi_2^\star, \sigma^\star) = (1.6,-0.8,0.4)$.]{Estimates of the number of changes, of the order of the autoregression, and RMSEs of the estimates of the autoregression parameters, for 100 AR(2) series with the parameters $(\phi_1^\star, \phi_2^\star, \sigma^\star) = (1.6,-0.8,0.4)$.}\label{table:phi1_16_phi2_-08_sigma04}
\end{table}

\begin{figure}[b]
\centering
\includegraphics[width=.55\textwidth]{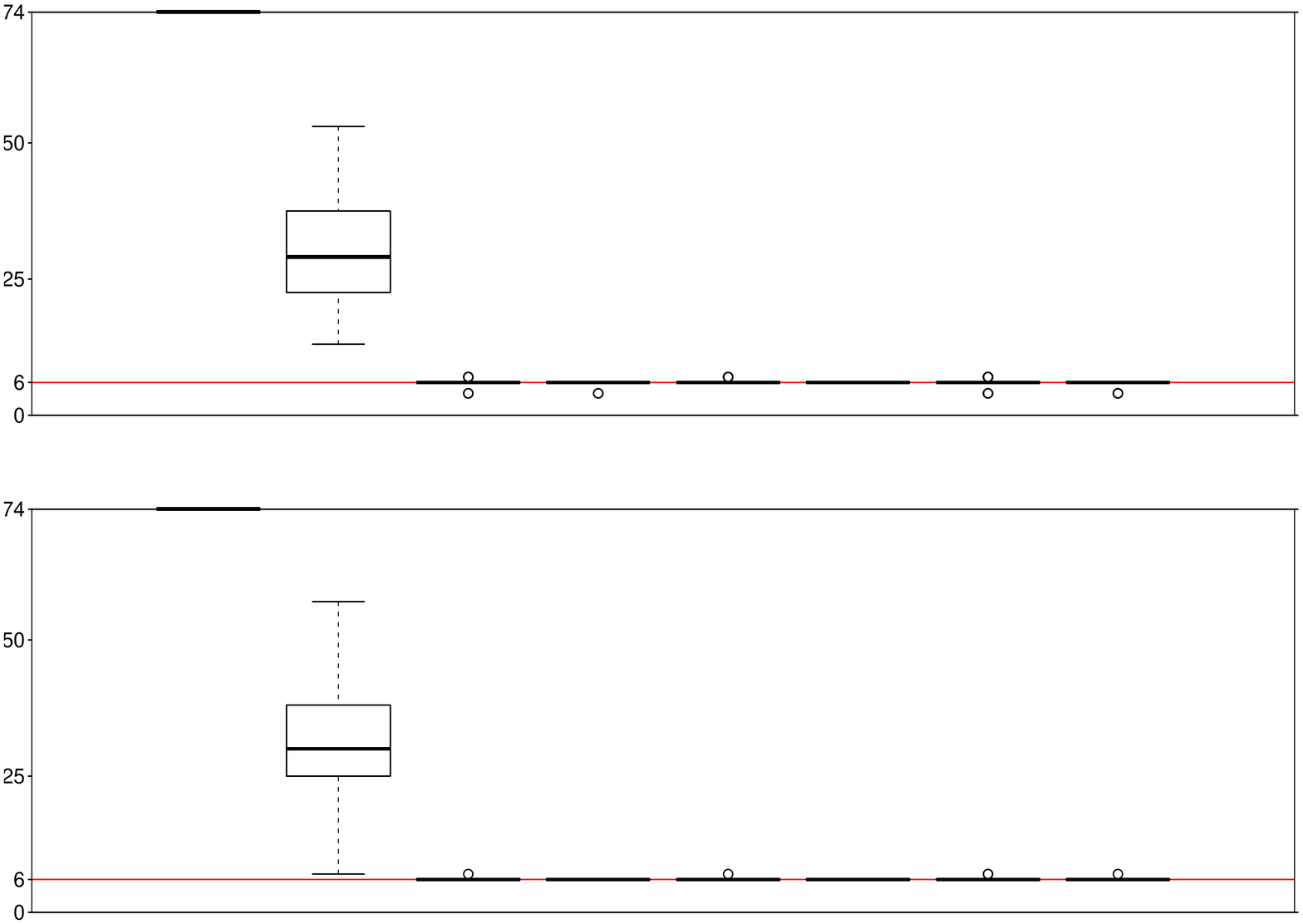}

\caption[Boîtes à moustaches du nombre de ruptures estimé, cas AR(2), $(\phi_1^\star, \phi_2^\star, \sigma^\star) = (1.6,-0.8,0.4)$.]{Boxplots of the estimates of the number of changes for 100 AR(2) series with the parameters $(\phi_1^\star, \phi_2^\star, \sigma^\star) = (1.6,-0.8,0.4)$. $n=7200$ (top) or $14400$ (bottom). In each plot, the estimates boxplots are in the following order (from left to right): $\widehat{m}^0_Y $, $\widehat{m}^0 $, $\widehat{m} $, $\widehat{m}_{PP} $, $\widehat{m}^\star $, $\widehat{m}_{PP}^\star $, $\widehat{m}^\prime $, $\widehat{m}_{PP}^\prime $. The true number of changes is equal to 6 (red horizontal line).}\label{fig:box:phi1_16_phi2_-08_sigma04}
\end{figure}

\begin{figure}[h]
\centering
\includegraphics[width=\textwidth]{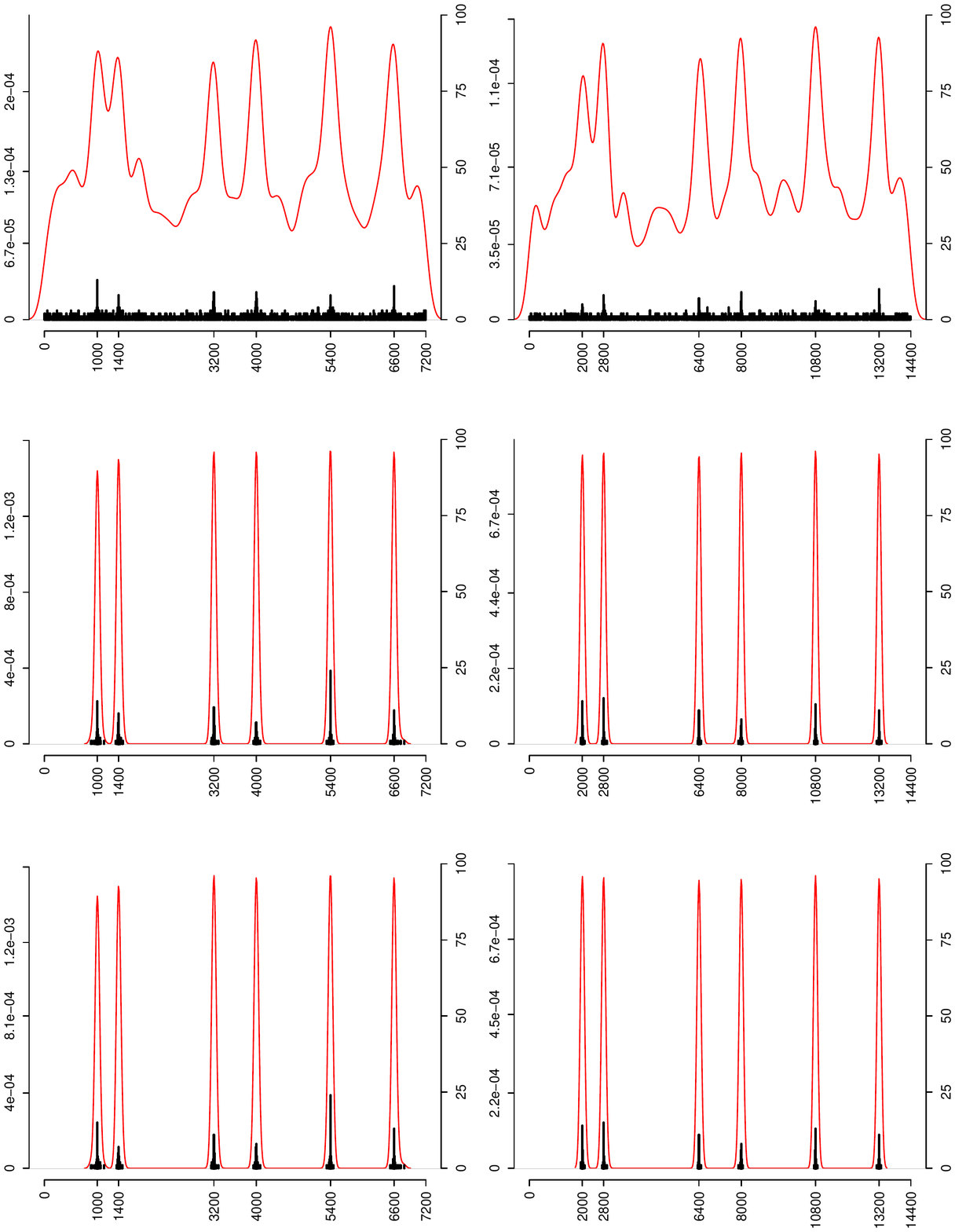}

\caption[Fréquence des instants de rupture estimés, cas AR(2), $(\phi_1^\star, \phi_2^\star, \sigma^\star) = (1.6,-0.8,0.4)$.]{Frequency plots of the change-point location estimates for 100 AR(2) series with the parameters $(\phi_1^\star, \phi_2^\star, \sigma^\star) = (1.6,-0.8,0.4)$. $n=7200$ (left) or $14400$ (right). Estimates: $\widehat{t}_n^0$ (top), $\widehat{t}_{n, PP}$ (middle), $\widehat{t}_{n, PP}^\prime$ (bottom). The black line represents the absolute frequency of each location between $1$ and $n$ in estimates (scale on right axis). The red line represents the Gaussian kernel density estimate of this dataset (scale on left axis).}\label{fig:dens:phi1_16_phi2_-08_sigma04}
\end{figure}

\FloatBarrier
\begin{table}
\centering
\begin{tabular}{|c||c|c|c|c|c||c|c|c|c|c|}
\hline
$n$ & \multicolumn{5}{c||}{7200} & \multicolumn{5}{c|}{14400} \\
\hline
\hline
estimate \textbackslash number of changes  & $<5$ & $5$ & $\mathbf{6} $ & $7$ & $>7$ & $<5$ & $5$ & $\mathbf{6} $ & $7$ & $>7$ \\
 \hline
 $\widehat{m}^0_Y$ & $0$ & $0$ & $\mathbf{0}$ & $0$ & $100$ & $0$ & $0$ & $\mathbf{0}$ & $0$ & $100$ \\
 \hline
$\widehat{m}^0$ & $0$ & $0$ & $\mathbf{51}$ & $20$ & $29$ & $0$ & $0$ & $\mathbf{60}$ & $17$ & $23$ \\
 \hline
$\widehat{m}$ & $3$ & $0$ & $\mathbf{96}$ & $3$ & $1$ & $0$ & $0$ & $\mathbf{98}$ & $2$ & $0$ \\
 \hline
$\widehat{m}_{PP}$ & $3$ & $0$ & $\mathbf{97}$ & $2$ & $1$ & $0$ & $0$ & $\mathbf{98}$ & $2$ & $0$ \\
 \hline
$\widehat{m}^\star$ & $0$ & $0$ & $\mathbf{97}$ & $3$ & $0$ & $0$ & $0$ & $\mathbf{99}$ & $1$ & $0$ \\
 \hline
$\widehat{m}_{PP}^\star$ & $0$ & $0$ & $\mathbf{98}$ & $2$ & $0$ & $0$ & $0$ & $\mathbf{99}$ & $1$ & $0$ \\
 \hline
$\widehat{m}^\prime$ & $0$ & $0$ & $\mathbf{97}$ & $3$ & $0$ & $0$ & $0$ & $\mathbf{99}$ & $1$ & $0$ \\
 \hline
$\widehat{m}_{PP}^\prime$ & $0$ & $0$ & $\mathbf{98}$ & $2$ & $0$ & $0$ & $0$ & $\mathbf{99}$ & $1$ & $0$ \\
 \hline
 \hline
 estimate \textbackslash order of the autoregression & 0 & 1 & $\mathbf{2}$ & 3 & $>3$ & 0 & 1 & $\mathbf{2}$ & 3 & $>3$ \\
 \hline
 $\widehat{p}^\prime$  & $0$ & $0$ & $\mathbf{59}$ & $25$ & $16$ & $0$ & $0$ & $\mathbf{45}$ & $30$ & $25$ \\
 \hline
 \hline
 $\widetilde{\phi}_{n,1}^{(2)}$ RMSE & \multicolumn{5}{c||}{$7.00 \cdot 10^{-2}$} & \multicolumn{5}{c|}{$6.44 \cdot 10^{-2}$} \\
 \hline
 $\widetilde{\phi}_{n,2}^{(2)}$ RMSE & \multicolumn{5}{c||}{$4.20 \cdot 10^{-2}$} & \multicolumn{5}{c|}{$3.68 \cdot 10^{-2}$} \\
 \hline
\end{tabular}
\caption[Résultats dans le cas AR(2), $(\phi_1^\star, \phi_2^\star, \sigma^\star) = (0.2,0.2,0.4)$.]{Estimates of the number of changes, of the order of the autoregression, and RMSEs of the estimates of the autoregression parameters, for 100 AR(2) series with the parameters $(\phi_1^\star, \phi_2^\star, \sigma^\star) = (0.2,0.2,0.4)$.}\label{table:phi1_02_phi2_02_sigma04}
\end{table}

\begin{figure}[b]
\centering
\includegraphics[width=.55\textwidth]{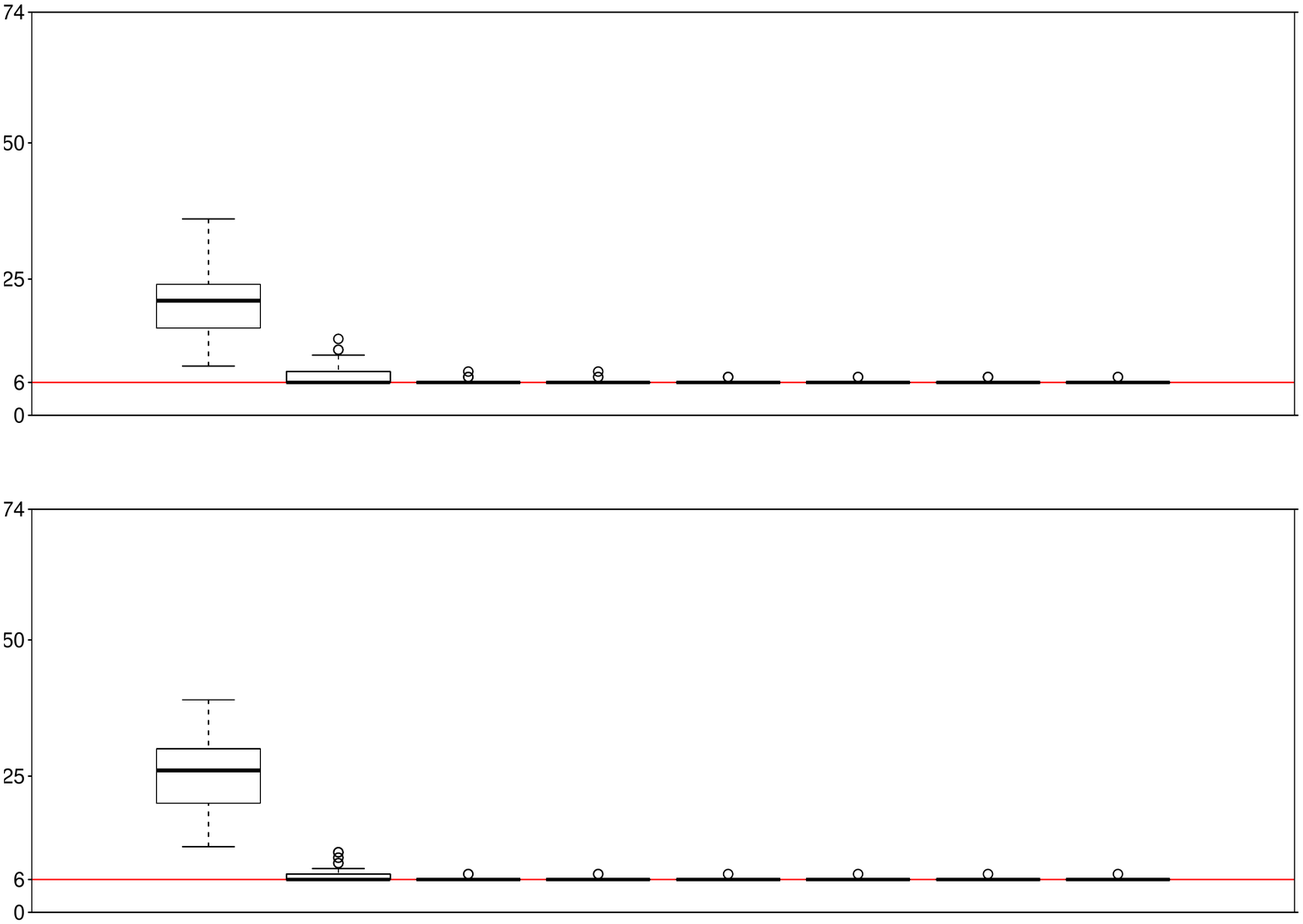}
\caption[Boîtes à moustaches du nombre de ruptures estimé, cas AR(2), $(\phi_1^\star, \phi_2^\star, \sigma^\star) = (0.2,0.2,0.4)$.]{Boxplots of the estimates of the number of changes for 100 AR(2) series with the parameters $(\phi_1^\star, \phi_2^\star, \sigma^\star) = (0.2,0.2,0.4)$. $n=7200$ (top) or $14400$ (bottom). In each plot, the estimates boxplots are in the following order (from left to right): $\widehat{m}^0_Y $, $\widehat{m}^0 $, $\widehat{m} $, $\widehat{m}_{PP} $, $\widehat{m}^\star $, $\widehat{m}_{PP}^\star $, $\widehat{m}^\prime $, $\widehat{m}_{PP}^\prime $. The true number of changes is equal to 6 (red horizontal line).}\label{fig:box:phi1_02_phi2_02_sigma04}
\end{figure}

\begin{figure}[h]
\centering
\includegraphics[width=\textwidth]{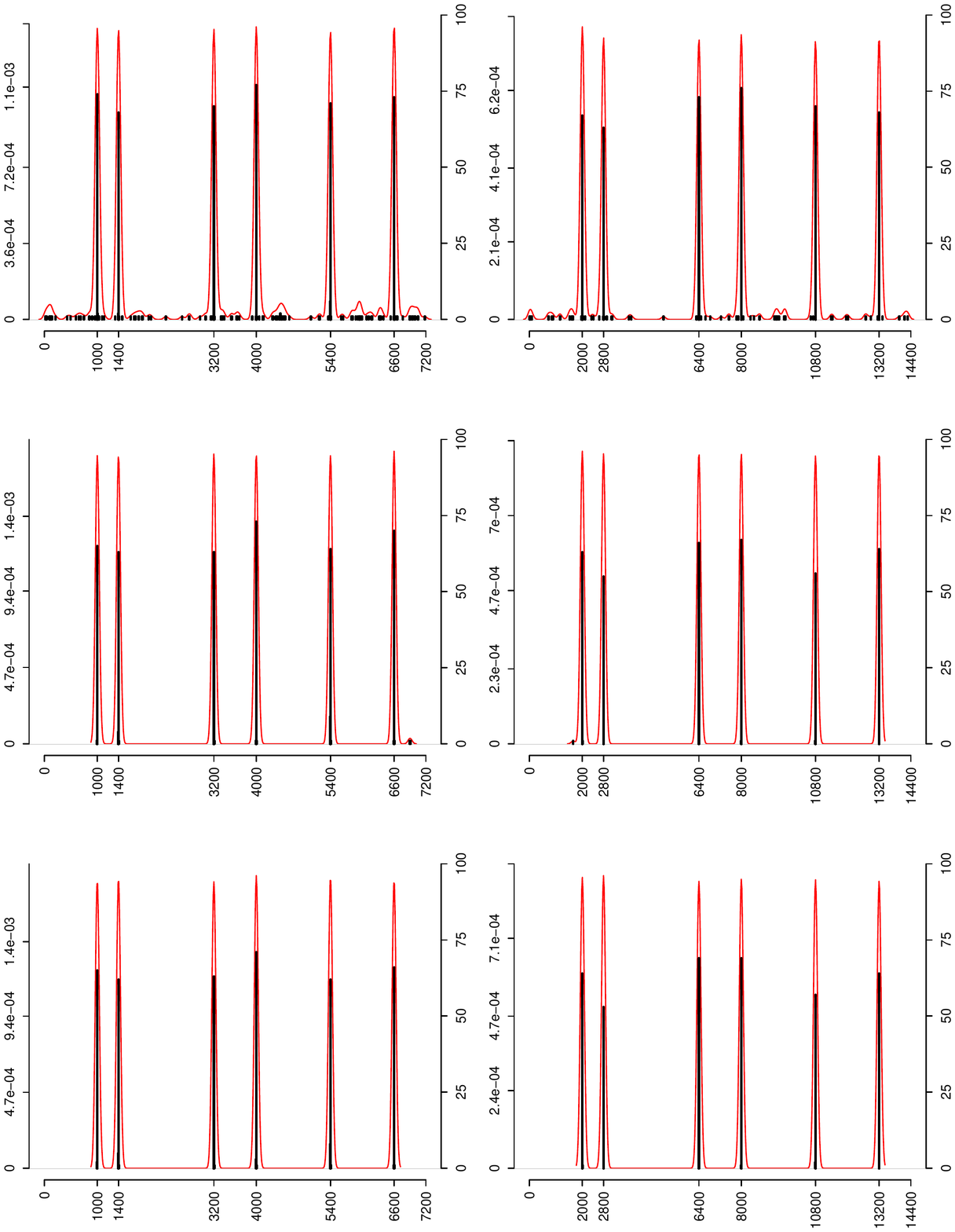}
\caption[Fréquence des instants de rupture estimés, cas AR(2), $(\phi_1^\star, \phi_2^\star, \sigma^\star) = (0.2,0.2,0.4)$.]{Frequency plots of the change-point location estimates for 100 AR(2) series with the parameters $(\phi_1^\star, \phi_2^\star, \sigma^\star) = (0.2,0.2,0.4)$. $n=7200$ (left) or $14400$ (right). Estimates: $\widehat{t}_n^0$ (top), $\widehat{t}_{n, PP}$ (middle), $\widehat{t}_{n, PP}^\prime$ (bottom). The black line represents the absolute frequency of each location between $1$ and $n$ in estimates (scale on right axis). The red line represents the Gaussian kernel density estimate of this dataset (scale on left axis).}\label{fig:dens:phi1_02_phi2_02_sigma04}
\end{figure}

\FloatBarrier
\begin{table}
\centering
\begin{tabular}{|c||c|c|c|c|c||c|c|c|c|c|}
\hline
$n$ & \multicolumn{5}{c||}{7200} & \multicolumn{5}{c|}{14400} \\
\hline
\hline
estimate \textbackslash number of changes  & $<5$ & $5$ & $\mathbf{6} $ & $7$ & $>7$ & $<5$ & $5$ & $\mathbf{6} $ & $7$ & $>7$ \\
 \hline
 $\widehat{m}^0_Y$ & $0$ & $0$ & $\mathbf{0}$ & $0$ & $100$ & $0$ & $0$ & $\mathbf{0}$ & $0$ & $100$ \\
 \hline
$\widehat{m}^0$ & $0$ & $0$ & $\mathbf{0}$ & $0$ & $100$ & $0$ & $0$ & $\mathbf{0}$ & $0$ & $100$ \\
 \hline
$\widehat{m}$ & $13$ & $0$ & $\mathbf{27}$ & $4$ & $56$ & $25$ & $0$ & $\mathbf{29}$ & $9$ & $37$ \\
 \hline
$\widehat{m}_{PP}$ & $13$ & $0$ & $\mathbf{28}$ & $3$ & $56$ & $25$ & $0$ & $\mathbf{33}$ & $5$ & $37$ \\
 \hline
$\widehat{m}^\star$ & $0$ & $1$ & $\mathbf{80}$ & $18$ & $1$ & $0$ & $0$ & $\mathbf{88}$ & $11$ & $1$ \\
 \hline
$\widehat{m}_{PP}^\star$ & $1$ & $0$ & $\mathbf{91}$ & $7$ & $1$ & $0$ & $0$ & $\mathbf{98}$ & $2$ & $0$ \\
 \hline
$\widehat{m}^\prime$ & $22$ & $0$ & $\mathbf{54}$ & $15$ & $9$ & $6$ & $0$ & $\mathbf{80}$ & $11$ & $3$ \\
 \hline
$\widehat{m}_{PP}^\prime$ & $22$ & $0$ & $\mathbf{66}$ & $4$ & $8$ & $6$ & $0$ & $\mathbf{90}$ & $3$ & $1$ \\
 \hline
 \hline
 estimate \textbackslash order of the autoregression & 0 & 1 & $\mathbf{2}$ & 3 & $>3$ & 0 & 1 & $\mathbf{2}$ & 3 & $>3$ \\
 \hline
 $\widehat{p}^\prime$  & $0$ & $0$ & $\mathbf{18}$ & $12$ & $70$ & $0$ & $0$ & $\mathbf{14}$ & $12$ & $74$ \\
 \hline
 \hline
 $\widetilde{\phi}_{n,1}^{(2)}$ RMSE & \multicolumn{5}{c||}{$3.44 \cdot 10^{-1}$} & \multicolumn{5}{c|}{$2.40 \cdot 10^{-1}$} \\
 \hline
 $\widetilde{\phi}_{n,2}^{(2)}$ RMSE & \multicolumn{5}{c||}{$2.41 \cdot 10^{-1}$} & \multicolumn{5}{c|}{$1.71 \cdot 10^{-1}$} \\
 \hline
\end{tabular}
\caption[Résultats dans le cas AR(2), $(\phi_1^\star, \phi_2^\star, \sigma^\star) = (0.2,0.6,0.4)$.]{Estimates of the number of changes, of the order of the autoregression, and RMSEs of the estimates of the autoregression parameters, for 100 AR(2) series with the parameters $(\phi_1^\star, \phi_2^\star, \sigma^\star) = (0.2,0.6,0.4)$.}\label{table:phi1_02_phi2_06_sigma04}
\end{table}

\begin{figure}[b]
\centering
\includegraphics[width=.55\textwidth]{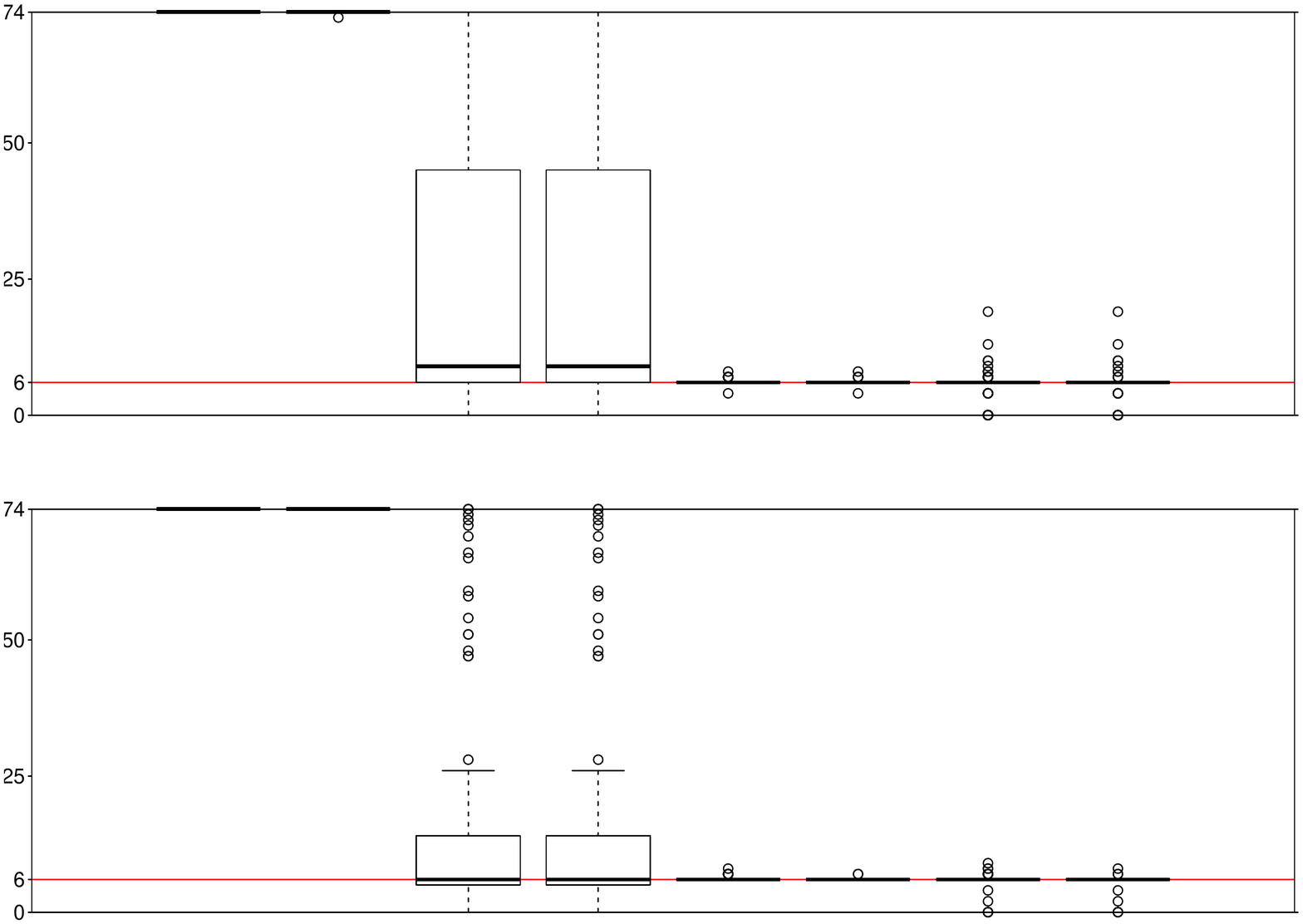}
\caption[Boîtes à moustaches du nombre de ruptures estimé, cas AR(2), $(\phi_1^\star, \phi_2^\star, \sigma^\star) = (0.2,0.6,0.4)$.]{Boxplots of the estimates of the number of changes for 100 AR(2) series with the parameters $(\phi_1^\star, \phi_2^\star, \sigma^\star) = (0.2,0.6,0.4)$. $n=7200$ (top) or $14400$ (bottom). In each plot, the estimates boxplots are in the following order (from left to right): $\widehat{m}^0_Y $, $\widehat{m}^0 $, $\widehat{m} $, $\widehat{m}_{PP} $, $\widehat{m}^\star $, $\widehat{m}_{PP}^\star $, $\widehat{m}^\prime $, $\widehat{m}_{PP}^\prime $. The true number of changes is equal to 6 (red horizontal line).}\label{fig:box:phi1_02_phi2_06_sigma04}
\end{figure}

\begin{figure}
\centering
\includegraphics[width=\textwidth]{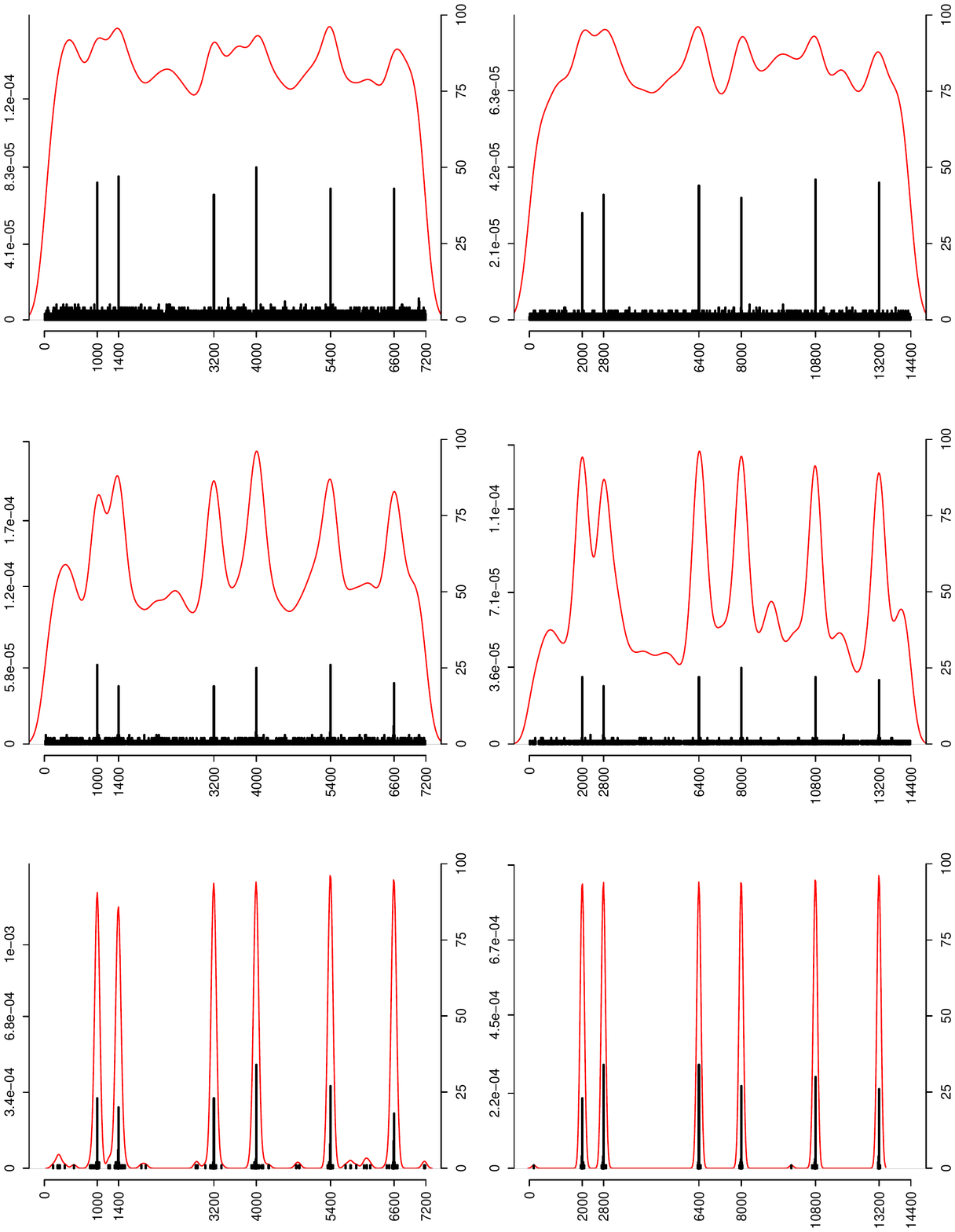}
\caption[Fréquence des instants de rupture estimés, cas AR(2), $(\phi_1^\star, \phi_2^\star, \sigma^\star) = (0.2,0.6,0.4)$.]{Frequency plots of the change-point location estimates for 100 AR(2) series with the parameters $(\phi_1^\star, \phi_2^\star, \sigma^\star) = (0.2,0.6,0.4)$. $n=7200$ (left) or $14400$ (right). Estimates: $\widehat{t}_n^0$ (top), $\widehat{t}_{n, PP}$ (middle), $\widehat{t}_{n, PP}^\prime$ (bottom). The black line represents the absolute frequency of each location between $1$ and $n$ in estimates (scale on right axis). The red line represents the Gaussian kernel density estimate of this dataset (scale on left axis).}\label{fig:dens:phi1_02_phi2_06_sigma04}
\end{figure}
\FloatBarrier
\begin{table}
\centering
\begin{tabular}{|c||c|c|c|c|c||c|c|c|c|c|}
\hline
$n$ & \multicolumn{5}{c||}{7200} & \multicolumn{5}{c|}{14400} \\
\hline
\hline
estimate \textbackslash number of changes  & $<5$ & $5$ & $\mathbf{6} $ & $7$ & $>7$ & $<5$ & $5$ & $\mathbf{6} $ & $7$ & $>7$ \\
 \hline
 $\widehat{m}^0_Y$ & $0$ & $0$ & $\mathbf{0}$ & $0$ & $100$ & $0$ & $0$ & $\mathbf{0}$ & $0$ & $100$ \\
 \hline
$\widehat{m}^0$ & $0$ & $0$ & $\mathbf{0}$ & $0$ & $100$ & $0$ & $0$ & $\mathbf{0}$ & $0$ & $100$ \\
 \hline
$\widehat{m}$ & $2$ & $0$ & $\mathbf{52}$ & $28$ & $18$ & $1$ & $0$ & $\mathbf{59}$ & $28$ & $12$ \\
 \hline
$\widehat{m}_{PP}$ & $3$ & $0$ & $\mathbf{85}$ & $10$ & $2$ & $1$ & $0$ & $\mathbf{96}$ & $3$ & $0$ \\
 \hline
$\widehat{m}^\star$ & $0$ & $0$ & $\mathbf{65}$ & $28$ & $7$ & $0$ & $0$ & $\mathbf{70}$ & $29$ & $1$ \\
 \hline
$\widehat{m}_{PP}^\star$ & $0$ & $0$ & $\mathbf{98}$ & $2$ & $0$ & $0$ & $0$ & $\mathbf{98}$ & $2$ & $0$ \\
 \hline
$\widehat{m}^\prime$ & $0$ & $0$ & $\mathbf{63}$ & $28$ & $9$ & $0$ & $0$ & $\mathbf{74}$ & $24$ & $2$ \\
 \hline
$\widehat{m}_{PP}^\prime$ & $0$ & $0$ & $\mathbf{100}$ & $0$ & $0$ & $0$ & $0$ & $\mathbf{99}$ & $1$ & $0$ \\
 \hline
 \hline
 estimate \textbackslash order of the autoregression & 0 & 1 & $\mathbf{2}$ & 3 & $>3$ & 0 & 1 & $\mathbf{2}$ & 3 & $>3$ \\
 \hline
 $\widehat{p}^\prime$  & $0$ & $0$ & $\mathbf{36}$ & $20$ & $44$ & $0$ & $0$ & $\mathbf{36}$ & $21$ & $43$ \\
 \hline
 \hline
 $\widetilde{\phi}_{n,1}^{(2)}$ RMSE & \multicolumn{5}{c||}{$1.11 \cdot 10^{-1}$} & \multicolumn{5}{c|}{$8.17 \cdot 10^{-2}$} \\
 \hline
 $\widetilde{\phi}_{n,2}^{(2)}$ RMSE & \multicolumn{5}{c||}{$5.16 \cdot 10^{-2}$} & \multicolumn{5}{c|}{$3.76 \cdot 10^{-2}$} \\
 \hline
\end{tabular}
\caption[Résultats dans le cas AR(2), $(\phi_1^\star, \phi_2^\star, \sigma^\star) = (0.4,0.2,0.2)$.]{Estimates of the number of changes, of the order of the autoregression, and RMSEs of the estimates of the autoregression parameters, for 100 AR(2) series with the parameters $(\phi_1^\star, \phi_2^\star, \sigma^\star) = (0.4,0.2,0.2)$.}\label{table:phi1_04_phi2_02_sigma02}
\end{table}

\begin{figure}[b]
\centering
\includegraphics[width=.55\textwidth]{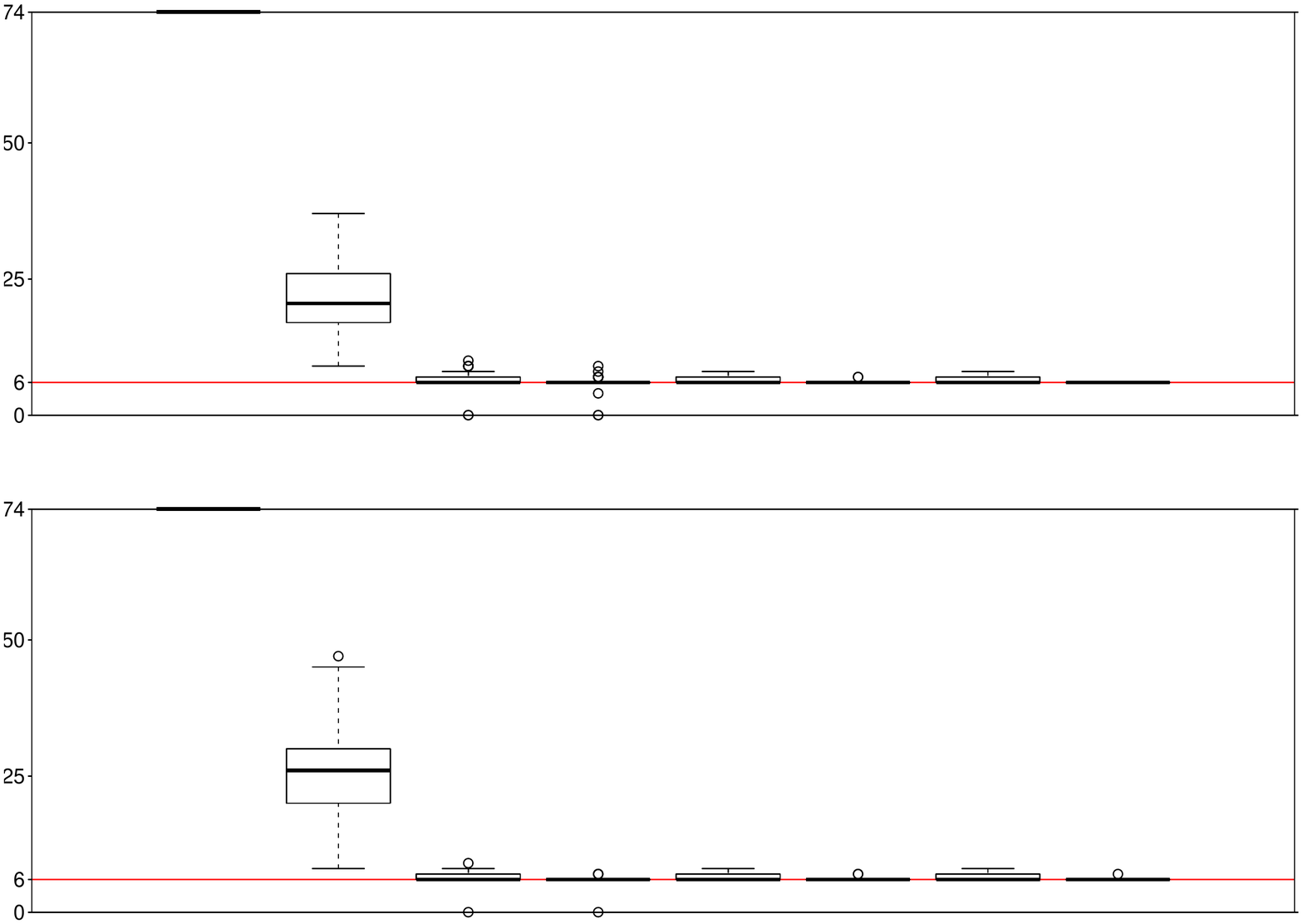}

\caption[Boîtes à moustaches du nombre de ruptures estimé, cas AR(2), $(\phi_1^\star, \phi_2^\star, \sigma^\star) = (0.4,0.2,0.2)$.]{Boxplots of the estimates of the number of changes for 100 AR(2) series with the parameters $(\phi_1^\star, \phi_2^\star, \sigma^\star) = (0.4,0.2,0.2)$. $n=7200$ (top) or $14400$ (bottom). In each plot, the estimates boxplots are in the following order (from left to right): $\widehat{m}^0_Y $, $\widehat{m}^0 $, $\widehat{m} $, $\widehat{m}_{PP} $, $\widehat{m}^\star $, $\widehat{m}_{PP}^\star $, $\widehat{m}^\prime $, $\widehat{m}_{PP}^\prime $. The true number of changes is equal to 6 (red horizontal line).}\label{fig:box:phi1_04_phi2_02_sigma02}
\end{figure}

\begin{figure}
\centering
\includegraphics[width=\textwidth]{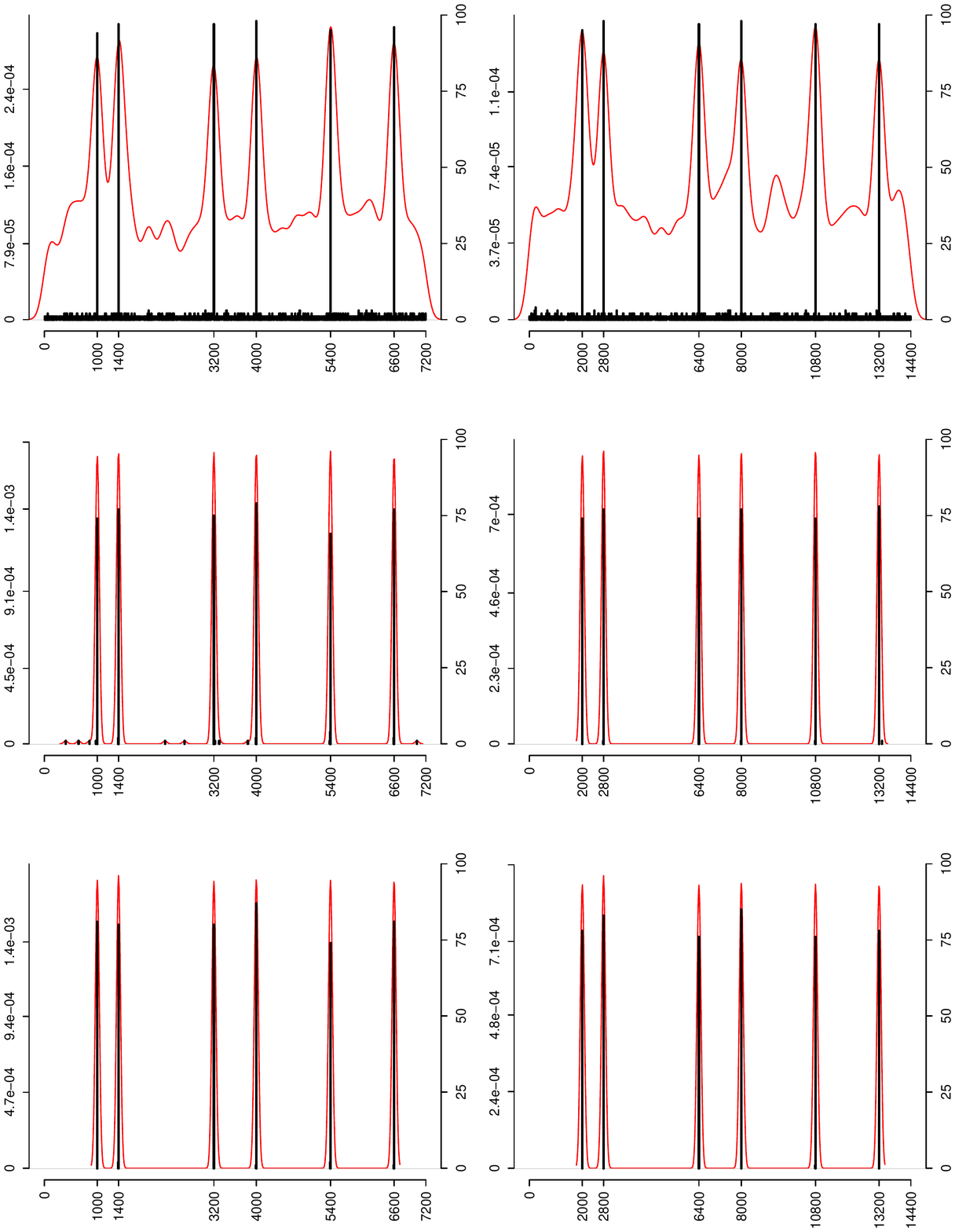}
\caption[Fréquence des instants de rupture estimés, cas AR(2), $(\phi_1^\star, \phi_2^\star, \sigma^\star) = (0.4,0.2,0.2)$.]{Frequency plots of the change-point location estimates for 100 AR(2) series with the parameters $(\phi_1^\star, \phi_2^\star, \sigma^\star) = (0.4,0.2,0.2)$. $n=7200$ (left) or $14400$ (right). Estimates: $\widehat{t}_n^0$ (top), $\widehat{t}_{n, PP}$ (middle), $\widehat{t}_{n, PP}^\prime$ (bottom). The black line represents the absolute frequency of each location between $1$ and $n$ in estimates (scale on right axis). The red line represents the Gaussian kernel density estimate of this dataset (scale on left axis).}\label{fig:dens:phi1_04_phi2_02_sigma02}
\end{figure}

\FloatBarrier
\begin{table}
\centering
\begin{tabular}{|c||c|c|c|c|c||c|c|c|c|c|}
\hline
$n$ & \multicolumn{5}{c||}{7200} & \multicolumn{5}{c|}{14400} \\
\hline
\hline
estimate \textbackslash number of changes  & $<5$ & $5$ & $\mathbf{6} $ & $7$ & $>7$ & $<5$ & $5$ & $\mathbf{6} $ & $7$ & $>7$ \\
 \hline
 $\widehat{m}^0_Y$ & $0$ & $0$ & $\mathbf{0}$ & $0$ & $100$ & $0$ & $0$ & $\mathbf{0}$ & $0$ & $100$ \\
 \hline
$\widehat{m}^0$ & $0$ & $0$ & $\mathbf{31}$ & $16$ & $53$ & $0$ & $0$ & $\mathbf{31}$ & $14$ & $55$ \\
 \hline
$\widehat{m}$ & $2$ & $0$ & $\mathbf{90}$ & $4$ & $4$ & $0$ & $0$ & $\mathbf{100}$ & $0$ & $0$ \\
 \hline
$\widehat{m}_{PP}$ & $2$ & $0$ & $\mathbf{92}$ & $2$ & $4$ & $0$ & $0$ & $\mathbf{100}$ & $0$ & $0$ \\
 \hline
$\widehat{m}^\star$ & $0$ & $0$ & $\mathbf{100}$ & $0$ & $0$ & $0$ & $0$ & $\mathbf{99}$ & $1$ & $0$ \\
 \hline
$\widehat{m}_{PP}^\star$ & $0$ & $0$ & $\mathbf{100}$ & $0$ & $0$ & $0$ & $0$ & $\mathbf{100}$ & $0$ & $0$ \\
 \hline
$\widehat{m}^\prime$ & $0$ & $0$ & $\mathbf{99}$ & $1$ & $0$ & $0$ & $0$ & $\mathbf{99}$ & $1$ & $0$ \\
 \hline
$\widehat{m}_{PP}^\prime$ & $0$ & $0$ & $\mathbf{99}$ & $1$ & $0$ & $0$ & $0$ & $\mathbf{100}$ & $0$ & $0$ \\
 \hline
 \hline
 estimate \textbackslash order of the autoregression & $<4$ & $4$ & $\mathbf{5}$ & $6$ & $>6$ & $<4$ & $4$ & $\mathbf{5}$ & $6$ & $>6$ \\
 \hline
 $\widehat{p}^\prime$  & $0$ & $0$ & $\mathbf{41}$ & $25$ & $34$ & $0$ & $0$ & $\mathbf{45}$ & $21$ & $34$ \\
 \hline
 \hline
 $\widetilde{\phi}_{n,1}^{(5)}$ RMSE & \multicolumn{5}{c||}{$1.01 \cdot 10^{-1}$} & \multicolumn{5}{c|}{$6.92 \cdot 10^{-2}$} \\
 \hline
 $\widetilde{\phi}_{n,2}^{(5)}$ RMSE & \multicolumn{5}{c||}{$4.36 \cdot 10^{-2}$} & \multicolumn{5}{c|}{$3.19 \cdot 10^{-2}$} \\
 \hline
  $\widetilde{\phi}_{n,3}^{(5)}$ RMSE & \multicolumn{5}{c||}{$3.54 \cdot 10^{-2}$} & \multicolumn{5}{c|}{$2.45 \cdot 10^{-2}$} \\
 \hline
 $\widetilde{\phi}_{n,4}^{(5)}$ RMSE & \multicolumn{5}{c||}{$2.48 \cdot 10^{-2}$} & \multicolumn{5}{c|}{$1.84 \cdot 10^{-2}$} \\
 \hline
 $\widetilde{\phi}_{n,5}^{(5)}$ RMSE & \multicolumn{5}{c||}{$3.72 \cdot 10^{-2}$} & \multicolumn{5}{c|}{$2.35 \cdot 10^{-2}$} \\
 \hline
\end{tabular}
\caption[Résultats dans le cas AR(5), $(\phi_1^\star, \phi_2^\star ,\phi_3^\star , \phi_4^\star , \phi_5^\star , \sigma^\star) = (0.5,0,0,0.5,-0.5,0.4)$.]{Estimates of the number of changes, of the order of the autoregression, and RMSEs of the estimates of the autoregression parameters, for 100 AR(5) series with the parameters $(\phi_1^\star, \phi_2^\star ,\phi_3^\star , \phi_4^\star , \phi_5^\star , \sigma^\star) = (0.5,0,0,0.5,-0.5,0.4)$.}\label{table:phi1_05_phi2_0_phi3_0_phi4_05_phi5_-05_sigma04}
\end{table}

\begin{figure}[b]
\centering
\includegraphics[width=.55\textwidth]{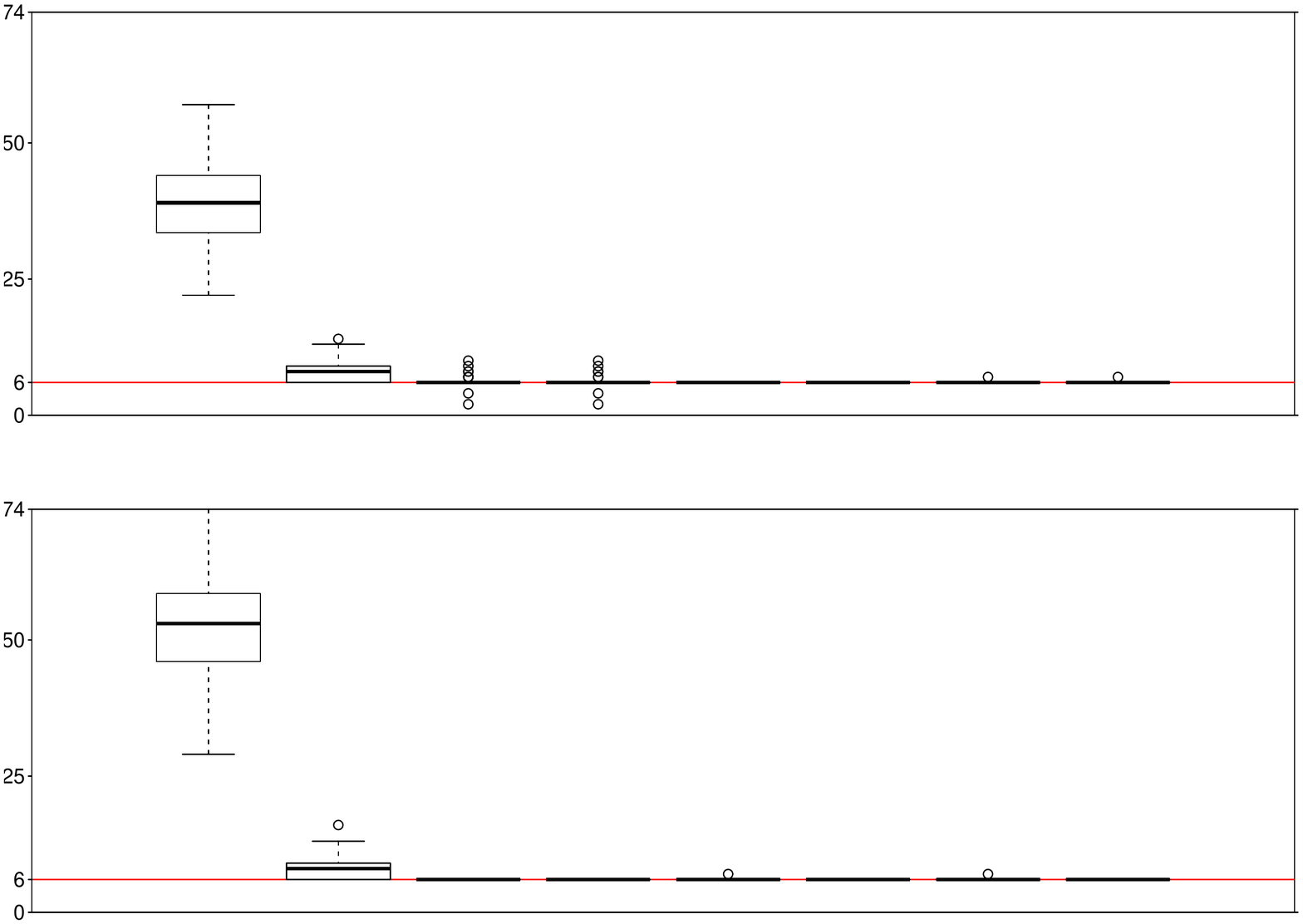}
\caption[Boîtes à moustaches du nombre de ruptures estimé, cas AR(5), $(\phi_1^\star, \phi_2^\star ,\phi_3^\star , \phi_4^\star , \phi_5^\star , \sigma^\star) = (0.5,0,0,0.5,-0.5,0.4)$.]{Boxplots of the estimates of the number of changes for 100 AR(5) series with the parameters $(\phi_1^\star, \phi_2^\star ,\phi_3^\star , \phi_4^\star , \phi_5^\star , \sigma^\star) = (0.5,0,0,0.5,-0.5,0.4)$. $n=7200$ (top) or $14400$ (bottom). In each plot, the estimates boxplots are in the following order (from left to right): $\widehat{m}^0_Y $, $\widehat{m}^0 $, $\widehat{m} $, $\widehat{m}_{PP} $, $\widehat{m}^\star $, $\widehat{m}_{PP}^\star $, $\widehat{m}^\prime $, $\widehat{m}_{PP}^\prime $. The true number of changes is equal to 6 (red horizontal line).}\label{fig:box:phi1_05_phi2_0_phi3_0_phi4_05_phi5_-05_sigma04}
\end{figure}

\begin{figure}[h]
\centering
\includegraphics[width=\textwidth]{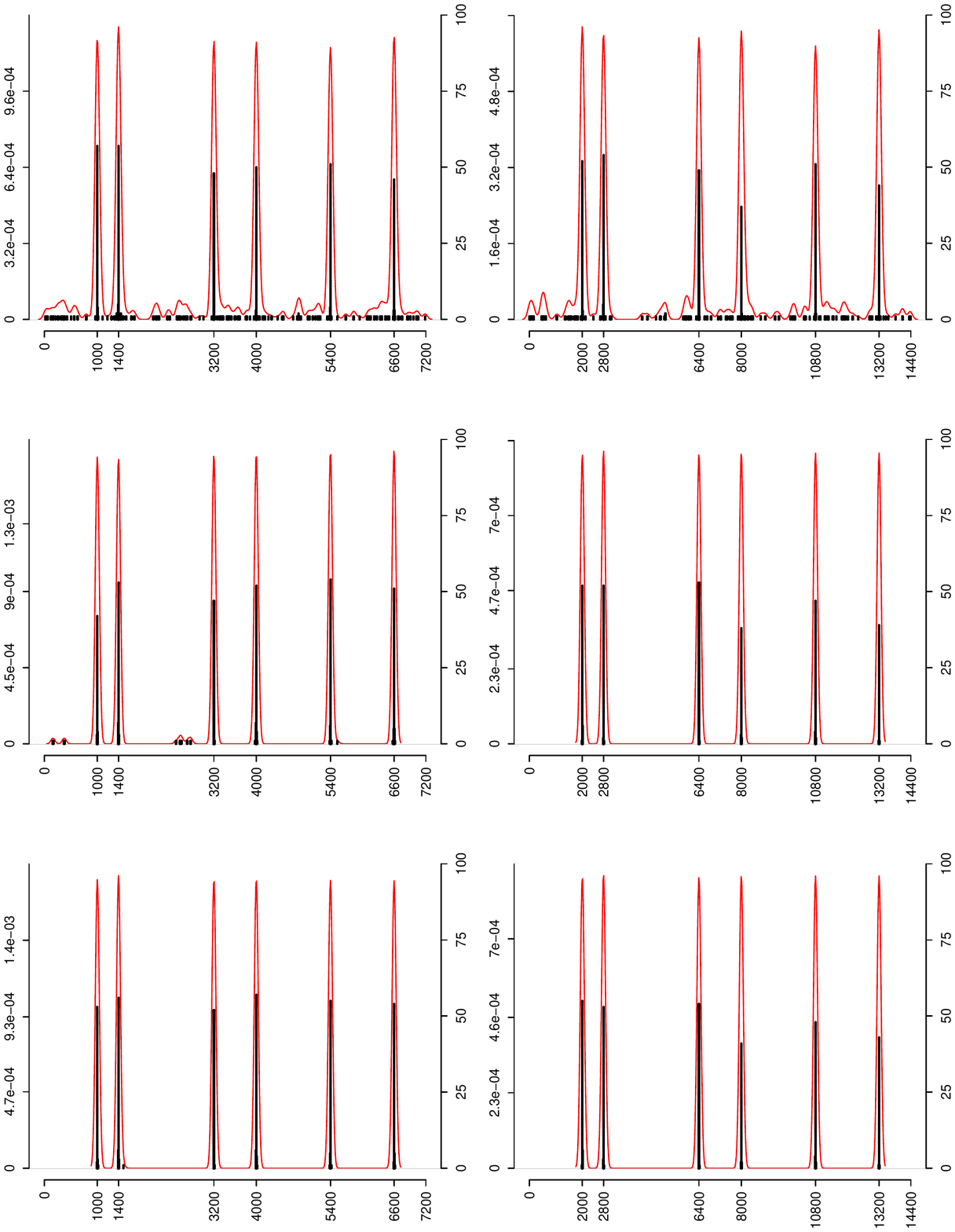}
\caption[Fréquence des instants de rupture estimés, cas AR(5), $(\phi_1^\star, \phi_2^\star ,\phi_3^\star , \phi_4^\star , \phi_5^\star , \sigma^\star) = (0.5,0,0,0.5,-0.5,0.4)$.]{Frequency plots of the change-point location estimates for 100 AR(5) series with the parameters $(\phi_1^\star, \phi_2^\star ,\phi_3^\star , \phi_4^\star , \phi_5^\star , \sigma^\star) = (0.5,0,0,0.5,-0.5,0.4)$. $n=7200$ (left) or $14400$ (right). Estimates: $\widehat{t}_n^0$ (top), $\widehat{t}_{n, PP}$ (middle), $\widehat{t}_{n, PP}^\prime$ (bottom). The black line represents the absolute frequency of each location between $1$ and $n$ in estimates (scale on right axis). The red line represents the Gaussian kernel density estimate of this dataset (scale on left axis).}\label{fig:dens:phi1_05_phi2_0_phi3_0_phi4_05_phi5_-05_sigma04}
\end{figure}

\begin{table}
\centering
\begin{tabular}{|c||c|c|c|c|c||c|c|c|c|c|}
\hline
$n$ & \multicolumn{5}{c||}{7200} & \multicolumn{5}{c|}{14400} \\
\hline
\hline
estimate \textbackslash number of changes  & $<5$ & $5$ & $\mathbf{6} $ & $7$ & $>7$ & $<5$ & $5$ & $\mathbf{6} $ & $7$ & $>7$ \\
 \hline
 $\widehat{m}^0_Y$ & $0$ & $0$ & $\mathbf{0}$ & $0$ & $100$ & $0$ & $0$ & $\mathbf{0}$ & $0$ & $100$ \\
 \hline
$\widehat{m}^0$ & $0$ & $0$ & $\mathbf{53}$ & $5$ & $42$ & $0$ & $0$ & $\mathbf{54}$ & $2$ & $44$ \\
 \hline
$\widehat{m}$ & $0$ & $0$ & $\mathbf{100}$ & $0$ & $0$ & $0$ & $0$ & $\mathbf{100}$ & $0$ & $0$ \\
 \hline
$\widehat{m}_{PP}$ & $0$ & $0$ & $\mathbf{100}$ & $0$ & $0$ & $0$ & $0$ & $\mathbf{100}$ & $0$ & $0$ \\
 \hline
$\widehat{m}^\star$ & $0$ & $0$ & $\mathbf{100}$ & $0$ & $0$ & $0$ & $0$ & $\mathbf{100}$ & $0$ & $0$ \\
 \hline
$\widehat{m}_{PP}^\star$ & $0$ & $0$ & $\mathbf{100}$ & $0$ & $0$ & $0$ & $0$ & $\mathbf{100}$ & $0$ & $0$ \\
 \hline
$\widehat{m}^\prime$ & $0$ & $0$ & $\mathbf{100}$ & $0$ & $0$ & $0$ & $0$ & $\mathbf{100}$ & $0$ & $0$ \\
 \hline
$\widehat{m}_{PP}^\prime$ & $0$ & $0$ & $\mathbf{100}$ & $0$ & $0$ & $0$ & $0$ & $\mathbf{100}$ & $0$ & $0$ \\
 \hline
 \hline
 estimate \textbackslash order of the autoregression & $<4$ & $4$ & $\mathbf{5}$ & $6$ & $>6$ & $<4$ & $4$ & $\mathbf{5}$ & $6$ & $>6$ \\
 \hline
 $\widehat{p}^\prime$  & $0$ & $0$ & $\mathbf{72}$ & $17$ & $11$ & $0$ & $0$ & $\mathbf{83}$ & $12$ & $5$ \\
 \hline
 \hline
 $\widetilde{\phi}_{n,1}^{(5)}$ RMSE & \multicolumn{5}{c||}{$2.99 \cdot 10^{-2}$} & \multicolumn{5}{c|}{$1.77 \cdot 10^{-2}$} \\
 \hline
 $\widetilde{\phi}_{n,2}^{(5)}$ RMSE & \multicolumn{5}{c||}{$1.24 \cdot 10^{-2}$} & \multicolumn{5}{c|}{$1.05 \cdot 10^{-2}$} \\
 \hline
  $\widetilde{\phi}_{n,3}^{(5)}$ RMSE & \multicolumn{5}{c||}{$1.25 \cdot 10^{-2}$} & \multicolumn{5}{c|}{$1.03 \cdot 10^{-2}$} \\
 \hline
 $\widetilde{\phi}_{n,4}^{(5)}$ RMSE & \multicolumn{5}{c||}{$1.28 \cdot 10^{-2}$} & \multicolumn{5}{c|}{$1.01 \cdot 10^{-2}$} \\
 \hline
 $\widetilde{\phi}_{n,5}^{(5)}$ RMSE & \multicolumn{5}{c||}{$1.29 \cdot 10^{-2}$} & \multicolumn{5}{c|}{$9.47 \cdot 10^{-3}$} \\
 \hline
\end{tabular}
\caption[Résultats dans le cas AR(5), $(\phi_1^\star, \phi_2^\star ,\phi_3^\star , \phi_4^\star , \phi_5^\star , \sigma^\star) = (0.5,0,0,0,-0.5,0.4)$.]{Estimates of the number of changes, of the order of the autoregression, and RMSEs of the estimates of the autoregression parameters, for 100 AR(5) series with the parameters $(\phi_1^\star, \phi_2^\star ,\phi_3^\star , \phi_4^\star , \phi_5^\star , \sigma^\star) = (0.5,0,0,0,-0.5,0.4)$.}\label{table:phi1_05_phi2_0_phi3_0_phi4_0_phi5_-05_sigma04}
\end{table}
\begin{figure}[b]
\centering
\includegraphics[width=.55\textwidth]{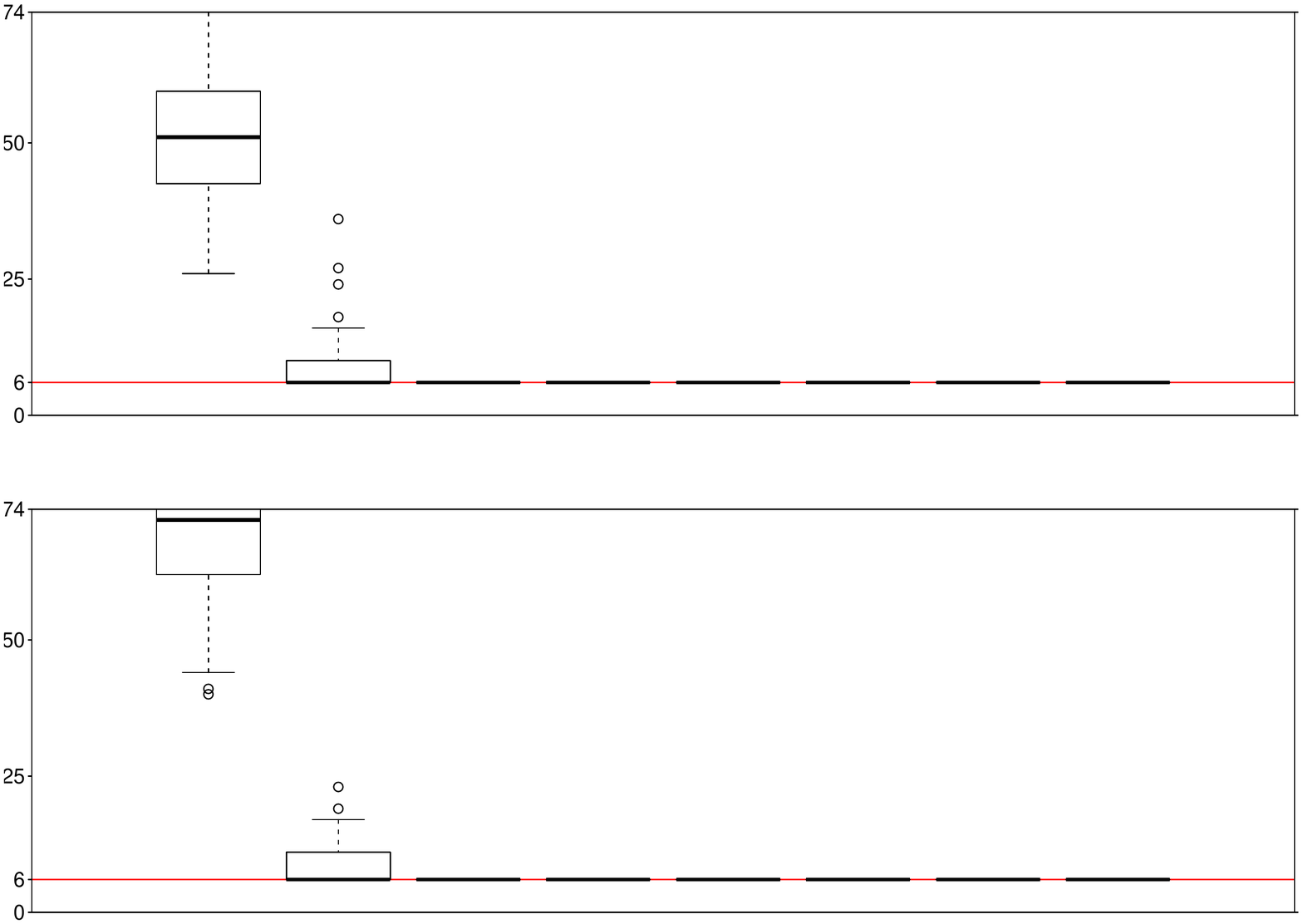}
\caption[Boîtes à moustaches du nombre de ruptures estimé, cas AR(5), $(\phi_1^\star, \phi_2^\star ,\phi_3^\star , \phi_4^\star , \phi_5^\star , \sigma^\star) = (0.5,0,0,0,-0.5,0.4)$.]{Boxplots of the estimates of the number of changes for 100 AR(5) series with the parameters $(\phi_1^\star, \phi_2^\star ,\phi_3^\star , \phi_4^\star , \phi_5^\star , \sigma^\star) = (0.5,0,0,0,-0.5,0.4)$. $n=7200$ (top) or $14400$ (bottom). In each plot, the estimates boxplots are in the following order (from left to right): $\widehat{m}^0_Y $, $\widehat{m}^0 $, $\widehat{m} $, $\widehat{m}_{PP} $, $\widehat{m}^\star $, $\widehat{m}_{PP}^\star $, $\widehat{m}^\prime $, $\widehat{m}_{PP}^\prime $. The true number of changes is equal to 6 (red horizontal line).}\label{fig:box:phi1_05_phi2_0_phi3_0_phi4_0_phi5_-05_sigma04}
\end{figure}
\begin{figure}[h]
\centering
\includegraphics[width=\textwidth]{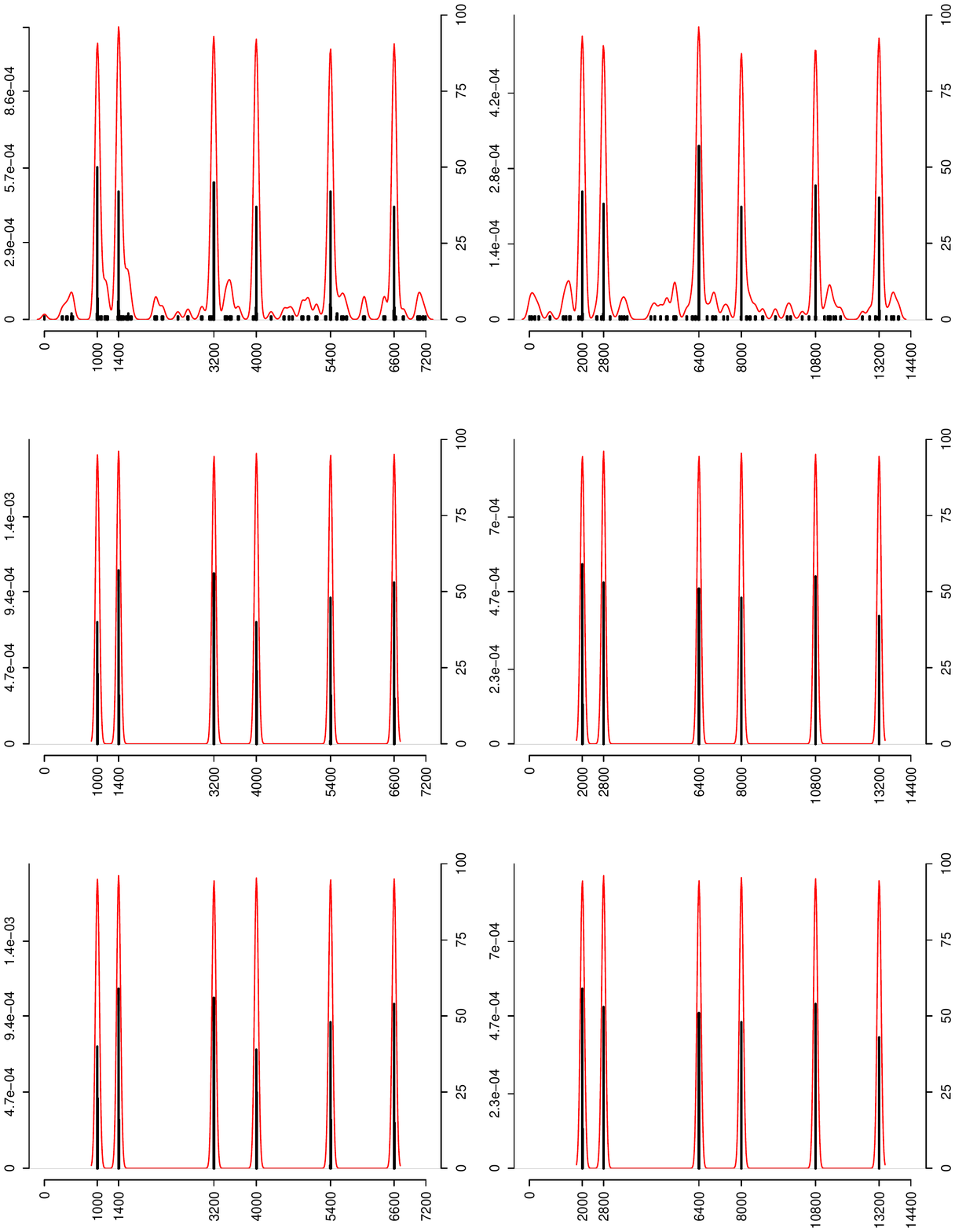}
\caption[Fréquence des instants de rupture estimés, cas AR(5), $(\phi_1^\star, \phi_2^\star ,\phi_3^\star , \phi_4^\star , \phi_5^\star , \sigma^\star) = (0.5,0,0,0,-0.5,0.4)$.]{Frequency plots of the change-point location estimates for 100 AR(5) series with the parameters $(\phi_1^\star, \phi_2^\star ,\phi_3^\star , \phi_4^\star , \phi_5^\star , \sigma^\star) = (0.5,0,0,0,-0.5,0.4)$. $n=7200$ (left) or $14400$ (right). Estimates: $\widehat{t}_n^0$ (top), $\widehat{t}_{n, PP}$ (middle), $\widehat{t}_{n, PP}^\prime$ (bottom). The black line represents the absolute frequency of each location between $1$ and $n$ in estimates (scale on right axis). The red line represents the Gaussian kernel density estimate of this dataset (scale on left axis).}\label{fig:dens:phi1_05_phi2_0_phi3_0_phi4_0_phi5_-05_sigma04}
\end{figure}
\clearpage
\section*{Acknowledgment}
I thank \'Emilie Lebarbier, Céline Lévy-Leduc and Stéphane Robin for comments that greatly improved this paper.
%
%

%% file: chapitres/arp/AppSec2.tex
\subsection{Proof of Proposition \ref{prop:corr_tcl}}\label{arp:proof:corr_tcl}
Since there is only a finite number of atypical values in the process $(x_i)$, 
Theorem 4 of \textcite{levy2011robust} still holds and gives that for all fixed $h\geq 1$:
\begin{equation*}
\sqrt{n-h}\left(\frac{Q_{n}^2 \left(x^+_h\right) - Q_{n}^2 \left(x^-_h\right)}{4}-\gamma(h)\right)
=\frac{1}{\sqrt{n-h}}\sum_{i=1}^{n-h}\psi(\nu_i,\nu_{i+h})+o_P(1)\; ,
\end{equation*}
where $\gamma$ denotes the autocovariance of $(\nu_i)$ and where for all $x$ and $y$,

\begin{multline*}
\psi : (x,y)\mapsto
\\ \left\{(\gamma(0)+\gamma(h)) \; \IF\left(\frac{x+y}{\sqrt{2(\gamma(0)+\gamma(h))}},Q,\Phi\right)
- (\gamma(0)-\gamma(h)) \; \IF\left(\frac{x-y}{\sqrt{2(\gamma(0)-\gamma(h))}},Q,\Phi\right)\right\}\; .
\end{multline*}

In this equation $\IF$ is defined by 

\begin{equation*}
\IF(x,Q,\Phi)
=c(\Phi)\left(\frac{1/4-\Phi(x+1/c(\Phi))
+\Phi(x-1/c(\Phi))}{\int_{\rset} \phi(y)\phi(y+1/c(\Phi))\rmd y}\right)\;,
\end{equation*}

where $c(\Phi)=1/(\sqrt{2}\Phi^{-1}(5/8))\approx 2.21914$, and $\Phi$ is here the cumulative distribution function of the standard normal distribution.
By Theorem 2 of \textcite{levy2011robust}, we obtain that 
$$
\frac{Q_{n}^2 \left(x^+_h\right) + Q_{n}^2 \left(x^-_h\right)}{4}-\gamma(0)=o_p(1)\;.
$$
Let $\widehat{\gamma}(0)=(Q_{n}^2 \left(x^+_h\right) + Q_{n}^2 \left(x^-_h\right))/4$ then

\begin{equation*}
\sqrt{n-h}\left(\widehat{\rho}(h)-\rho(h)\right)
=\frac{\widehat{\gamma}(0)^{-1}}{\sqrt{n-h}}\sum_{i=1}^{n-h}\psi(\nu_i,\nu_{i+h})+o_P(1)\; ,
\end{equation*}

In order to prove a central limit theorem for 
$\boldsymbol{\widehat{\rho}}_{1:(p+1)}$, it is enough to prove
by the Cramér-Wold device \parencite[Theorem 29.4]{billingsley1995probability}, that for any $a_k$ in $\rset$:
\begin{equation*}
\frac{\widehat{\gamma}(0)^{-1}}{\sqrt{n}}\sum_{i=1}^{n-(p+1)}\sum_{k=1}^{p+1} a_k\psi(\nu_i,\nu_{i+k})
\end{equation*}
converges in distribution to a centered Gaussian rv. 
By Lemma 13 of \textcite{levy2011robust},
$\psi$ is of Hermite rank 2. Hence, the Hermite rank of the linear combination of the $\psi$ function is 
larger than 2. Thus, Condition (2.40) of Theorem 4 in \textcite{arcones1994limit} is satisfied
and the quantity $n^{-1/2}\sum_{i=1}^{n-(p+1)}\sum_{k=1}^{p+1} a_k\psi(\nu_i,\nu_{i+k})$ 
converges in distribution to a centered Gaussian rv with variance $\tilde{\sigma}^2$ given by
$$
\tilde{\sigma}^2=\PE\left[\left(\sum_{k=1}^{p+1} a_k\psi(\nu_1,\nu_{k+1})\right)^2\right]+2\sum_{\ell\geq 1}
\PE\left[\left(\sum_{k=1}^{p+1} a_k\psi(\nu_1,\nu_{k+1})\right)\left(\sum_{k=1}^{p+1} a_k\psi(\nu_{\ell+1},\nu_{k+\ell+1})\right)\right]\;.
$$
By Slutsky's lemma,
$
\sqrt{n-h}\left(\widehat{\rho}(h)-\rho(h)\right)
$
converges in distribution to a centered Gaussian rv with variance $\gamma(0)\tilde{\sigma}^2$.
Since $\left(x_i - \mathbb{E}x_i\right)$ is an ARMA($p$,1) process, with autoregressive parameters $\phi_1^\star, \dots ,\phi_p^\star$, 
we get, by \textcite[Chapter 11, Paragraph 2]{azencott}, that $R_p$, as defined in \eqref{eq:matrixR}, is invertible, and
\begin{equation*}
\Phi^\star = R_p^{-1}\boldsymbol{\rho}_{2:(p+1)}^T\;,
\end{equation*}
where $\boldsymbol{\rho}_{2:(p+1)}$ is defined in \eqref{eq:rho}.

Let $g:\mathcal{D}\subset\mathbb{R}^{p+1}\rightarrow\mathbb{R}^p$ defined by
\begin{equation}\label{eq:def_g}
g(u) = \left(u_{|j-i-1|}\mathbf{1}_{j-i-1\neq 0} + \mathbf{1}_{j-i-1=0}\right)_{1\leq i,j\leq p}^{-1}\left(u_2,\dots ,u_{p+1}\right)^T \; .
\end{equation}
Hence, 
\begin{eqnarray*}
\Phi^\star & = & g(\boldsymbol{\rho}_{1:(p+1)}) \; ,\\
\widetilde{\Phi}_n^{(p)} & = & g(\boldsymbol{\widetilde{\rho}_n}_{1:(p+1)})\; .
\end{eqnarray*}
By \eqref{eq:ANrho} and the delta method:
\begin{equation*}
\sqrt{n}\left(\widetilde{\Phi}_n^{(p)} - \Phi^\star\right) \underset{n\rightarrow\infty}{\rightarrow} \mathcal{N}\left(0, \nabla g (\boldsymbol{\rho}_{1:(p+1)})^T V \nabla g (\boldsymbol{\rho}_{1:(p+1)})\right) \; ,
\end{equation*}
$\nabla g (\boldsymbol{\rho}_{1:(p+1)})$ being the Jacobian matrix of $g$ in $\boldsymbol{\rho}_{1:(p+1)}$. Let us determine $\nabla g (u)\cdot h$ for all $u\in\mathcal{D},h\in\mathbb{R}^{p+1}$. Using \eqref{eq:def_g}, we get
\begin{equation*}
\left(h_{|j-i-1|}\mathbf{1}_{j-i-1\neq 0}\right)_{1\leq i,j\leq p}g(u) + \left(u_{|j-i-1|}\mathbf{1}_{j-i-1\neq 0} + \mathbf{1}_{j-i-1=0}\right)_{1\leq i,j\leq p}\nabla g(u)\cdot h = \left( h_2,\dots ,h_{p+1}\right)^T \; .
\end{equation*}
Applied to $u=\boldsymbol{\rho}_{1:(p+1)}$, we have
\begin{equation*}
R_p \nabla g \left(\boldsymbol{\rho}_{1:(p+1)} \right)\cdot h = \left(h_{j+1} - \sum_{j=1}^i \phi_{i+1-j}^\star h_j + \sum_{j=1}^{p-i-1}\phi_{i+j+1}^\star h_j\right)_{1\leq j \leq p} \; ,
\end{equation*}
and then $\nabla g \left(\boldsymbol{\rho}_{1:(p+1)}\right) = R_p^{-1}M$, where $M$ is defined in \eqref{eq:Jacobian}.


%% file: chapitres/arp/AppSec3.tex
\subsection{Proof of Proposition \ref{Prop:Segment}}\label{subsec:prop:Segment}

In the sequel, we need the following definitions, notations and remarks. Observe that \eqref{eq:bkw} can be rewritten as follows:
\begin{equation}\label{eq:modele_matriciel}
z = \sum_{r=1}^p \phi_r^\star B^r z + T\left(\boldsymbol{t}_n^\star\right) \boldsymbol{\delta}^\star + \epsilon\;,
\end{equation}
where
\begin{equation}\label{eq:YXE}
z = \left( \begin{array}{c} z_1 \\\vdots \\z_{n} \end{array}\right)\;,
\qquad
B^r z = \left( \begin{array}{c} z_{1-r} \\ \vdots \\ z_{n-r} \end{array}\right)\;, 
\qquad
\boldsymbol{\delta}^\star = \left( \begin{array}{c} \delta^\star_0 \\ \vdots \\ \delta^\star_{m} \end{array}\right)\;, 
\qquad 
\epsilon = \left( \begin{array}{c} \epsilon_1 \\ \vdots \\ \epsilon_n \end{array}\right)\;,
\end{equation}
where $\delta_k^\star = (1-\sum_{r=1}^p\phi_r^\star) \mu^\star_k$, for $0 \leq k \leq m$,
and $T\left(\boldsymbol{t}\right)$ is an $n \times (m+1)$ matrix where the $k$th column is $( \underset{t_{k-1}}{\underbrace{0,\dots , 0}} \; \underset{t_k-t_{k-1}}{\underbrace{1,\dots ,1}} \; \underset{n - t_k}{\underbrace{0,\dots ,0}} )^T$.

Let us define the exact and estimated decorrelated series by
\begin{eqnarray}\label{eq:decor}
w^\star & = & z - \sum_{r=1}^p\phi_r^\star B^r z\;, \\
\overline{w} & = & z - \sum_{r=1}^p\overline{\phi}_{r,n} B^r z\label{eq:overline_x}\;.
\end{eqnarray}
where $\overline{\Phi}_n = \left(\phi_{r,n}\right)_{1\leq r\leq p}$.

For any vector subspace $E$ of $\mathbb{R}^{n}$, let $\pi_E$ denote the orthogonal projection of $\mathbb{R}^{n}$ on $E$.
Let also $\Vert \cdot \Vert$ be the Euclidean norm on $\mathbb{R}^{n}$, $\langle \cdot , \cdot \rangle$ the canonical scalar product on  $\mathbb{R}^{n}$ and 
$\Vert \cdot \Vert_{\infty}$ the sup norm. 

For $x$ a vector of $\rset^n$ and $\boldsymbol{t}\in\mathcal{A}_{n,m}$, let

\begin{equation}\label{eq:Jnm}
J_{n,m}\left(x,\boldsymbol{t}\right)
= \frac{1}{n} \left( \Vert \pi_{E_{ \boldsymbol{t}_n^\star } }\left( x\right) \Vert^2 - \Vert \pi_{ E_{\boldsymbol{t}}} \left( x \right) \Vert^2 \right)\;,
\end{equation}
written $J_n\left(x,\boldsymbol{t}\right)$ in the sequel for notational simplicity. In \eqref{eq:Jnm}, $E_{\boldsymbol{t}_n^\star}$ and $E_{\boldsymbol{t}}$ correspond to the
linear subspaces of $\mathbb{R}^{n}$ generated by the columns of $T\left(\boldsymbol{t}_n^\star\right)$ and $T\left(\boldsymbol{t}\right)$, respectively. We shall use the same decomposition
as the one introduced in \textcite{LM}:
\begin{equation}\label{eq:Jn_decomp}
J_n\left(x,\boldsymbol{t}\right) =  K_n\left(x,\boldsymbol{t}\right) + V_n\left(x,\boldsymbol{t}\right) + W_n\left(x,\boldsymbol{t}\right)\;,
\end{equation}
where
\begin{eqnarray*}
K_n\left(x,\boldsymbol{t}\right) & = & \frac{1}{n}\left\Vert \left(\pi_{E_{\boldsymbol{t}_n^\star}}- \pi_{E_{\boldsymbol{t}}}\right)\mathbb{E}x\right\Vert^2\;, \\
V_n\left(x,\boldsymbol{t}\right) & = & \frac{1}{n}\left(\left\Vert \pi_{E_{\boldsymbol{t}_n^\star}} \left( x-\mathbb{E}x\right)\right\Vert^2 - \left\Vert \pi_{E_{\boldsymbol{t}}} \left( x-\mathbb{E}x\right)\right\Vert^2\right)\;, \\
W_n\left(x,\boldsymbol{t}\right) & = & \frac{2}{n}\left( \left\langle \pi_{E_{\boldsymbol{t}_n^\star}} \left( x-\mathbb{E}x\right), \pi_{E_{\boldsymbol{t}_n^\star}} \left( \mathbb{E}x \right)\right\rangle - \left\langle \pi_{E_{\boldsymbol{t}}} \left( x-\mathbb{E}x\right), \pi_{E_{\boldsymbol{t}}} \left( \mathbb{E}x \right)\right\rangle  \right)\;.
\end{eqnarray*}

We shall also use the following notations:
\begin{eqnarray}
\underline{\lambda} & = & \underset{1\leq k\leq m}{\min} \left\vert \delta_k^\star-\delta_{k-1}^\star \right\vert\;,\label{eq:underline_lambda}\\
\overline{\lambda} & = & \underset{1\leq k\leq m}{\max} \left\vert \delta_k^\star-\delta_{k-1}^\star \right\vert\;,\label{eq:overline_lambda}\\
\Delta_{\boldsymbol{\tau}^\star} & = & \underset{1\leq k\leq m+1}{\min} \left(\tau_k^\star - \tau_{k-1}^\star\right)\;,\label{eq:Delta_tau*}\\
\mathcal{C}_{\alpha,\gamma,n} & = & \left\lbrace \boldsymbol{t} \in \mathcal{A}_{n,m}; \alpha\underline{\lambda}^{-2}\leq \Vert \boldsymbol{t} - \boldsymbol{t}_n^\star\Vert\leq n\gamma\Delta_{\boldsymbol{\tau}^\star}\right\rbrace\;,\label{eq:C_alpha_gamma_n}\\
\mathcal{C}_{\alpha,\gamma,n}' & = & \mathcal{C}_{\alpha,\gamma,n}\cap \left\lbrace \boldsymbol{t}\in\mathcal{A}_{n,m}; \forall k = 1,\dots, m, t_k\geq t_{n,k}^\star\right\rbrace\;,
\label{eq:C'_alpha_gamma_n} \\
\mathcal{C}_{\alpha,\gamma,n}'\left(\mathcal{I}\right) & = & \left\lbrace \boldsymbol{t}\in \mathcal{C}_{\alpha,\gamma,n}' ; \right. \nonumber\\
 & & \left. \forall k\in\mathcal{I}, \alpha\underline{\lambda}^{-2}\leq t_k - t_{n,k}^\star\leq n\gamma\Delta_{\boldsymbol{\tau}^\star} \textit{ and } \forall k\notin \mathcal{I}, t_k - t_{n,k}^\star < \alpha\underline{\lambda}^{-2}\right\rbrace\label{eq:C'alpha_gamma_n_I}\;,
\end{eqnarray}
for any $\alpha>0$, $0<\gamma <1/2$ and $\mathcal{I}\subset\left\lbrace 1,\dots ,m\right\rbrace$.
We shall also need the following lemmas in order to prove Proposition \ref{Prop:Segment} which are proved below.
\begin{lemma}\label{lem:rateXY}
Let $\left(z_{1-p} ,\dots ,z_n\right)$ be defined by \eqref{eq:modele_new} or \eqref{eq:bkw}, then, for all $r=0,\dots ,p$:
\begin{equation*}
\Vert B^r z \Vert = O_P\left(n^{1/2}\right)\;,
\end{equation*}
as $n$ tends to infinity, where $B^r z$ is defined in \eqref{eq:YXE}.
\end{lemma}

\begin{lemma}\label{lem:BoundedUnifBound}
Let $\left(z_{1-p} ,\dots ,z_n\right)$ be defined by \eqref{eq:modele_new} or \eqref{eq:bkw} then,
for all $\boldsymbol{t}\in\mathcal{A}_{n,m}$,
\begin{equation*}
\left\vert J_n \left(\overline{w}, \boldsymbol{t}\right) - J_n \left(w^\star, \boldsymbol{t}\right)\right\vert \leq \frac{2}{n} \sum_{r=1}^p \left| \phi_r^\star-\overline{\phi}_{r,n}\right|\left\Vert B^r z \right\Vert \left(p\left| \phi_r^\star-\overline{\phi}_{r,n}\right|\left\Vert B^r z \right\Vert + 2 \left\Vert w^\star\right\Vert\right) = O_P \left(n^{-1/2}\right) \;,
\end{equation*}
where $J_n$ is defined in \eqref{eq:Jnm}, $Bz$ and $z$ are defined in \eqref{eq:YXE}, $w^\star$ is defined in \eqref{eq:decor} and $\overline{w}$ is defined in \eqref{eq:overline_x}.
\end{lemma}

\begin{lemma}\label{lem:consistency}
Under the assumptions of Proposition \ref{Prop:Segment},
$\Vert\boldsymbol{\overline{\tau}_n}-\boldsymbol{\tau}^\star\Vert_{\infty}$ converges in probability to $0$, as $n$ tends to infinity.
\end{lemma}

\begin{lemma}\label{lem:almost72LM}
Under the assumptions of Proposition \ref{Prop:Segment} and for any $\alpha>0$, $0<\gamma <1/2$ and $\mathcal{I}\subset\left\lbrace 1,\dots ,m\right\rbrace$, 
\begin{equation*}
P\left(\min_{\boldsymbol{t}\in \mathcal{C}_{\alpha,\gamma,n}'\left(\mathcal{I}\right)} \left(\frac{1}{2} K_n\left(w^\star,\boldsymbol{t}\right) + V_n\left(w^\star,\boldsymbol{t}\right) + W_n\left(w^\star,\boldsymbol{t}\right)\right)\leq 0 \right) \longrightarrow 0\;,\; \textrm{as } n\to\infty\;,
\end{equation*}
where $\mathcal{C}_{\alpha,\gamma,n}'\left(\mathcal{I}\right)$ is defined in \eqref{eq:C'alpha_gamma_n_I} and $w^\star$ is defined in \eqref{eq:decor}.
\end{lemma}

\begin{lemma}\label{lem:almost72LM2}
Under the assumptions of Proposition \ref{Prop:Segment} and for any $\alpha>0$, $0<\gamma <1/2$ and $\mathcal{I}\subset\left\lbrace 1,\dots ,m\right\rbrace$, 
\begin{equation*}
P\left(\min_{\boldsymbol{t}\in \mathcal{C}_{\alpha,\gamma,n}'\left(\mathcal{I}\right)} J_n\left(\overline{w},\boldsymbol{t}\right)\leq 0\right) \longrightarrow 0
\;,\; \textrm{as } n\to\infty\;,
\end{equation*}
where $\mathcal{C}_{\alpha,\gamma,n}'\left(\mathcal{I}\right)$ is defined in \eqref{eq:C'alpha_gamma_n_I} and $\overline{w}$ is defined in \eqref{eq:overline_x}.
\end{lemma}

\begin{lemma}\label{lem:rateT}
Under the assumptions of Proposition \ref{Prop:Segment},
\begin{equation*}
\Vert\boldsymbol{\widehat{\tau}_n}(z, \overline{\Phi}_n)-\boldsymbol{\tau}^\star \Vert_{\infty} = O_P\left(n^{-1}\right)\;.
\end{equation*}
\end{lemma}

\begin{proof}[Proof of Lemma \ref{lem:rateXY}]
Without loss of generality, assume $\left(z_{1-p} ,\dots ,z_n \right)$ is defined by \eqref{eq:bkw}. %
$
\Vert B^r z\Vert^2 = \sum_{i=1-r}^{n-r}z_i^2 
$
then Markov inequality implies that
$
\Vert B^r z\Vert^2=O_P (n).
$
\end{proof}

\begin{proof}[Proof of Lemma \ref{lem:BoundedUnifBound}]
By \eqref{eq:decor}, $\overline{w}=w^\star+\sum_{r=1}^p(\phi_r^\star-\overline{\phi}_{r,n})B^r z$. 
We get
\begin{multline*}
\left\Vert \pi_{E_{\boldsymbol{t}}} (\overline{w})\right\Vert^2 - \left\Vert \pi_{E_{\boldsymbol{t}}} (w^\star)\right\Vert^2 = 
\left\Vert \pi_{E_{\boldsymbol{t}}} (w^\star) + \sum_{r=1}^p(\phi_r^\star-\overline{\phi}_{r,n})\pi_{E_{\boldsymbol{t}}} (B^r z)\right\Vert^2 - \left\Vert \pi_{E_{\boldsymbol{t}}} (w^\star)\right\Vert^2 \\
=  \left\Vert \sum_{r=1}^p(\phi_r^\star-\overline{\phi}_{r,n})\pi_{E_{\boldsymbol{t}}} (B^r z)\right\Vert^2 + 2 \sum_{r=1}^p (\phi_r^\star-\overline{\phi}_{r,n})\left\langle \pi_{E_{\boldsymbol{t}}} (w^\star), \pi_{E_{\boldsymbol{t}}} (B^r z)\right\rangle \\
 \leq \sum_{r=1}^p p(\phi_r^\star-\overline{\phi}_{r,n})^2\left\Vert \pi_{E_{\boldsymbol{t}}} (B^r z)\right\Vert^2 + 2 \sum_{r=1}^p (\phi_r^\star-\overline{\phi}_{r,n})\left\langle \pi_{E_{\boldsymbol{t}}} (w^\star), \pi_{E_{\boldsymbol{t}}} (B^r z)\right\rangle \\
 \leq \sum_{r=1}^p  (\phi_r^\star-\overline{\phi}_{r,n}) \left(p(\phi_r^\star-\overline{\phi}_{r,n}) \left\Vert \pi_{E_{\boldsymbol{t}}} (B^r z)\right\Vert^2 + 2 \left\langle \pi_{E_{\boldsymbol{t}}} (w^\star), \pi_{E_{\boldsymbol{t}}} (B^r z)\right\rangle\right) \\
 \leq \sum_{r=1}^p  (\phi_r^\star-\overline{\phi}_{r,n}) \left\langle \pi_{E_{\boldsymbol{t}}} (B^r z) , p(\phi_r^\star-\overline{\phi}_{r,n}) \pi_{E_{\boldsymbol{t}}} (B^r z) + 2\pi_{E_{\boldsymbol{t}}} (w^\star)\right\rangle  \;.
\end{multline*}

The Cauchy-Schwarz inequality and the $1$-Lipschitz property of projections give
\begin{equation*}
\left|\left\Vert \pi_{E_{\boldsymbol{t}}} (\overline{w})\right\Vert^2 - \left\Vert \pi_{E_{\boldsymbol{t}}} (w^\star)\right\Vert^2 \right| \leq \sum_{r=1}^p \left| \phi_r^\star-\overline{\phi}_{r,n}\right|\left\Vert B^r z \right\Vert \left(p\left| \phi_r^\star-\overline{\phi}_{r,n}\right|\left\Vert B^r z \right\Vert + 2 \left\Vert w^\star\right\Vert\right)
\end{equation*}
The conclusion follows from \eqref{eq:Jnm}, \eqref{eq:hypRhoRate} and Lemma \ref{lem:rateXY}.
\end{proof}

\begin{proof}[Proof of Lemma \ref{lem:consistency}]
\textcite[proof of Theorem 3]{LM} give the following bounds for any $\boldsymbol{t} \in\mathcal{A}_{n,m} $:
\begin{eqnarray}\label{eq:LMbounds}
K_n\left(w^\star,\boldsymbol{t} \right) & \geq & \underline{\lambda}^2 \min\left(\frac{1}{n}\max_{1\leq k \leq m}\left\vert t_k - t_{n,k}^\star\right\vert , \Delta_{\boldsymbol{\tau}^\star}\right)\;,\\
V_n\left(w^\star,\boldsymbol{t}\right) & \geq & -\frac{2\left(m+1\right)}{n\Delta_n}\left(\max_{1\leq s\leq n} \left(\sum_{i=1}^s \epsilon_i\right)^2 + \max_{1\leq s\leq n} \left(\sum_{i=n-s}^{n} \epsilon_i\right)^2\right)\;,\\
\left\vert  W_n\left(w^\star,\boldsymbol{t} \right) \right\vert & \leq & \frac{3\left(m+1\right)^2 \overline{\lambda}}{n} \left(\max_{1\leq s\leq n} \left\vert\sum_{i=1}^s \epsilon_i\right\vert + \max_{1\leq s\leq n} \left\vert \sum_{i=n-s}^{n} \epsilon_i\right\vert \right)\;,
\end{eqnarray}
where $\underline{\lambda}$, $\overline{\lambda}$ anf $\Delta_{\boldsymbol{\tau}^\star}$ are defined in~(\ref{eq:underline_lambda}--\ref{eq:Delta_tau*}).
For any $\alpha>0$, define, as in the proof of Theorem~3 of \textcite{LM},
\begin{equation}\label{eq:Cn_alpha}
\mathcal{C}_{n,\alpha} = \left\lbrace \boldsymbol{t}\in \mathcal{A}_{n,m} ; \left\Vert \boldsymbol{t} - \boldsymbol{t}_{n}^\star\right\Vert_{\infty} \geq n\alpha\right\rbrace \;.
\end{equation}
For $0<\alpha<\Delta_{\boldsymbol{\tau}^\star}$, we have:
\begin{eqnarray*}
P\left(\left\Vert \boldsymbol{\widehat{t}}_n(z, \overline{\Phi}_n) - \boldsymbol{t}_{n}^\star \right\Vert_{\infty} \geq n\alpha\right)  & \leq & P \left( \min_{\boldsymbol{t}\in\mathcal{C}_{n,\alpha}} J_n \left(\overline{w}, \boldsymbol{t} \right) \leq 0\right)\\
 & \leq &P \left( \min_{\boldsymbol{t}\in\mathcal{C}_{n,\alpha}} \left(J_n \left(\overline{w}, \boldsymbol{t} \right) - J_n \left(w^\star, \boldsymbol{t}\right) \right)\leq -\alpha\underline{\lambda}^2 \right)\\
 & + & P \left( \min_{\boldsymbol{t}\in\mathcal{C}_{n,\alpha}} \left(V_n\left(w^\star, \boldsymbol{t}\right) + W_n\left(w^\star, \boldsymbol{t}\right)\right) \leq  -\alpha\underline{\lambda}^2 \right) \\
 & \leq & P \left( \min_{\boldsymbol{t}\in\mathcal{C}_{n,\alpha}} \left(J_n \left(\overline{w}, \boldsymbol{t} \right) - J_n \left(w^\star, \boldsymbol{t} \right)\right)\leq -\alpha\underline{\lambda}^2 \right)\\
 & + & P \left( \max_{1\leq s\leq n}{\left(\sum_{i=1}^s \epsilon_i\right)^2} + \max_{1 \leq s \leq n} \left(\sum_{i=n-s}^{n} \epsilon_i\right)^2 \geq c\underline{\lambda}^2 n\Delta_n \alpha \right)\\
 & + & P \left( \max_{1 \leq s \leq n}{\left\vert\sum_{i=1}^s \epsilon_i\right\vert} + \max_{1 \leq s \leq n} \left\vert\sum_{i=n-s}^{n} \epsilon_i\right\vert \geq c\underline{\lambda}^2 n \alpha \overline{\lambda}^{-1} \right)
\end{eqnarray*}
for some positive constant $c$. The last two terms of this sum go to $0$ when $n$ goes to infinity~\parencite[proof of Theorem 3]{LM}. To show that the first term shares the same property, it suffices to show that $J_n \left(\overline{w}, \boldsymbol{t}\right) - J_n\left(w^\star,\boldsymbol{t}\right)$ is bounded uniformly in $\boldsymbol{t}$ by a sequence of rv's which converges to $0$ in probability. This result holds by Lemma \ref{lem:BoundedUnifBound}.
\end{proof}

\begin{proof}[Proof of Lemma \ref{lem:almost72LM}]
Using \textcite[Equations (64--66)]{LM}, one can show the bound (73) of \textcite{LM} on 
$$P\left(\min_{\boldsymbol{t}\in \mathcal{C}_{\alpha,\gamma,n}'\left(\mathcal{I}\right)} \left(K_n\left(w^\star,\boldsymbol{t}\right) + V_n\left(w^\star,\boldsymbol{t}\right) + W_n\left(w^\star,\boldsymbol{t}\right)\right)\leq 0 \right).$$
 Using the same arguments, we have the same bound on 
$$P\left(\min_{\boldsymbol{t} \in \mathcal{C}_{\alpha,\gamma,n}'\left(\mathcal{I}\right)} \left(\frac{1}{2} K_n\left(w^\star,\boldsymbol{t} \right) + V_n\left(w^\star,\boldsymbol{t} \right) + W_n\left(w^\star,\boldsymbol{t} \right)\right)\leq 0 \right).$$
We conclude using \textcite[Equations (67--71)]{LM}.
\end{proof}

\begin{proof}[Proof of Lemma \ref{lem:almost72LM2}]
By (\ref{eq:Jn_decomp}),
\begin{multline*}
P\left(\min_{\boldsymbol{t}\in \mathcal{C}_{\alpha,\gamma,n}'\left(\mathcal{I}\right)} J_n\left(\overline{w},\boldsymbol{t}\right)\leq 0\right)
\leq P\left(\min_{\boldsymbol{t}\in \mathcal{C}_{\alpha,\gamma,n}'\left(\mathcal{I}\right)} \left(J_n\left(\overline{w},\boldsymbol{t}\right) - J_n\left(w^\star,\boldsymbol{t}\right) + \frac{1}{2} K_n\left(w^\star,\boldsymbol{t}\right)\right)\leq 0\right)\\
 +  P\left(\min_{\boldsymbol{t}\in \mathcal{C}_{\alpha,\gamma,n}'\left(\mathcal{I}\right)} \left(\frac{1}{2} K_n\left(w^\star,\boldsymbol{t}\right) + V_n\left(w^\star,\boldsymbol{t}\right) + W_n\left(w^\star,\boldsymbol{t}\right)\right)\leq 0 \right)\;.
\end{multline*}
By Lemma \ref{lem:almost72LM}, the conclusion thus follows if
\begin{equation*}
P \left( \min_{ \boldsymbol{t} \in \mathcal{C}_{\alpha,\gamma,n}' \left(\mathcal{I}\right) } 
\left(J_n \left( \overline{w}, \boldsymbol{t} \right) 
- J_n \left( w^\star, \boldsymbol{t} \right) 
+ \frac{1}{2} K_n \left( w^\star , \boldsymbol{t} \right)\right) \leq 0 \right ) 
 \longrightarrow 0\;, \textrm{ as } n\to\infty\;.
\end{equation*}
Since $\underset{\boldsymbol{t}\in \mathcal{C}_{\alpha,\gamma,n}'\left(\mathcal{I}\right)}{\min} K_n\left(w^\star,\boldsymbol{t}\right) \geq \left(1-\gamma\right)\Delta_{\boldsymbol{\tau}^\star}\alpha$ \parencite[Equation (65)]{LM}, 
$$P\left( \min_{\boldsymbol{t}\in \mathcal{C}_{\alpha,\gamma,n}'\left(\mathcal{I}\right)} \left( J_n\left(\overline{w},\boldsymbol{t}\right)-J_n\left(w^\star,\boldsymbol{t}\right)+\frac{1}{2} K_n\left(w^\star,\boldsymbol{t}\right)\right)\leq 0\right)$$
$$ \leq P\left(\min_{\boldsymbol{t}\in \mathcal{C}_{\alpha,\gamma,n}'\left(\mathcal{I}\right)} \left(J_n\left(\overline{w},\boldsymbol{t} \right)-J_n\left(w^\star,\boldsymbol{t}\right)\right) \leq \frac{1}{2} \left(\gamma-1\right)\Delta_{\boldsymbol{\tau}^\star}\alpha\right)\;, $$
 and we conclude by Lemma \ref{lem:BoundedUnifBound}.
\end{proof}

\begin{proof}[Proof of Lemma \ref{lem:rateT}]For notational simplicity, $\boldsymbol{\widehat{t}}_n(z, \overline{\Phi}_n)$ will be replaced by $\boldsymbol{\overline{t}}_n$ in this proof.
Since for any $\alpha>0$,
$$
P\left(\Vert  \boldsymbol{\overline{t}}_n - \boldsymbol{t}_n^\star\Vert_\infty< \alpha \underline{\lambda}^{-2}\right)
=P( \Vert\boldsymbol{\overline{t}}_n - \boldsymbol{t}_n^\star \Vert_\infty\leq n \gamma \Delta_{\boldsymbol{\tau}^\star} ) - P( \boldsymbol{\overline{t}}_n \in C_{\alpha , \gamma ,n })\;,
$$
it is enough, by Lemma \ref{lem:consistency}, to prove that
\begin{equation*}
P\left(\boldsymbol{\overline{t}}_n \in \mathcal{C}_{\alpha,\gamma,n}\right) \longrightarrow 0\;,\; \textrm{as } n\to\infty\;,
\end{equation*}
for all $\alpha>0$ and $0<\gamma<1/2 $.
Since
$  \mathcal{C}_{\alpha,\gamma,n} = \underset{\mathcal{I}\subset\lbrace 1,\dots ,m\rbrace}{\bigcup} \mathcal{C}_{\alpha,\gamma,n}\cap \left\lbrace \boldsymbol{t}\in\mathcal{A}_{n,m}; \forall k \in \mathcal{I}, t_k\geq t_{n,k}^\star\right\rbrace $, we shall only study one set in the union without loss of generality and prove that
\begin{equation*}
P\left(\boldsymbol{\overline{t}}_n \in \mathcal{C}_{\alpha,\gamma,n}'\right) \longrightarrow 0\;,\; \textrm{as } n\to\infty\;,
\end{equation*}
where $\mathcal{C}_{\alpha,\gamma,n}'$ is defined in (\ref{eq:C'_alpha_gamma_n}). Since
$\mathcal{C}_{\alpha,\gamma,n}' = \underset{\mathcal{I}\subset\lbrace 1,\dots ,m\rbrace}{\bigcup} \mathcal{C}_{\alpha,\gamma,n}'\left(\mathcal{I}\right)$,
we shall only study one set in the union without loss of generality and prove that
\begin{equation*}
P\left(\boldsymbol{\overline{t}}_n \in \mathcal{C}_{\alpha,\gamma,n}'\left(\mathcal{I}\right)\right) \longrightarrow 0\;,\; \textrm{as } n\to\infty\;.
\end{equation*}
Since 
$$ P\left(\boldsymbol{\overline{t}}_n \in \mathcal{C}_{\alpha,\gamma,n}'\left(\mathcal{I}\right)\right) \leq P\left(\min_{\boldsymbol{t} \in \mathcal{C}_{\alpha,\gamma,n}'\left(\mathcal{I}\right)} J_n\left(\overline{w},\boldsymbol{t} \right)\leq 0\right)\;,$$
the proof is complete by Lemma \ref{lem:almost72LM2}.

\end{proof}

\begin{proof}[Proof of Proposition \ref{Prop:Segment}]For notational simplicity, $\boldsymbol{\widehat{\delta}}_n(z, \overline{\Phi}_n)$ will be replaced by $\boldsymbol{\overline{\delta}}_n$ in this proof.
By Lemma \ref{lem:rateT}, the last result to show is
\begin{equation*}
\Vert \boldsymbol{\overline{\delta}_n}-\boldsymbol{\delta}^\star\Vert = O_P\left(n^{-1/2}\right)\;,
\end{equation*}
that is, for all $k$,
$
\overline{\delta}_{n,k} - \delta_k^\star = O_P\left(n^{-1/2}\right).
$
By \eqref{eq:decor} and \eqref{eq:overline_x},
\begin{eqnarray*}
\overline{\delta}_{n,k} & = & \frac{1}{\overline{t}_{n,k+1}-\overline{t}_{n,k}} \sum_{i=\overline{t}_{n,k}+1}^{\overline{t}_{n,k+1}}\overline{w}_i
  =  \frac{1}{n\left(\overline{\tau}_{n,k+1}-\overline{\tau}_{n,k}\right)}\left( \sum_{i=\overline{t}_{n,k}+1}^{\overline{t}_{n,k+1}}w_i^\star + \sum_{r=1}^p \left( \phi_r^\star - \overline{\phi}_{r,n}\right) \sum_{i=\overline{t}_{n,k}+1}^{\overline{t}_{n,k+1}} z_{i-r} \right)\;.
\end{eqnarray*}
By the Cauchy-Schwarz inequality,
\begin{eqnarray*}
\left\vert \sum_{i=\overline{t}_{n,k}+1}^{\overline{t}_{n,k+1}}z_{i-r} \right\vert & \leq & \left(\overline{t}_{n,k+1}-\overline{t}_{n,k}\right)^{1/2}
\left(z_{\overline{t}_{n,k}+1-r}^2+ \dots +z_{\overline{t}_{n,k+1}-r}^2\right)^{1/2} 
  \leq  n^{1/2} \left\Vert Bz \right\Vert
  =  O_P\left(n\right)\;,
\end{eqnarray*}
where the last equality comes from Lemma \ref{lem:rateXY}. Hence by \eqref{eq:hypRhoRate} and Lemma \ref{lem:rateT},
\begin{eqnarray*}
\overline{\delta}_{n,k} & = & \frac{1}{n\left(\overline{\tau}_{n,k+1}-\overline{\tau}_{n,k}\right)} \sum_{i=\overline{t}_{n,k}+1}^{\overline{t}_{n,k+1}}w_i^\star + O_P\left(n^{-1/2}\right)\\
 & = & \frac{1}{n\left(\overline{\tau}_{n,k+1}-\overline{\tau}_{n,k}\right)} \left(\sum_{i=\overline{t}_{n,k}+1}^{\overline{t}_{n,k+1}}\mathbb{E} w_i^\star + \sum_{i=\overline{t}_{n,k}+1}^{\overline{t}_{n,k+1}} \epsilon_i\right) + O_P\left(n^{-1/2}\right),
\end{eqnarray*}
where the last equality comes from (\ref{eq:modele_matriciel}) and (\ref{eq:decor}).

Let us now prove that 
\begin{equation}\label{eq:clt_aleatoire}
\frac{1}{n\left(\overline{\tau}_{n,k+1}-\overline{\tau}_{n,k}\right)} \sum_{i=\overline{t}_{n,k}+1}^{\overline{t}_{n,k+1}} \epsilon_i = O_P\left(n^{-1/2}\right)\;.
\end{equation}
By Lemma \ref{lem:consistency}, $n^{-1}\left(\overline{\tau}_{n,k+1}-\overline{\tau}_{n,k}\right)^{-1}=O_P(n^{-1})$. Moreover,
\begin{equation}\label{eq:eps_decomp}
\sum_{i=\overline{t}_{n,k}+1}^{\overline{t}_{n,k+1}} \epsilon_i = \sum_{i=t_{n,k}^\star+1}^{t_{n,k+1}^\star} \epsilon_i \pm \sum_{i=\overline{t}_{n,k}+1}^{t_{n,k}^\star} \epsilon_i \pm \sum_{i=t_{n,k+1}^\star}^{\overline{t}_{n,k+1}+1} \epsilon_i\;.
\end{equation}
By the Central limit theorem, the first term in the right-hand side of (\ref{eq:eps_decomp}) is $O_P(n^{1/2})$. By using the Cauchy-Schwarz inequality, we get that 
the second term of (\ref{eq:eps_decomp}) satisfies:
$|\sum_{i=\overline{t}_{n,k}+1}^{t_{n,k}^\star} \epsilon_i|\leq |t_{n,k}^\star-\overline{t}_{n,k}|^{1/2} \left(\sum_{i=1}^n  \epsilon_i^2\right)^{1/2}=O_P(1) O_P(n^{1/2})=O_P(n^{1/2})$, by Lemma 
\ref{lem:rateT}. The same holds for the last term in the right-hand side of \eqref{eq:eps_decomp}, which gives \eqref{eq:clt_aleatoire}.

Hence,
\begin{eqnarray*}
\overline{\delta}_{n,k} - \delta_k^\star & = & \frac{1}{n\left(\overline{\tau}_{n,k+1}-\overline{\tau}_{n,k}\right)} \sum_{i=\overline{t}_{n,k}+1}^{\overline{t}_{n,k+1}}\left(\mathbb{E} w_i^\star - \delta_k^\star\right) + O_P\left(n^{-1/2}\right)\\
 & = &  \frac{1}{n\left(\overline{\tau}_{n,k+1}-\overline{\tau}_{n,k}\right)} \sum_{i\in\left\lbrace\overline{t}_{n,k}+1,\dots ,\overline{t}_{n,k+1}\right\rbrace \setminus \left\lbrace t_{n,k}^\star+1,\dots ,t_{n,k+1}^\star\right\rbrace}\left(\mathbb{E} w_i^\star - \delta_k^\star\right) + O_P\left(n^{-1/2}\right)\;,
\end{eqnarray*}
and then
\begin{eqnarray*}
\left\vert \overline{\delta}_{n,k} - \delta_k^\star \right\vert & \leq & \frac{1}{n\left(\overline{\tau}_{n,k+1}-\overline{\tau}_{n,k}\right)} \sharp\left\lbrace\overline{t}_{n,k}+1,\dots ,\overline{t}_{n,k+1}\right\rbrace \setminus \left\lbrace t_{n,k}^\star+1,\dots ,t_{n,k+1}^\star\right\rbrace \max_{l= 0,\dots,m}\left\vert \delta_l^\star - \delta_k^\star\right\vert \\
 & + & O_P\left(n^{-1/2}\right)\;.
\end{eqnarray*}

We conclude by using Lemma \ref{lem:rateT} to get $\sharp\left\lbrace\overline{t}_{n,k}+1,\dots ,\overline{t}_{n,k+1}\right\rbrace \setminus \left\lbrace t_{n,k}^\star+1,\dots ,t_{n,k+1}^\star\right\rbrace = O_P\left(1\right)$ and Lemma \ref{lem:consistency} to get $\left(\overline{\tau}_{n,k+1}-\overline{\tau}_{n,k}\right)^{-1} = O_P\left(1\right)$.
\end{proof}

\subsection{Proof of Proposition \ref{Prop:Segment2}}\label{subsec:prop:Segment2}

The connection between Models~\eqref{eq:modele_new} and~\eqref{eq:bkw} is made by the following lemmas.

\begin{lemma} \label{Lem:YZ}
 Let $(y_0, \dots y_n)$ be defined by \eqref{eq:modele_new} and let
 \begin{eqnarray}
 v^\star_i & = & y_i - \sum_{r=1}^p \phi_r^\star y_{i-1}\label{eq:v_star},\\
 \Delta^\star_i & = & \begin{cases}
-\left(\mu^\star_k  - \mu^\star_{k-1}\right)\sum_{s=r}^p\phi_s^\star \textit{ if } i = t_{n,k}^\star+r \textit{ and } 1\leq r \leq p\\
0, \textit{ otherwise, }
\end{cases} \label{eq:delta_star}
 \end{eqnarray} where the $\mu_k^\star$'s are defined in \eqref{eq:modele_new}, then the process 
\begin{equation}\label{eq:vw_star}
 w^\star_i = v^\star_i + \Delta^\star_i
\end{equation}
equals $z_i - \sum_{r=1}^p z_{i-r}$ where $(z_{1-p}, \dots , z_n)$ verify \eqref{eq:bkw}. Such a process $(z_{1-p}, \dots , z_n)$ can be constructed recursively as

\begin{equation}\label{eq:z_rec}
\begin{cases}
z_i & =  y_i \textit{ for } 1-p\leq i\leq 0\\
z_i & =  w^\star_i + \sum_{r=1}^p \phi_r^\star z_{i-r} \textit{ for } i>0.
\end{cases}
\end{equation}
\end{lemma}

\begin{lemma}\label{Lem:YZbar}
 Let $(y_{1-p}, \dots , y_n)$ be defined by \eqref{eq:modele_new} and let $z$ be defined by (\ref{eq:v_star}-- \ref{eq:z_rec}). Then
\begin{equation}\label{eq:vw}
\overline{w}_i = \overline{v}_i + \overline{\Delta}_i
\end{equation}
where
\begin{eqnarray}
\overline{v}_i & = & y_i - \sum_{r=1}^p\overline{\phi}_{r,n} y_{i-r}\label{eq:v_bar}\\
\overline{w}_i & = & z_i - \sum_{r=1}^p\overline{\phi}_{r,n} z_{i-r}\label{eq:w_bar}\\
\overline{\Delta}_i & = & \Delta^\star_i + \sum_{r=1}^p\left(\phi_r^\star - \overline{\phi}_{r,n}\right)\left(z_{i-r}-y_{i-r}\right)\; .\label{eq:Delta_bar}
\end{eqnarray}
\end{lemma}

\begin{lemma} \label{Lem:Delta_order}
 Let $\overline{\Delta}=(\overline{\Delta}_i)_{0\leq i\leq n}$ as defined in \eqref{eq:Delta_bar}. 
Then $\left\Vert\overline{\Delta}\right\Vert = O_P\left(1\right)$.
\end{lemma}

\begin{proof}[Proof of Lemma \ref{Lem:YZ}.]
Let $z$ being defined by \eqref{eq:z_rec}. Using \eqref{eq:vw_star}, we get, for all $0\leq k \leq m , t_{n,k}^\star<i\leq t_{n,k+1}^\star$
\begin{multline*}
\left(z_i - \mu_k^\star \right) - \sum_{r=1}^p\phi_r^\star \left(z_{i-r} - \mu_k^\star\right) = \left(y_i - \mu_k^\star\right) - \sum_{r=1}^p\phi_r^\star \left(y_{i-r} - \mu_k^\star\right) + \Delta_i^\star \\
 = \left(y_i - \mu_k^\star\right) - \sum_{r=1}^{p} \phi_r^\star \left(y_{i-1} - (\mu_{k-1}^\star \mathbf{1}_{r\geq i-t_{n,k}^\star} + \mu_k^\star \mathbf{1}_{r< i-t_{n,k}^\star})\right)
\end{multline*}
This expression equals $\left(y_i - \PE \left(y_i\right)\right) - \sum_{r=1}^p\phi_r^\star \left(y_{i-r} - \PE\left(y_{i-r}\right)\right) = \eta_i - \sum_{r=1}^p \phi_r^\star \eta_{i-r} = \epsilon_i$ by \eqref{eq:modele_new} and \eqref{eq:arp}. Then $z$ satisfies \eqref{eq:bkw}.
\end{proof}
The proof of Lemma \ref{Lem:YZbar} is straightforward.
\begin{proof}[Proof of Lemma \ref{Lem:Delta_order}]
\eqref{eq:Delta_bar} can be written as
$$
\overline{\Delta} = \Delta^\star + \sum_{r=1}^p\left(\phi_r^\star - \overline{\phi}_{r,n}\right)\left(B^r y-B^r z\right)
$$
where $\Delta^\star=\left(\Delta_i^\star\right)_{1\leq i\leq n}$, $B^r y = \left(y_{i-r}\right)_{1\leq i\leq n}$  and $B^r z$ is defined in \eqref{eq:YXE}. By the triangle inequality,
\begin{equation}\label{eq:triangular_Delta_bar}
\left\Vert\overline{\Delta}\right\Vert \leq \left\Vert\Delta^\star\right\Vert + \sum_{r=1}^p \left\vert\phi_r^\star - \overline{\phi}_{r,n} \right\vert \left(\left\Vert B^r y\right\Vert+\left\Vert B^r z\right\Vert\right).
\end{equation}
Since $\left\Vert\Delta^\star\right\Vert$ is constant in $n$ it is bounded.
The conclusion follows from \eqref{eq:triangular_Delta_bar}, \eqref{eq:hypRhoRate} and Lemma \ref{lem:rateXY}.
\end{proof}

\begin{proof}[Proof of Proposition \ref{Prop:Segment2}]
Let $y$, $z$, $\overline{v}$, $\overline{w}$ and $\overline{\Delta}$ be defined in Lemma \ref{Lem:YZbar}.

Using \eqref{eq:Jnm} and Lemma \ref{Lem:YZbar}, we get
\begin{equation*}
J_n\left(\overline{v},\boldsymbol{t}\right) = J_n\left(\overline{w},\boldsymbol{t}\right) + J_n\left(\overline{\Delta},\boldsymbol{t}\right) - \frac{2}{n}\left(\left\langle \pi_{E_{\boldsymbol{t}_n^\star}}\left(\overline{w}\right),\pi_{E_{\boldsymbol{t}_n^\star}}\left(\overline{\Delta}\right)\right\rangle - \left\langle \pi_{E_{\boldsymbol{t}}}\left(\overline{w}\right),\pi_{E_{\boldsymbol{t}}}\left(\overline{\Delta}\right)\right\rangle\right).
\end{equation*}
By the Cauchy-Schwarz inequality and the $1$-Lipschitz property of projections, we have
\begin{eqnarray*}
\left\vert J_n\left(\overline{\Delta},\boldsymbol{t} \right) \right\vert & \leq & 2 \Vert \overline{\Delta} \Vert^2 ,\\
\left\vert \left\langle \pi_{E_{\boldsymbol{t}_n^\star}}\left(\overline{w}\right),\pi_{E_{\boldsymbol{t}_n^\star}}\left(\overline{\Delta}\right)\right\rangle - \left\langle \pi_{E_{\boldsymbol{t}}}\left(\overline{w}\right),\pi_{E_{\boldsymbol{t}}}\left(\overline{\Delta}\right)\right\rangle \right\vert & \leq & 2 \Vert \overline{\Delta} \Vert \Vert \overline{w} \Vert. 
\end{eqnarray*}
Note that $\overline{w} = z - \sum_{r=1}^p \overline{\phi}_{r,n} B^r z$ thus by the triangle inequality

\begin{equation*}
\Vert \overline{w}\Vert \leq \Vert z \Vert + \sum_{r=1}^p \vert \overline{\phi}_{r,n} \vert \Vert B^r z \Vert \; .
\end{equation*}

Since $\vert \overline{\phi}_{r,n} \vert = O_P\left(1\right)$ for all $1\leq r\leq p$, we deduce from Lemma \ref{lem:rateXY} that $\Vert \overline{w}\Vert=O_P \left(n^{1/2}\right)$. Since, by Lemma \ref{Lem:Delta_order}, $\Vert \overline{\Delta} \Vert = O_P\left(1\right)$, we
obtain that

\begin{equation}\label{eq:controleunif}
\sup_{\boldsymbol{t}} \left\lbrace J_n\left(\overline{\Delta},\boldsymbol{t}\right) - \frac{2}{n}\left(\left\langle \pi_{E_{\boldsymbol{t}_n^\star}}\left(\overline{w}\right),\pi_{E_{\boldsymbol{t}_n^\star}}\left(\overline{\Delta}\right)\right\rangle - \left\langle \pi_{E_{\boldsymbol{t}}}\left(\overline{w}\right),\pi_{E_{\boldsymbol{t}}}\left(\overline{\Delta}\right)\right\rangle\right) \right\rbrace =O_P\left(n^{-1/2}\right).
\end{equation}
For $0<\alpha<\Delta_{\boldsymbol{\tau}^\star}$, using \eqref{eq:Jn_decomp} and \eqref{eq:Cn_alpha}, we get:
\begin{eqnarray*}
P\left(\left\Vert \boldsymbol{\overline{t}}_n - \boldsymbol{t}^\star \right\Vert_{\infty} \geq \alpha\right)  & \leq & P \left( \min_{\boldsymbol{t}\in\mathcal{C}_{n,\alpha}} J_n \left(\overline{v}, \boldsymbol{t} \right) \leq 0\right)\\
  & \leq & P \left( \min_{\boldsymbol{t} \in\mathcal{C}_{n,\alpha}} \left\lbrace J_n\left(\overline{w},\boldsymbol{t}\right) + J_n\left(\overline{\Delta},\boldsymbol{t}\right)  \right.\right. \\
  & & \ \ - \left.\left. \frac{2}{n}\left(\left\langle \pi_{E_{\boldsymbol{t}_n^\star}}\left(\overline{w}\right),\pi_{E_{\boldsymbol{t}_n^\star}}\left(\overline{\Delta}\right)\right\rangle - \left\langle \pi_{E_{\boldsymbol{t}}}\left(\overline{w}\right),\pi_{E_{\boldsymbol{t}}}\left(\overline{\Delta}\right)\right\rangle\right)\right\rbrace \leq 0 \right)\\
 & \leq & P \left( \min_{\boldsymbol{t}\in\mathcal{C}_{n,\alpha}} \left\lbrace K_n\left(\overline{w},\boldsymbol{t}\right) + V_n\left(\overline{w},\boldsymbol{t}\right) + W_n\left(\overline{w},\boldsymbol{t}\right)+ J_n\left(\overline{\Delta},\boldsymbol{t}\right) \right. \right. \\
  & & \ \ - \left. \left. \frac{2}{n}\left(\left\langle \pi_{E_{\boldsymbol{t}_n^\star}}\left(\overline{w}\right),\pi_{E_{\boldsymbol{t}_n^\star}}\left(\overline{\Delta}\right)\right\rangle - \left\langle \pi_{E_{\boldsymbol{t}}}\left(\overline{w}\right),\pi_{E_{\boldsymbol{t}}}\left(\overline{\Delta}\right)\right\rangle\right)\right\rbrace \leq 0 \right)\\
  & \leq & P \left( \min_{\boldsymbol{t}\in\mathcal{C}_{n,\alpha}} \left\lbrace \frac{1}{2}K_n\left(\overline{w},\boldsymbol{t}\right) + V_n\left(\overline{w},\boldsymbol{t}\right) + W_n\left(\overline{w},\boldsymbol{t}\right)\right\rbrace \leq 0 \right)\\
   & & + P \left( \min_{\boldsymbol{t}\in\mathcal{C}_{n,\alpha}} \left\lbrace \frac{1}{2}K_n\left(\overline{w},\boldsymbol{t}\right) + J_n\left(\overline{\Delta},\boldsymbol{t}\right) \right. \right. \\
   & & \ \ - \left. \left. \frac{2}{n}\left(\left\langle \pi_{E_{\boldsymbol{t}_n^\star}}\left(\overline{w}\right),\pi_{E_{\boldsymbol{t}_n^\star}}\left(\overline{\Delta}\right)\right\rangle - \left\langle \pi_{E_{\boldsymbol{t}}}\left(\overline{w}\right),\pi_{E_{\boldsymbol{t}}}\left(\overline{\Delta}\right)\right\rangle\right)\right\rbrace \leq 0 \right).
\end{eqnarray*}
Following the proof of Lemma \ref{lem:consistency}, one can prove that $$P \left( \underset{\boldsymbol{t}\in\mathcal{C}_{n,\alpha}}{\min} \left\lbrace\frac{1}{2}K_n\left(\overline{w},\boldsymbol{t}\right) + V_n\left(\overline{w},\boldsymbol{t}\right) + W_n\left(\overline{w},\boldsymbol{t}\right)\right\rbrace \leq 0\right) \underset{n\rightarrow\infty}{\longrightarrow} 0 \; .$$
Using \eqref{eq:LMbounds}, we get that
\begin{eqnarray*}
P \left( \min_{\boldsymbol{t}\in\mathcal{C}_{n,\alpha}} \left\lbrace\frac{1}{2}K_n\left(\overline{w},\boldsymbol{t}\right) +  J_n\left(\overline{\Delta},\boldsymbol{t}\right) - \frac{2}{n}\left(\left\langle \pi_{E_{\boldsymbol{t}_n^\star}}\left(\overline{w}\right),\pi_{E_{\boldsymbol{t}_n^\star}}\left(\overline{\Delta}\right)\right\rangle - \left\langle \pi_{E_{\boldsymbol{t}}}\left(\overline{w}\right),\pi_{E_{\boldsymbol{t}}}\left(\overline{\Delta}\right)\right\rangle\right)\right\rbrace \leq 0\right) & & \\
 \leq P \left( \frac{1}{2}\underline{\lambda}^2 \alpha + \min_{\boldsymbol{t}\in\mathcal{C}_{n,\alpha}}  \left\{ J_n\left(\overline{\Delta},\boldsymbol{t}\right) - \frac{2}{n}\left(\left\langle \pi_{E_{\boldsymbol{t}_n^\star}}\left(\overline{w}\right),\pi_{E_{\boldsymbol{t}_n^\star}}\left(\overline{\Delta}\right)\right\rangle - \left\langle \pi_{E_{\boldsymbol{t}}}\left(\overline{w}\right),\pi_{E_{\boldsymbol{t}}}\left(\overline{\Delta}\right)\right\rangle\right)\right\} \leq 0\right) & & \\
\end{eqnarray*}
which goes to zero when $n$ goes to infinity by \eqref{eq:controleunif}.

Then Lemma \ref{lem:consistency} still holds if $y$ is defined by \eqref{eq:modele_new}.
To show the rate of convergence, we use the same decomposition. As in the proof of Lemma \ref{lem:rateT}, $P\left(\underset{\boldsymbol{t} \in \mathcal{C}_{\alpha,\gamma,n}'\left(\mathcal{I}\right)}{\min} J_n\left(\overline{v},\boldsymbol{t}\right)\leq 0\right)\underset{n\rightarrow \infty}{\longrightarrow} 0$ for all $\alpha>0$ and $0<\gamma <1/2$ is a sufficient condition 
for proving that $P\left(\boldsymbol{\widehat{t}}_n(y,\overline{\rho}_n)\in \mathcal{C}_{\alpha,\gamma,n}\right) \longrightarrow_{n\rightarrow\infty} 0$, which allows us to conclude on the rate of convergence of the estimated change-points. Note that
\begin{eqnarray*}
P\left(\min_{\boldsymbol{t} \in \mathcal{C}_{ \alpha, \gamma, n}'\left(\mathcal{I}\right)} J_n\left(\overline{v},\boldsymbol{t}\right) \leq 0\right)  & \leq & P \left( \min_{\boldsymbol{t} \in \mathcal{C}_{\alpha,\gamma,n}'} 
\left\lbrace\frac{1}{2}K_n\left( \overline{w}, \boldsymbol{t} \right) + V_n \left( \overline{w}, \boldsymbol{t}\right) + W_n \left( \overline{w}, \boldsymbol{t} \right)\right\rbrace \leq 0\right)\\
   & + & P \left( \frac{1}{2}\underline{\lambda}^2 \alpha + J_n\left(\overline{\Delta},\boldsymbol{t}\right) \right. \\
   & - & \left. \frac{2}{n}\left(\left\langle \pi_{E_{\boldsymbol{t}_n^\star}}\left(\overline{w}\right),\pi_{E_{\boldsymbol{t}_n^\star}}\left(\overline{\Delta}\right)\right\rangle - \left\langle \pi_{E_{\boldsymbol{t}}}\left(\overline{w}\right),\pi_{E_{\boldsymbol{t}}}\left(\overline{\Delta}\right)\right\rangle\right) \leq 0 \right).
\end{eqnarray*}
In the latter equation, the second term of the right-hand side goes to zero as $n$ goes to infinity by \eqref{eq:controleunif}.
The first term of the right-hand side goes to zero when $n$ goes to infinity by following the same line of reasoning as the one of Lemma \ref{lem:almost72LM2}. This concludes the proof of Proposition \ref{Prop:Segment2}.
\end{proof}
